\documentclass[10pt,english,journal]{IEEEtran}
\usepackage[T1]{fontenc}
\usepackage[latin9]{inputenc}
\usepackage{geometry}
\usepackage{booktabs}
\usepackage[table,xcdraw]{xcolor}
\usepackage{longtable}
\usepackage{colortbl}
\usepackage{multirow}
\usepackage{multicol}
\usepackage{ifpdf}
\usepackage{cite}
\usepackage{comment}
\usepackage[pdftex]{graphicx}
\usepackage{epstopdf}
\DeclareGraphicsExtensions{.eps}

\usepackage{array}
\usepackage{tikz}
\usetikzlibrary{mindmap, shapes, arrows.meta, positioning, shapes.geometric}
\newcolumntype{P}[1]{>{\centering\arraybackslash}p{#1}}
\usepackage{lettrine}
\usepackage{amsmath}
\DeclareMathOperator*{\argmax}{arg\,max}

\usepackage{mwe}
\usepackage{amsmath,amsfonts,amssymb}
\usepackage{enumitem}   
\usepackage{subcaption}
\usepackage{algorithm}
\usepackage{algorithmic}
\usepackage{amsthm}

\usepackage[acronym,nomain,nonumberlist]{glossaries}

\newacronym{3gpp}{3GPP}{3rd Generation Partnership Project}          
\newacronym{4g}{4G}{Fourth Generation} 
\newacronym{5g}{5G}{Fifth Generation} 
\newacronym{6g}{6G}{Sixth Generation}
\newacronym{a2c}{A2C}{Advantage Actor Critic}
\newacronym{ai}{AI}{Artificial Intelligence}
\newacronym{aml}{AML}{Adversarial Machine Learning}
\newacronym{ann}{ANN}{Artificial Neutral Networks}
\newacronym{api}{API}{Application Programming Interface}
\newacronym{b5g}{B5G}{Beyond Fifth-Generation}
\newacronym{bbu}{BBU}{Baseband Unit} 
\newacronym{bs}{BS}{Base Station}
\newacronym{capex}{CAPEX}{CAPital EXpenditures} 
\newacronym{cicd}{CI/CD}{Continuous Integration and Delivery} 
\newacronym{cnn}{CNN}{Convolutional Neural Network}
\newacronym{cp}{CP}{Control Plane}
\newacronym{ct}{CT}{Continuous Training}
\newacronym{cu}{CU}{Central Unit}
\newacronym{dag}{DAG}{Directed Acyclic Graph}
\newacronym{dapp}{dApp}{Distributed Application}
\newacronym{darpa}{DARPA}{Defense Advanced Research Projects Agency}
\newacronym{devops}{DevOps}{DEVelopment and IT Operations}
\newacronym{dlift}{DeepLIFT}{Deep Learning Important FeaTures}
\newacronym{dl}{DL}{Deep Learning}
\newacronym{dnn}{DNN}{Deep Neural Network}
\newacronym{dqn}{DQN}{Deep Q-Network}
\newacronym{drl}{DRL}{Deep Reinforcement Learning}
\newacronym{du}{DU}{Distributed Unit}
\newacronym{embb}{eMBB}{enhanced Mobile Broadband}
\newacronym{enb}{eNB}{eNodeB}
\newacronym{eni}{ENI}{Experiential Networked Intelligence}
\newacronym{etsi}{ETSI}{European Telecommunications Standards Institute}
\newacronym{fl}{FL}{Federated Learning}
\newacronym{gan}{GAN}{Generative Adversarial Network}
\newacronym{gnb}{gNB}{gNodeB}
\newacronym{gnn}{GNN}{Graph Neural Network}
\newacronym{he}{HE}{Horizon Europe}
\newacronym{hitl}{\textcolor{blue}{HITL}}{\textcolor{blue}{Human-in-the-Loop}}
\newacronym{ieee}{IEEE}{Institute of Electrical and Electronics Engineers}
\newacronym{ig}{IG}{Integrated Gradients}
\newacronym{iot}{IoT}{Internet of Things}
\newacronym{isg}{ISG}{Industry Specification Group}
\newacronym{kl}{KL}{Kullback-Leibler}
\newacronym{lime}{LIME}{Local Interpretable Model-Agnostic Explanations}
\newacronym{llm}{LLM}{Large Language Model}
\newacronym{lo}{LO}{Log-Odds}
\newacronym{lstm}{LSTM}{Long Short Term Memory}
\newacronym{mac}{MAC}{Medium Access Control} 
\newacronym{mdp}{MDP}{Markov Decision Process}
\newacronym{mec}{MEC}{Multi-access Edge Computing}
\newacronym{mlops}{MLOps}{ML system operations}
\newacronym{ml}{ML}{Machine Learning} 
\newacronym{mmtc}{mMTC}{massive Machine Type Communications}
\newacronym{mm}{MM}{Mobility Management}
\newacronym{mno}{MNO}{Mobile Network Operator}
\newacronym{mr}{MR}{Machine Reasoning}
\newacronym{mvno}{MVNO}{Mobile Virtual Network Operator}
\newacronym{near rt ric}{Near RT RIC}{Near Real-Time RAN Intelligent Controller}
\newacronym{nesy}{\textcolor{blue}{NeSy}}{\textcolor{blue}{Neuro-Symbolic}}
\newacronym{nfv}{NFV}{Network Function Virtualization} 
\newacronym{ng ran}{NG RAN}{New Generation RAN}
\newacronym{non rt ric}{Non RT RIC}{Non Real-Time RAN Intelligent Controller} 
\newacronym{nr-mac}{NR-MAC}{New Radio Medium Access Control}
\newacronym{ns}{NS}{Network Slicing}
\newacronym{o-cloud}{O-Cloud}{Open Cloud}
\newacronym{o-cu-cp}{O-CU-CP}{Open RAN Central Unit Control Plane} 
\newacronym{o-cu-up}{O-CU-UP}{Open RAN Central Unit User Plane}
\newacronym{o-cu}{O-CU}{Open RAN Central Unit} 
\newacronym{o-du}{O-DU}{Open RAN Distributed Unit}
\newacronym{o-ran}{O-RAN}{Open Radio Access Network}
\newacronym{o-ru}{O-RU}{Open RAN Radio Unit}
\newacronym{opex}{OPEX}{OPerational EXpenditures}
\newacronym{osc}{OSC}{Open RAN Software Community}
\newacronym{pca}{PCA}{Principal Component Analysis}
\newacronym{pdcch}{PDCCH}{Physical Downlink Control Channel}
\newacronym{pdcp}{PDCP}{Packet Data Control Protocol}
\newacronym{pdsch}{PDSCH}{Physical Downlink Shared Channel}
\newacronym{phy}{PHY}{Physical}
\newacronym{pnf}{PNF}{Physical Network Function}
\newacronym{pucch}{PUCCH}{Physical Uplink Control Channel}
\newacronym{pusch}{PUSCH}{Physical Uplink Shared Channel}
\newacronym{qoe}{QoE}{Quality of Experience}
\newacronym{qos}{QoS}{Quality of Service}
\newacronym{qot}{QoT}{Quality of Transport}
\newacronym{r2}{R2}{R-squared}
\newacronym{ran}{RAN}{Radio Access Network}
\newacronym{rapp}{rApp}{Non RT RIC Applicaiton}
\newacronym{rat}{RAT}{Radio Access Technologies}
\newacronym{reco}{ReCo}{Relative Consistency}
\newacronym{ric}{RIC}{RAN Intelligent Controller}
\newacronym{rl}{RL}{Reinforcement Learning}
\newacronym{rm}{RM}{Resource Management}
\newacronym{rnn}{RNN}{Recurrent Neural Network}
\newacronym{rss}{RSS}{Received Signal Strenght}
\newacronym{rt}{RT}{Real Time}
\newacronym{ru}{RU}{Radio Unit}
\newacronym{scm}{SCM}{Structural Causal Models}
\newacronym{sdn}{SDN}{Software Defined Network}
\newacronym{shap}{SHAP}{SHapley Additive exPlanations}
\newacronym{sla}{SLA}{Service Level Agreement}
\newacronym{smo}{SMO}{Service Management and Orchestration}
\newacronym{sm}{SM}{Spectrum Management}
\newacronym{svm}{SVM}{Support Vector Machines}
\newacronym{tti}{TTI}{Transmission Time Interval}
\newacronym{ue}{UE}{User Equipment}
\newacronym{up}{UP}{User Plane}
\newacronym{urllc}{URLLC}{Ultra Reliable Low Latency Communication}
\newacronym{vbbu}{vBBU}{Virtual BBU}
\newacronym{vnf}{VNF}{Virtual Network Function}
\newacronym{vo-cu}{vO-CU}{virtual Open RAN Central Unit} 
\newacronym{vo-du}{vO-DU}{virtual Open RAN Distributed Unit}
\newacronym{vran}{vRAN}{Virtual RAN}
\newacronym{wg}{WG}{Working Group}
\newacronym{xai}{XAI}{eXplainable AI}
\newacronym{xapp}{xApp}{Near RT RIC Application}
\newacronym{xr}{XR}{eXtended Reality}

\newcommand*\xor{\oplus}

\usepackage{amsthm}
\theoremstyle{definition}
\newtheorem{definition}{Definition}[section]

\title{Explainable \acrshort{ai} in 6G \gls{o-ran}: A Tutorial and Survey on Architecture, Use Cases, \\ Challenges, and Future Research}

\author{
Bouziane Brik,~\IEEEmembership{Senior~Member,~IEEE}, 
Hatim~Chergui,~\IEEEmembership{Senior~Member,~IEEE},
Lanfranco~Zanzi,~\IEEEmembership{Member,~IEEE}, Francesco~Devoti,~\IEEEmembership{Member,~IEEE},
Adlen Ksentini,~\IEEEmembership{Senior~Member,~IEEE},
Muhammad~Shuaib~Siddiqui,
Xavier~Costa-P\'erez,~\IEEEmembership{Senior~Member,~IEEE}, and~Christos~Verikoukis,~\IEEEmembership{Senior~Member,~IEEE}

\IEEEcompsocitemizethanks{\IEEEcompsocthanksitem  B. Brik is with the Computer Science Department, College of Computing and Informatics, University of Sharjah, UAE(e-mail: bbrik@sharjah.ac.ae).}
\IEEEcompsocitemizethanks{\IEEEcompsocthanksitem H. Chergui and M.-S. Siddiqui are with i2CAT Foundation (e-mail: chergui@ieee.org, shuaib.siddiqui@i2cat.net).}
\IEEEcompsocitemizethanks{\IEEEcompsocthanksitem L. Zanzi and F. Devoti are with NEC Laboratories Europe, 69115 Heidelberg, Germany (e-mails: lanfranco.zanzi@neclab.eu, francesco.devoti@neclab.eu).}
\IEEEcompsocitemizethanks{\IEEEcompsocthanksitem A. Ksentini is with EURECOM, France (e-mail: adlen.ksentini@eurecom.fr).}
\IEEEcompsocitemizethanks{\IEEEcompsocthanksitem X. Costa-P\'erez is with NEC Laboratories Europe, i2CAT Foundation, and ICREA, 08034 Barcelona, Spain (e-mail: xavier.costa@neclab.eu).}
\IEEEcompsocitemizethanks{\IEEEcompsocthanksitem C. Verikoukis is with the University of Patras, ATHENA/ISI, and IQUADRAT Informatica, Barcelona 08006, Spain (e-mail: cveri@ceid.upatras.gr).}
}

\begin{document}

\maketitle
\begin{abstract}
The recent \gls{o-ran} specifications promote the evolution of \glsunset{ran}\gls{ran} architecture by function disaggregation, adoption of open interfaces, and instantiation of a hierarchical closed-loop control architecture managed by \glspl{ric} entities.
This paves the road to novel data-driven network management approaches based on programmable logic. Aided by \gls{ai} and \gls{ml}, novel solutions targeting traditionally unsolved \gls{ran} management issues can be devised.
Nevertheless, the adoption of such smart and autonomous systems is limited by the current inability of human operators to understand the decision process of such \gls{ai}/\gls{ml} solutions, affecting their trust in such novel tools.   
\gls{xai} aims at solving this issue, enabling human users to better understand and effectively manage the emerging generation of artificially intelligent schemes, reducing the \emph{human-to-machine} barrier.
In this survey, we provide a summary of the \gls{xai} methods and metrics before studying their deployment over the \gls{o-ran} Alliance \gls{ran} architecture along with its main building blocks. We then present various use-cases and discuss the automation of \gls{xai} pipelines for \gls{o-ran} as well as the underlying security aspects. We also review some projects/standards that tackle this area. 
Finally, we identify different challenges and research directions that may arise from the heavy adoption of \gls{ai}/\gls{ml} decision entities in this context, focusing on how \gls{xai} can help to interpret, understand, and improve trust in \gls{o-ran} operational networks.

\end{abstract}

\section{Introduction}

\glsresetall

\subsection{Context and Motivation}

\IEEEPARstart{6}{G} wireless networks are growing to revolutionize the way we connect, communicate, and share information, catalyzing smart services and innovative applications~\cite{chowdhury20206g, xie20216g, 6g, 6g1}. \glsunset{6g}\gls{6g} is expected to transform mobile communication networks from the \gls{iot} to "connected intelligence", by leveraging \gls{ai} techniques and connecting billions of devices and people~\cite{6g_tuto_survey1, 6g_tuto_survey2, 6g_tuto_survey3, zsm}. The promise of immense connected devices, ultra-low latency, low energy footprint, and extremely high data rates is expected to enhance the sustainability, connectivity, and trustworthiness of the next-generation mobile network, and support the development of innovative applications, such as truly immersive \gls{xr}, smart grid 2.0, high-fidelity mobile hologram, and Industry 5.0~\cite{6g4, 6g5, 6g7,GE_}.

The co-existence of such a variety of applications, along with their specific requirements, demands a versatile mobile network capable of accommodating and guaranteeing the expected performances by means of accurate and smart management of network components and resources~\cite{ran_5g, ns_adl, ns_wu} across different technological domains, i.e., \gls{ran}, core network, cloud, and edge. To this end, both industry and academia are leveraging \gls{ns}, \gls{sdn}, and \gls{nfv} paradigms to transform the mobile ecosystem into a more intelligent, energy-efficient, virtual, and software-focused ecosystem~\cite{zakka, SDN, net_sli, ns_b5g}.

In this context, a global initiative was formed, consisting of over $200$ companies from the telecommunication industry, who collaborated under the umbrella of the \gls{o-ran} Alliance to introduce a novel \gls{ran} architectural design for the forthcoming generation of mobile networks (\glsunset{b5g}\gls{b5g} and \gls{6g})~\cite{oran1}\cite{oran2}.
The core concept of \gls{o-ran} revolves around the disaggregation of traditional \gls{ran} system functionalities and their conversion into software components, known as Virtual Network Functions \gls{vnf}, which are interconnected through standardized and open interfaces.
Additionally, \gls{o-ran} introduces a novel hierarchical \gls{ric} architecture~\cite{oran_architecture}, which includes two main building blocks namely \gls{non rt ric}~\cite{non-rt-ric} and \gls{near rt ric}~\cite{near-rt-ric}, designed to enhance the capabilities and flexibility of the RAN ecosystem.
The \gls{non rt ric} is responsible for executing non-time-critical functions and tasks, such as policy management,

\begin{table}[t!]
\caption*{List of Acronyms}
\label{tab:acro1}
\begin{tabular}{ll}
\acrshort{3gpp} & \acrlong{3gpp} \\
\acrshort{4g} & \acrlong{4g} \\
\acrshort{5g} & \acrlong{5g} \\
\acrshort{6g} & \acrlong{6g} \\
\acrshort{a2c} & \acrlong{a2c} \\
\acrshort{ai} & \acrlong{ai} \\
\acrshort{ann} & \acrlong{ann} \\
\acrshort{aml} & \acrlong{aml} \\
\acrshort{api} & \acrlong{api} \\
\acrshort{b5g} & \acrlong{b5g} \\
\acrshort{bbu} & \acrlong{bbu} \\
\acrshort{bs} & \acrlong{bs} \\
\acrshort{capex} & \acrlong{capex} \\
\acrshort{cicd} & \acrlong{cicd} \\
\acrshort{cnn} & \acrlong{cnn} \\
\acrshort{cp} & \acrlong{cp} \\
\acrshort{ct} & \acrlong{ct} \\
\acrshort{cu} & \acrlong{cu} \\
\acrshort{dag} & \acrlong{dag} \\
\acrshort{dapp} & \acrlong{dapp} \\
\acrshort{darpa} & \acrlong{darpa} \\
\acrshort{devops} & \acrlong{devops} \\
\acrshort{dlift} & \acrlong{dlift} \\
\acrshort{dl} & \acrlong{dl} \\
\acrshort{dnn} & \acrlong{dnn} \\
\acrshort{dqn} & \acrlong{dqn} \\
\acrshort{drl} & \acrlong{drl} \\
\acrshort{du} & \acrlong{du} \\
\acrshort{embb} & \acrlong{embb} \\
\acrshort{enb} & \acrlong{enb} \\
\acrshort{eni} & \acrlong{eni} \\
\acrshort{etsi} & \acrlong{etsi} \\
\acrshort{fl} & \acrlong{fl} \\
\acrshort{gan} & \acrlong{gan} \\
\acrshort{gnb} & \acrlong{gnb} \\
\acrshort{gnn} & \acrlong{gnn} \\
\acrshort{he} & \acrlong{he} \\
\acrshort{ieee} & \acrlong{ieee} \\
\acrshort{ig} & \acrlong{ig} \\
\acrshort{iot} & \acrlong{iot} \\
\acrshort{isg} & \acrlong{isg} \\
\acrshort{kl} & \acrlong{kl} \\
\acrshort{lime} & \acrlong{lime} \\
\acrshort{llm} & \acrlong{llm} \\
\acrshort{lo} & \acrlong{lo} \\
\acrshort{lstm} & \acrlong{lstm} \\
\acrshort{mac} & \acrlong{mac} \\
\acrshort{mdp} & \acrlong{mdp} \\
\acrshort{mec} & \acrlong{mec} \\
\acrshort{mlops} & \acrlong{mlops} \\
\acrshort{ml} & \acrlong{ml} \\
\acrshort{mmtc} & \acrlong{mmtc} \\
\acrshort{mm} & \acrlong{mm} \\
\acrshort{mno} & \acrlong{mno} \\
\acrshort{mr} & \acrlong{mr} \\
\acrshort{mvno} & \acrlong{mvno} \\
\acrshort{near rt ric} & \acrlong{near rt ric} \\
\acrshort{nfv} & \acrlong{nfv} \\
\acrshort{ng ran} & \acrlong{ng ran} \\
\acrshort{non rt ric} & \acrlong{non rt ric} \\
\acrshort{nr-mac} & \acrlong{nr-mac} \\
\acrshort{ns} & \acrlong{ns} \\
\acrshort{o-cloud} & \acrlong{o-cloud} \\
\acrshort{o-cu-cp} & \acrlong{o-cu-cp} \\
\acrshort{o-cu-up} & \acrlong{o-cu-up} \\
\acrshort{o-cu} & \acrlong{o-cu} \\
\acrshort{o-du} & \acrlong{o-du} \\
\end{tabular}
\vspace{-1cm}
\end{table}

\begin{table}[t!]
\begin{tabular}{ll}
\acrshort{o-ran} & \acrlong{o-ran} \\
\acrshort{o-ru} & \acrlong{o-ru} \\
\acrshort{opex} & \acrlong{opex} \\
\acrshort{osc} & \acrlong{osc} \\
\acrshort{pca} & \acrlong{pca} \\
\acrshort{pdcch} & \acrlong{pdcch} \\
\acrshort{pdcp} & \acrlong{pdcp} \\
\acrshort{pdsch} & \acrlong{pdsch} \\
\acrshort{phy} & \acrlong{phy} \\
\acrshort{pnf} & \acrlong{pnf} \\
\acrshort{pucch} & \acrlong{pucch} \\
\acrshort{pusch} & \acrlong{pusch} \\
\acrshort{qoe} & \acrlong{qoe} \\
\acrshort{qos} & \acrlong{qos} \\
\acrshort{qot} & \acrlong{qot} \\
\acrshort{r2} & \acrlong{r2} \\
\acrshort{ran} & \acrlong{ran} \\
\acrshort{rapp} & \acrlong{rapp} \\
\acrshort{rat} & \acrlong{rat} \\
\acrshort{reco} & \acrlong{reco} \\
\acrshort{ric} & \acrlong{ric} \\
\acrshort{rl} & \acrlong{rl} \\
\acrshort{rm} & \acrlong{rm} \\
\acrshort{rnn} & \acrlong{rnn} \\
\acrshort{rss} & \acrlong{rss} \\
\acrshort{rt} & \acrlong{rt} \\
\acrshort{ru} & \acrlong{ru} \\
\acrshort{scm} & \acrlong{scm} \\
\acrshort{sdn} & \acrlong{sdn} \\
\acrshort{shap} & \acrlong{shap} \\
\acrshort{sla} & \acrlong{sla} \\
\acrshort{smo} & \acrlong{smo} \\
\acrshort{sm} & \acrlong{sm} \\
\acrshort{svm} & \acrlong{svm} \\
\acrshort{tti} & \acrlong{tti} \\
\acrshort{ue} & \acrlong{ue} \\
\acrshort{up} & \acrlong{up} \\
\acrshort{urllc} & \acrlong{urllc} \\
\acrshort{vbbu} & \acrlong{vbbu} \\
\acrshort{vnf} & \acrlong{vnf} \\
\acrshort{vo-cu} & \acrlong{vo-cu} \\
\acrshort{vo-du} & \acrlong{vo-du} \\
\acrshort{vran} & \acrlong{vran} \\
\acrshort{wg} & \acrlong{wg} \\
\acrshort{xai} & \acrlong{xai} \\
\acrshort{xapp} & \acrlong{xapp} \\
\acrshort{xr} & \acrlong{xr} \\
\end{tabular}
\end{table}

\noindent network optimization, and long-term analytics, while the \gls{near rt ric} focuses on time-critical operations and tasks that require low latency and quick decision-making.

It is easy to claim that \gls{ai} will play a critical role in the development and implementation of future network management operations, pursuing better network performance, cost savings, and enhanced customer experience~\cite{ai_networking}\cite{fl_iot}\cite{ai_networking1}. In this context, \gls{o-ran} envisions \gls{ric} entities to support programmable-based functions and logics, featured by the heavy usage of \gls{ai} techniques, in particular, \gls{ml} and \gls{dl}, to ease the development of intelligent and flexible \gls{ran} applications and reduce operational complexity~\cite{bonati2022intelligent}. Among others, the \gls{ai}-based \glspl{ric} aim to tackle traditionally hard-to-solve aspects of the \gls{ran} domain, such as spectrum management, mobility, radio resource assignment and scheduling, admission control, link management, and power allocation~\cite{tl, sulaiman2022coordinated}. This is particularly beneficial in the \gls{6g} landscape when considering various vertical industries and their corresponding networking requirements.
Having said that, the recent European Union (EU)'s AI Act establishes XAI regulation that mandates transparency and human oversight for high-risk AI-driven systems, such as future 6G networks \cite{european2021fosterinapproachtoai}. Moreover, the United States (US) focuses on maintaining global AI competitiveness while fostering trustworthy systems, with initiatives like the National AI Initiative Act \cite{MaintainingAmericanLeadershipinArtificialIntelligence}. The United Kingdom (UK)'s approach falls between the EU and US models, emphasizing responsible innovation and practical guidance, such as the ICO and Alan Turing Institute's AI decision explanation framework, alongside ambitions for global AI leadership \cite{UkNAtionalAIStrategyActionPlan2022,AlanTuringCommonRegCapAI}.

In this context, the widespread adoption of \gls{ai} techniques in future 6G \gls{o-ran} should be accompanied by mechanisms that verify and explain the black-box models' decisions in a systematic and objective fashion ~\cite{xai-iot}, especially when they lead to \gls{sla} violations~\cite{sla} or failures. This urges designers to clearly identify the operational boundaries of \gls{ai}/\gls{ml} models, characterize and understand their behaviour, and prioritize faithful and trustworthy decision-making processes to enable automated network service management while leaving the quality of service unaffected. On that account, new approaches are required to provide explainable and understandable decisions~\cite{sla1}. \gls{xai} is an emerging paradigm that aims to shed light on the decision process that is performed by closed (black box) \gls{ai} models. The main objective of \gls{xai} is to create a transparent and human-understandable model (white box) that clarifies the internal processes of \gls{ai} models, e.g., by determining the contribution of each input feature to an \gls{ai} decision or prediction~\cite{surv5}. \gls{xai} is crucial to demonstrate the accuracy, fairness, and transparency of \gls{ai} models that drive decisions and operations in the network, thereby instilling trust and confidence in the deployment of \gls{ai}-powered components in the \gls{o-ran} ecosystem by businesses and organizations~\cite{sabra, surv2}.

\begin{table*}[t!]
\caption{Existing surveys on \gls{o-ran}, \gls{xai}, \gls{xai} for \gls{b5g}. \textbf{H: High}, \textbf{M: Medium}, and \textbf{L: Low}.}
\label{tab:exis_surv}
\begin{center}
\begin{tabular}{|l|
>{\columncolor[HTML]{FFCE93}}c |c|c|c|
>{\columncolor[HTML]{96FFFB}}c |
>{\columncolor[HTML]{FFCE93}}c |l|}
\hline
\multicolumn{1}{|c|}{\cellcolor[HTML]{EFEFEF}\textbf{Works}} &
  \multicolumn{1}{l|}{\cellcolor[HTML]{EFEFEF}\textbf{\gls{ai}}} &
  \multicolumn{1}{l|}{\cellcolor[HTML]{EFEFEF}\textbf{\gls{xai}}} &
  \multicolumn{1}{l|}{\cellcolor[HTML]{EFEFEF}\textbf{ \rotatebox{90}{\gls{b5g}/\gls{6g} support}}} &
  \multicolumn{1}{l|}{\cellcolor[HTML]{EFEFEF} \rotatebox{90}{\textbf{\gls{o-ran} Architecture}}} &
  \multicolumn{1}{l|}{\cellcolor[HTML]{EFEFEF} \rotatebox{90}{\textbf{\gls{o-ran} use cases}}} &
  \cellcolor[HTML]{EFEFEF} \rotatebox{90}{\textbf{\begin{tabular}[c]{@{}c@{}}Future research directions\\ (\gls{b5g}, \gls{o-ran}, or \gls{xai})\end{tabular}}} &
  \multicolumn{1}{c|}{\cellcolor[HTML]{EFEFEF}\textbf{Contribution}} \\ \hline \hline
\cite{polese} &
  \cellcolor[HTML]{96FFFB}L &
  \cellcolor[HTML]{96FFFB}L &
  \cellcolor[HTML]{FFCE93}H &
  \cellcolor[HTML]{FFCE93}H &
  L &
  H &
    \begin{tabular}[c]{@{}l@{}}A tutorial on \gls{o-ran} framework, by describing recent specifications in terms of \\  architecture, design, and open interfaces.\end{tabular} \\ \hline

\cite{sur-oran} &
  \cellcolor[HTML]{96FFFB}L &
  \cellcolor[HTML]{96FFFB}L &
  \cellcolor[HTML]{FFCE93}H &
  \cellcolor[HTML]{FFCE93}H &
  L &
  H &
    \begin{tabular}[c]{@{}l@{}}A short survey on \gls{o-ran}'s architecture, benefits, shortcomings, and future \\ directions.\end{tabular} \\ \hline
    
\cite{niknam2020} &
  \cellcolor[HTML]{9AFF99}M &
  \cellcolor[HTML]{96FFFB}L &
  \cellcolor[HTML]{FFCE93}H &
  \cellcolor[HTML]{FFCE93}H &
  \cellcolor[HTML]{FFCE93}H &
  H &
  \begin{tabular}[c]{@{}l@{}}A concise paper on \gls{o-ran} architecture. It designed a \gls{dl}-based resource allocation scheme \\ and discussed future directions.\end{tabular} \\ \hline
\cite{dis_oran} &
  \cellcolor[HTML]{9AFF99}M &
  \cellcolor[HTML]{96FFFB}L &
  \cellcolor[HTML]{FFCE93}H &
  \cellcolor[HTML]{FFCE93}H &
  \cellcolor[HTML]{FFCE93}H &
  H &
    \begin{tabular}[c]{@{}l@{}}A short survey on \gls{o-ran}'s architecture, benefits, and future directions. It showed \\ the deployment of \gls{dl}-based scenarios in \gls{o-ran}.\end{tabular} \\ \hline
    \begin{tabular}[c]{@{}l@{}}\cite{oran_evolu1} \cite{oran_evolu2}\end{tabular} &
  \cellcolor[HTML]{96FFFB}L &
  \cellcolor[HTML]{96FFFB}L &
  \cellcolor[HTML]{FFCE93}H &
  \cellcolor[HTML]{FFCE93}H &
  L &
  \cellcolor[HTML]{96FFFB}L &
    \begin{tabular}[c]{@{}l@{}}Short review papers that discussed the evolution of \gls{ran} architectures towards \gls{o-ran} \\ in terms of functionality and implementation.\end{tabular} \\ \hline
    \cite{oran_mec_son_ns} &
  \cellcolor[HTML]{96FFFB}L &
  \cellcolor[HTML]{96FFFB}L &
  \cellcolor[HTML]{FFCE93}H &
  \cellcolor[HTML]{9AFF99}M &
  L &
  \cellcolor[HTML]{96FFFB}L &
  \begin{tabular}[c]{@{}l@{}}A concise paper discussed the integration of emergent \gls{b5g} concepts with \gls{o-ran}, \\ such as network slicing and \gls{mec}.\end{tabular} \\ \hline
  \cite{Our_surv} &
  H &
  \cellcolor[HTML]{96FFFB}L &
  \cellcolor[HTML]{FFCE93}H &
  \cellcolor[HTML]{FFCE93}H &
  \cellcolor[HTML]{FFCE93}H &
  H &
  \begin{tabular}[c]{@{}l@{}}A survey on \gls{dl}/\gls{ml}-based solutions for \gls{ran}/\gls{o-ran}. It includes \gls{o-ran} architecture\\  description along with its use cases as well as future directions and open challenges.\end{tabular} \\ \hline
  \cite{surv1}\cite{surv4} &
  H &
  \cellcolor[HTML]{FFCE93}H &
  \cellcolor[HTML]{9AFF99}M &
  \cellcolor[HTML]{96FFFB}L &
  L &
  H &
  \begin{tabular}[c]{@{}l@{}}A review of \gls{xai} approaches, in terms of their algorithmic aspects, classifications, \\ application domains, and future research directions.\end{tabular} \\ \hline
  \cite{surv2} &
  H &
  \cellcolor[HTML]{FFCE93}H &
  \cellcolor[HTML]{9AFF99}M &
  \cellcolor[HTML]{96FFFB}L &
  L &
  H &
  \begin{tabular}[c]{@{}l@{}}A review of the main principles and practice of \gls{xai}. In particular, the specific pattern\\ recognition models of machine learning are targeted.\end{tabular} \\ \hline
  \cite{surv3} &
  H &
  \cellcolor[HTML]{FFCE93}H &
  \cellcolor[HTML]{9AFF99}M &
  \cellcolor[HTML]{96FFFB}L &
  L &
  H &
  \begin{tabular}[c]{@{}l@{}}A review on a set of key measurement metrics, which can help to measure and evaluate\\ any explainable \gls{ai} system.\end{tabular} \\ \hline
\cite{surv5} &
  H &
  \cellcolor[HTML]{FFCE93}H &
  \cellcolor[HTML]{FFCE93}H &
  \cellcolor[HTML]{96FFFB}L &
  L &
  \cellcolor[HTML]{9AFF99}M &
  \begin{tabular}[c]{@{}l@{}}A survey on the use of \gls{xai} for \gls{b5g}/\gls{6g} networks. It addresses how to design \gls{xai} \\ systems for \gls{b5g} use cases, as well as future research directions in such context.\end{tabular} \\ \hline
  \cite{surv6} &
  H &
  \cellcolor[HTML]{FFCE93}H &
  \cellcolor[HTML]{FFCE93}H &
  \cellcolor[HTML]{96FFFB}L &
  L &
  \cellcolor[HTML]{96FFFB}L &
  \begin{tabular}[c]{@{}l@{}}A review of \gls{dl}-based solutions in PHY and MAC layers and their performance \\ vs \gls{xai} trade-off.\end{tabular} \\ \hline

  \cite{comcom_sur} &
  H &
  \cellcolor[HTML]{FFCE93}H &
  \cellcolor[HTML]{FFCE93}H &
  \cellcolor[HTML]{96FFFB}L &
  L &
  \cellcolor[HTML]{96FFFB}L &
  \begin{tabular}[c]{@{}l@{}}A review of existing \gls{xai} techniques and their applicability to deal with \\ the \gls{6g} network challenges. \end{tabular} \\ \hline
 
  \cite{thulitha} &
  H &
  \cellcolor[HTML]{FFCE93}H &
  \cellcolor[HTML]{FFCE93}H &
  \cellcolor[HTML]{96FFFB}L &
  L &
  \cellcolor[HTML]{9AFF99}M &
  \begin{tabular}[c]{@{}l@{}}A survey on the application of \gls{xai} to the security aspects of \gls{b5g} networks as well \\ as future research directions.\end{tabular} \\ \hline

\textbf{This survey} &
  \textbf{H} &
  \cellcolor[HTML]{FFCE93}\textbf{H} &
  \cellcolor[HTML]{FFCE93}\textbf{H} &
  \cellcolor[HTML]{FFCE93}\textbf{H} &
  \cellcolor[HTML]{FFCE93}\textbf{H} &
  \textbf{H} &
  \textbf{\begin{tabular}[c]{@{}l@{}}A comprehensive survey on the use of \gls{xai} to design transparent and trustworthy\\
  \gls{o-ran} architecture, covering architectural aspects, use cases, projects, \\standardization approaches, and future research directions.\end{tabular}} \\ \hline
\end{tabular}
\end{center}
\end{table*}

\subsection{Review of Existing Related Surveys}
Several studies already addressed the novel \gls{o-ran} architecture, highlighting its novel approach and investigating potential benefits and drawbacks.
In~\cite{sur-oran}, the authors provided a short review of both advantages and limitations of \gls{o-ran}, focusing on the \gls{o-ran} architecture and its main modules. The authors conducted a community survey on the benefits of \gls{o-ran} among $95$ researchers from all around the world. Most of them agreed on the fact that \gls{o-ran} will be the foundation of next-generation networks. Finally, the authors discussed the benefits, current shortcomings, and future research directions of \gls{o-ran}. 
Similarly,~\cite{niknam2020} described the \gls{o-ran} architecture and its key concepts. In addition, the authors present a novel \gls{dl}-based scheme for radio resource assignment, validating their performance using data collected from real mobile network deployments. The authors conclude their work by discussing open challenges and future research opportunities.
Another review study is provided by~\cite{dis_oran}. The authors showcase how a \gls{dl}-based scenario can be deployed on top of the \gls{o-ran} architecture, highlighting the main advantages and shortcomings of \gls{o-ran}.

The evolution of \gls{ran} architectures towards the \gls{o-ran} proposal both in terms of functionality and implementation is discussed in~\cite{oran_evolu1}~\cite{oran_evolu2}. In the same context, the support of \gls{b5g} key concepts, such as network slicing and \gls{mec}, by the \gls{o-ran} architecture is elaborated by~\cite{oran_mec_son_ns}\cite{mec2}\cite{mec1}\cite{mec3}.\\
In our previous work~\cite{Our_surv}, we proposed a survey study on the \gls{o-ran} architecture, discussing the evolution of \gls{ran} architectures, and comparing different studies based on various perspectives. We focused our review on existing \gls{ai}-based schemes dealing with the \gls{ran} challenges, and show how these schemes can be supported by \gls{o-ran} by considering the deployment of two realistic \gls{dl}-based case studies.
Similarly, in~\cite{polese}, the authors provided a tutorial on the \gls{o-ran} framework describing recent specifications in terms of architecture, design, and open interfaces. They also discuss the main open research challenges and the new innovation possibilities in the \gls{o-ran} architecture, focusing on \gls{ai} and deep learning.   

Besides, the \gls{xai} topic is attracting interest from research and industry domains. Currently, \gls{xai} is one of the main programs of the \gls{darpa}, expected to design efficiently the "third-wave \gls{ai} systems"~\cite{darpa}. In~\cite{surv1}\cite{surv4}, the authors reviewed and analyzed several \gls{xai} approaches focusing on algorithmic aspects, classifications, and application domains, identifying several still open challenges and key future research directions. The main principles and practices of \gls{xai} are summarized in~\cite{surv2}. In particular, the authors target the specific pattern recognition models of machine learning in order to enhance the understanding of such models for industry practitioners (data scientists). 
In~\cite{surv3}, the authors discussed a set of key measurement metrics that can help evaluate explainable \gls{ai} systems. 
In \gls{6g} networks context, the authors of~\cite{surv5} discussed the use of \gls{xai}, targeting different \gls{6g} use cases (e.g., Industry $5.0$). Similarly, in~\cite{comcom_sur} the authors highlight existing tools in addition to their use to deal with \gls{6g} network challenges, discussing how to integrate \gls{xai} into \gls{6g} networks architecture through a real mobile traffic prediction use-case, and validating their findings on realistic traffic data. Conversely, the authors of~\cite{surv6} focused on \gls{xai} methods in low protocol layers of mobile networks, e.g., \gls{phy} and \gls{mac}. In the same context, the authors of~\cite{thulitha} describe the application of \gls{xai} related to security aspects, discussing how \gls{xai} can improve the interpretation of \gls{ai}-based models for a wide range of security use-cases related to \gls{b5g}/\gls{6g} networks.  

Table~\ref{tab:exis_surv} summarizes the main topics discussed along the above works, and compares their contributions with respect to our work, in order to provide an easy understanding of the differentiation features with respect to the state-of-the-art.
Despite the presence of several survey papers discussing \gls{xai} and \gls{o-ran}, there is a lack of comprehensive surveys jointly investigating \gls{xai} and \gls{o-ran} aspects able to effectively explore the potential of \gls{xai} for developing responsible, trustworthy, and transparent \gls{ai}-powered \gls{o-ran} architecture. In addition, although the integration of \gls{xai} with \gls{b5g} networks has been addressed e.g., in ~\cite{surv5}\cite{surv6}, such studies do not focus either on the \gls{ran} part or consider the novel \gls{o-ran} architecture. 
Therefore, a comprehensive survey of \gls{xai} and its potential in designing the future \gls{o-ran} is greatly needed to guide the practitioners as well as researchers.  

\begin{figure*}[h!]
	\centering
	\includegraphics[height=5.5in,width=7.5in]{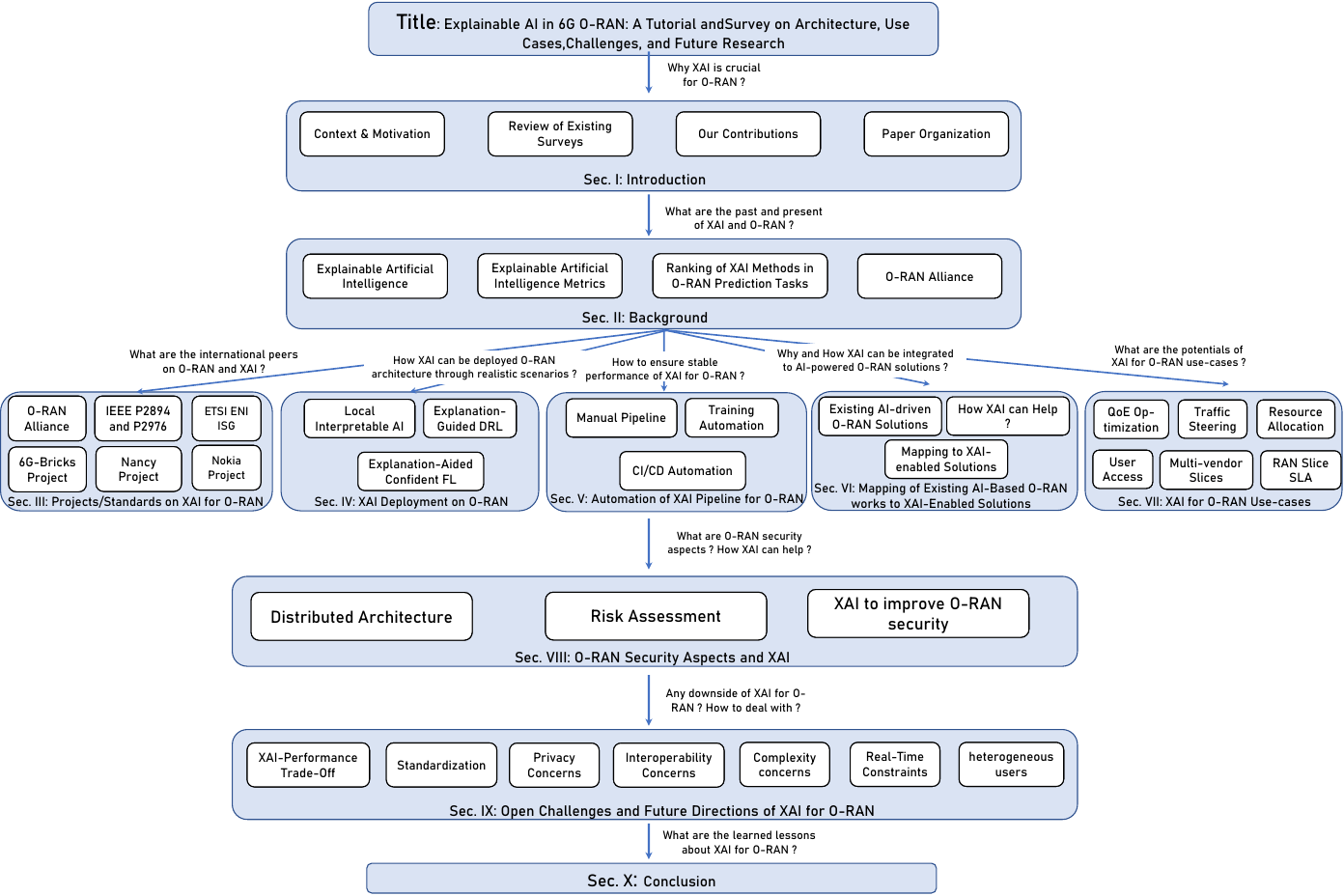}
	\caption{The taxonomy of the article.}
	\label{fig:stc_paper}
\end{figure*}

\subsection{Main Contributions}
The contributions of this paper can be summarized as follows:
\begin{itemize}
    \item \textit{Bridging the gap between \gls{o-ran} and \gls{xai}:} Existing surveys on \gls{o-ran} focused on its enabling technologies, such as hierarchical \gls{ran} Intelligent Controller, open interfaces, and programmable functions. To the best of our knowledge, there is no survey addressing the potential of human and \gls{o-ran} interactions, through \gls{xai} systems. Similarly, existing surveys on \gls{xai} targeted different \gls{xai} approaches and their taxonomies, and more recently their applications to \gls{b5g}/\gls{6g} networks. However, discussions on the potential of \gls{xai} for \gls{o-ran} are still missing. Therefore, this survey paper aims to bridge this gap by jointly exploring the key benefits of the introduction of \gls{xai} to \gls{o-ran}. 

    \item \textit{A comprehensive survey of \gls{xai} deployment on top of \gls{o-ran}:} Existing works studied both \gls{o-ran} and \gls{xai} separately, i.e. no work has combined both paradigms in its study. Hence, in this survey paper, we study the promising deployment of \gls{xai} on top of the \gls{ai}-enabled \gls{o-ran}. This includes \gls{o-ran} architecture as well as \gls{o-ran} use cases. Furthermore, We study the mapping of existing \gls{o-ran} research solutions to \gls{xai}-supported solutions.
    
    \item \textit{A depth analysis of \gls{xai} automation for \gls{o-ran}:} We provide an exhaustive analysis of how to automate the whole \gls{xai} pipeline on top of \gls{o-ran}, in order to ensure stable performance of deployed \gls{xai} techniques. To the best of our knowledge, no existing work has discussed the automation of \gls{xai} process for \gls{o-ran}. We also design new architectures showing the automated deployment of \gls{xai}, for different levels of automation. 

    \item \textit{\gls{o-ran} Security aspects and \gls{xai}:} We present the key findings of the official security risk assessments conducted on the \gls{o-ran} environment. We explore the potential of \gls{xai} to significantly improve the security layer of \gls{o-ran}, and how it could be used to build interpretable security threat detection mechanisms. Additionally, we discuss how \gls{xai} can help establish trust among stakeholders.

    \item \textit{Identifying New \gls{xai}-related Issues and Promising Research Directions:} Integrating \gls{xai} with \gls{o-ran} will rise new issues, which should also be considered in future research studies. Thus, we exhaustively discuss new open challenges, along with future research directions. 
\end{itemize}

\subsection{Paper Organization}
As shown in Fig.~\ref{fig:stc_paper}, the survey is organized as follows. Section.~\ref{sec:background} provides background information related to the topics considered in the survey.\\
Section.~\ref{sec:projects} presents the main ongoing projects and standards that are working to promote the adoption of \gls{ai}/\gls{ml} techniques in \gls{o-ran}, and show how they can be enhanced by \gls{xai}. Section.~\ref{sec:xai} describes how \gls{xai} methods and models can be deployed on top of the \gls{o-ran} architecture, considering four realistic deployment scenarios, and communication interfaces. Section.~\ref{sec:automation} details an automation pipeline for \gls{xai} model training and deployment in the \gls{o-ran} context, involving multiple architectural components and communication interfaces.\\
Section~\ref{sec:existingXAI-works} gives a literature review of existing recent works, which leverage XAI techniques for the 6G O-RAN architecture, while Section.~\ref{sec:xai_oran} gives a literature review of existing related works in the field, focusing on \gls{ai} techniques targeting \gls{ran} optimization and highlighting how these works can be mapped to \gls{xai}-enabled solutions to optimize multiple performances. Section.~\ref{sec:oran_use_cases} provides an overview of \gls{o-ran} use cases taken from the literature and standard documentation, highlighting how \gls{xai} could bring benefits to the considered scenarios.\\ 
Section.~\ref{sec:security} provides an overview of security issues related to the \gls{o-ran} architecture, focusing on \gls{xai}-related aspects. Section.~\ref{sec:open_challenges} highlights and discusses still open challenges along with their future research directions to deal with them. Finally, section.~\ref{sec:conclusion} concludes this paper. Note that the used acronyms in this paper are described in the \emph{List of Acronyms}, in alphabetical order, for ease of reference.
\section{Background}
\label{sec:background}
This section provides background information on \gls{xai} and \gls{o-ran} topics which are required to fully understand the potential of \gls{xai} techniques in the \gls{o-ran} domain. 
Firstly, we describe the main concepts, techniques, and emergent applications of \gls{xai}. Secondly, we present the \gls{o-ran} architecture along with its main modules as designed by the \gls{o-ran} Alliance. 


\subsection{\acrfull{xai}}
\label{sec:background_xai}
In this subsection, we provide the background on \gls{xai} and its main concepts, applications, and ongoing studies.
\subsubsection{Definitions and Key Concepts}

\begin{definition}[\gls{xai}\label{def:xai}]
\gls{xai} comprises methods and tools that help interpret, understand, and trust \gls{ai}-based model results~\cite{surv1, surv2}, using objective metrics (see Subsection \ref{subsec: metrics}). These tools assist in identifying and mitigating biases by revealing which features significantly impact predictions. This understanding promotes fairness and guides developers in refining algorithms, data, and features to improve model performance.
\end{definition}

In other words, \gls{xai} aims to build a white-box model that provides insights into the inner workings of underlying \gls{ml}/\gls{ai} black-box models. This helps characterize model fairness, accuracy, and transparency in \gls{ai}-enabled decisions, which is vital for businesses and organizations to have confidence and trust when deploying \gls{ai} models~\cite{surv2}. More specifically, \gls{xai} leverages concepts of \textit{explainability} and \textit{interpretability} to expose information about the internal mechanisms of \gls{ai} models.

\begin{definition}[Explainability and Interpretability\label{def:xai_and_int}]
\textit{Explainability} refers to the extent to which the internal mechanisms of a machine learning model can be understood in human terms. It involves providing insights into how a model makes decisions, often through tools and methods that elucidate the model's logic. \textit{Interpretability}, on the other hand, is the degree to which a human can consistently predict the model's output given a set of inputs. The key difference is that explainability focuses on the underlying workings and reasons behind a model's decisions, whereas interpretability emphasizes the ability to understand and predict a model's behavior based on its input-output relationship.
\end{definition}

XAI enables to identify which features influence model predictions the most, shedding light on potential biases encoded in the data or model architecture.

\begin{definition}[Bias in telecom operations\label{def:bias}]
Bias in Deep Neural Network (DNN) predictions refers to systematic errors that lead to unfair outcomes for certain groups. This might be caused by i) \textit{Unbalanced and Biased Data}, where the training dataset is not representative of the diverse network states, e.g., having more SLA violation samples leads to false alarms predictions; and ii) \textit{Inappropriate Feature Selection}, where using features that encode sensitive attributes or their proxies can make a  model inclined to allocate more resources to a service regardless of e.g., the traffic.
\end{definition}

\gls{xai} models incorporate the so-called \textit{explanation user interface} to generate a user-understandable explanation and/or interpretation of the rationale behind decisions taken by the model. Most \gls{ai} models can be translated into an equivalent \gls{xai} counterpart, at the expense of integrating additional layers supporting the explanation user interface on top of the deployed model. Based on the design of the explanation user interface, the \gls{xai} model can provide both explainability and interpretability or only one, depending on the target human user~\cite{surv1}.

{\setlength\arrayrulewidth{0.1pt}
\begin{table*}
\caption{\gls{xai} Taxonomy. A: Agnostic, S: Specific.}
\label{tab:xai_taxonomy}
\centering
\begin{tabular}{|P{.08\textwidth}|P{.08\textwidth}|P{.08\textwidth}|P{.12\textwidth}|P{.04\textwidth}|P{.4\textwidth}|c|}
\hline 
\cellcolor[HTML]{EFEFEF}\textbf{Transparency} & \cellcolor[HTML]{EFEFEF}\textbf{Explain. Basis} & \cellcolor[HTML]{EFEFEF}\textbf{Technique} & \cellcolor[HTML]{EFEFEF}\textbf{Algorithm} & \cellcolor[HTML]
{EFEFEF}\textbf{Agno.} & \cellcolor[HTML]{EFEFEF}\textbf{Pros and Cons} & \cellcolor[HTML]{EFEFEF}\textbf{Reference}\tabularnewline
\hline 
\hline 
\multirow{11}{*}{\shortstack{Black-Box \\ Models}} & \multirow{6}{*}{Attributions} & \multirow{4}{*}{Gradient} & Saliency Maps &  A & Pros: Simplicity, visual interpretability, widely applicable. Cons: Lack of context, sensitivity to input perturbations, limited to input gradients. & \cite{saliency, saliency2}\tabularnewline
\cline{4-7} \cline{5-7} \cline{6-7} \cline{7-7}
 &  &  & Gradient x Input & A & Pros: Simplicity, direct relevance, feature importance ranking. Cons: Input scaling sensitivity, limited to linear relationships, potential for misleading interpretations. &\cite{inputgrad, inputgrad2}\tabularnewline
\cline{4-7} \cline{5-7} \cline{6-7} \cline{7-7}
 &  &  & Integrated Gradients & A & Pros: Baseline comparison, path-based attribution, completeness, and sensitivity. Cons: Computationally intensive, baseline selection challenge, linearity assumption. & \cite{intgrad, intgrad2}\tabularnewline
\cline{4-7} \cline{5-7} \cline{6-7} \cline{7-7}
 &  &  & Smooth Gradient & A & Pros: Noise reduction, robustness to adversarial examples, gradient visualization. Cons: Interpretation challenges, hyperparameter sensitivity, computational overhead. & \cite{smoothgrad, smoothgrad2}
 \tabularnewline
\cline{4-7} \cline{5-7} \cline{6-7} \cline{7-7}
 &  &  & Epsilon-LRP & S & Pros: Deep model interpretability, conceptual clarity, attribution preservation. Cons: Complexity, parameter tuning, vulnerability to network architecture. & \cite{lrp, lrp2} \tabularnewline
\cline{3-7} \cline{4-7} \cline{5-7} \cline{6-7} \cline{7-7}
 &  & \multirow{2}{*}{Perturbation} & SHAP  & A & Pros: Theoretical grounding based on game theory, global and local interpretability, and consistency. Cons: Computational complexity, high-dimensional data challenge, model approximation dependency. & \cite{shap, shap2}\tabularnewline
\cline{4-7} \cline{5-7} \cline{6-7} \cline{7-7}
 &  &  & DeepLIFT & A & Pros: Model-agnostic, captures interactions, relevance conservation. Cons: Computational overhead, baseline selection challenge, interpretation complexity. & \cite{deeplift} \tabularnewline
\cline{4-7} \cline{5-7} \cline{6-7} \cline{7-7}
 &  &  & Occlusion & A & Pros: Intuitive visual interpretation, robustness to model architecture, spatial localization. Cons: Computational expense, coarseness of occlusion, interpretation subjectivity. & \cite{occlusion, occlusion2}\tabularnewline
\cline{3-7} \cline{4-7} \cline{5-7} \cline{6-7} \cline{7-7}
  &  & Importance Weights & GNNExplainer & A & Pros: Graph-specific interpretability, node and edge importance, feature relevance analysis. Cons: Complexity, model-specific, interpretation scalability. & \cite{gnnexp} \tabularnewline
\cline{2-7} \cline{3-7} \cline{4-7} \cline{5-7} \cline{6-7} \cline{7-7}  
 & \multirow{2}{*}{Surrogates} & Local Techniques & LIME 
& A & Pros: Model-agnostic, local interpretability, simplicity. Cons: Interpretability limitation, instability, assumes linearity. & \cite{lime, lime2} \tabularnewline
\cline{3-7} \cline{4-7} \cline{5-7} \cline{6-7} \cline{7-7} 
 &  & Global Techniques & TREPAN  & S & Pros: Decision tree interpretability, human-readable explanations, transparent model behavior. Cons: Limited to decision tree models, model-specific, interpretation scalability. & \cite{trepan}\tabularnewline
 \cline{3-7} \cline{4-7} \cline{5-7} \cline{6-7} \cline{7-7}
 &  & Rule-Based & RuleFit  & S & Pros: combines decision trees and linear regression to provide interpretable insights into the model's decision-making process. Cons: may struggle to model highly intricate or complex nonlinear patterns in the data. & \cite{rule_fit,rule_fit1}\tabularnewline
 \cline{2-7} \cline{3-7} \cline{4-7} \cline{5-7} \cline{6-7} \cline{7-7}
 & RL Reward & Rule-based & Reward Shaping 
& S & Pros: Provides explicit guidance to the RL/DRL agent, allowing it to focus on desired behaviors. Cons: Can introduce biases if the reward shaping is not carefully designed. & \cite{reward_shaping, reward_shaping2} \tabularnewline
 \cline{2-7} \cline{3-7} \cline{4-7} \cline{5-7} \cline{6-7} \cline{7-7}
 & RL State & Model-based & Attention Mechanisms 
& S & Pros: Offers transparency by showing which parts of the input state the RL/DRL agent attends to. Cons: Attention mechanisms do not explicitly explain the agent's internal reasoning or decision-making process. & \cite{attention, attention2} \tabularnewline 
\cline{2-7} \cline{3-7} \cline{4-7} \cline{5-7} \cline{6-7} \cline{7-7}
& Symbolic  & \multicolumn{2}{c|}{Machine Reasoning} & S & Pros: Provides human-interpretable explanations for model decisions and shows explicit reasoning behind decisions, enhancing transparency. Cons: Requires expertise, computationally expensive, and may struggle with uncertain or probabilistic information. & \cite{MR, mr2} \tabularnewline
\cline{2-7} \cline{3-7} \cline{4-7} \cline{5-7} \cline{6-7} \cline{7-7}
& Transformers' Attention Head  & \multicolumn{2}{c|}{Attention Flow Analysis} & S & Pros: Provides interpretability, enables fine-grained analysis, helps improve models, and offers domain-specific insights. Cons: Complexity, lack of unique interpretations, limited context, challenges in generalization. & \cite{attention_analysis,AttnRLP} \tabularnewline
\cline{2-7} \cline{3-7} \cline{4-7} \cline{5-7} \cline{6-7} \cline{7-7} 
 & Visual  & \multicolumn{2}{c|}{SCM}  & S & Pros: Causal understanding, intuitive visualization, identifying confounding variables. Cons: Limited to causal modeling, simplified representation, expert knowledge required. & \cite{visual,scm}\tabularnewline
\cline{2-7} \cline{3-7} \cline{4-7} \cline{5-7} \cline{6-7} \cline{7-7}
 & Text  & \multicolumn{2}{c|}{Caption Generation} & S & Pros: Contextual understanding, language comprehension, multimodal interpretation. Cons: Subjectivity and ambiguity, lack of fine-grained control, reliance on training data. & \cite{text, Caption}\tabularnewline
\cline{2-7} \cline{3-7} \cline{4-7} \cline{5-7} \cline{6-7} \cline{7-7} 
 & Graph & \multicolumn{2}{c|}{Knowledge Graphs} & A & Pros: Structured representation, relationship understanding, integration, and interoperability. Cons: Knowledge acquisition and maintenance, incompleteness and accuracy, limited context and ambiguity.  & \cite{kg_book,kg}\tabularnewline
\hline 
\multirow{7}{*}{\shortstack{Transparent \\Models}} & \multicolumn{3}{c|}{Logistic / Linear Regression}  & \multirow{7}{*}{S} & \multirow{7}{*}{\shortstack[l]{Pros: Explainability, trust and accountability, debugging and \\ error analysis. Cons: Performance limitations, vulnerability to \\ adversarial attacks.}} &\cite{ml_book}\tabularnewline
\cline{2-4}\cline{7-7}  
 & \multicolumn{3}{c|}{Decision Trees} & & & \cite{xgboost}\tabularnewline
\cline{2-4}\cline{7-7}  
 & \multicolumn{3}{c|}{K-Nearest Neighbors} & & & \cite{knear}\tabularnewline
\cline{2-4}\cline{7-7}  
 & \multicolumn{3}{c|}{Rule-Based Learners} & & & \cite{rule}\tabularnewline
\cline{2-4}\cline{7-7}
 & \multicolumn{3}{c|}{Generative Additive Models} & &  & \cite{gen}\tabularnewline
\cline{2-4}\cline{7-7} 
 & \multicolumn{3}{c|}{Bayesian Models} & & & \cite{bayes}\tabularnewline
\cline{2-4}\cline{7-7} 
 & \multicolumn{3}{c|}{Self-Explainable Neural Networks} & & & \cite{senn}\tabularnewline
\hline 
\end{tabular}
\end{table*}
}

\subsubsection{Taxonomy of \gls{xai} Techniques, Applications, and Stakeholders}
There are several existing taxonomies in the \gls{xai} realm, which can complement and/or overlap each other. Table \ref{tab:xai_taxonomy} describes an \gls{xai} taxonomy that is mainly inspired by~\cite{surv1, surv2}, and is based on the following three main criteria:
\begin{itemize}
    \item \textit{Model Transparency:} \gls{xai} models can be classified based on the target \gls{ml} models' transparency. In this regard, models are classified as interpretable or complex. Interpretable models are by themselves understandable for human users. In other words, such models are able to provide the rationale behind their decisions in an interpretable way to users~\cite{surv1}. Several proposed works succeeded in interpreting some relatively low-complex \gls{ml} models, including logistic/linear regression, decision trees, K-Nearest neighbors, rule-based learners, etc.~\cite{surv1}. On the other hand, more complex models such as deep neural networks, in order to be interpretable, have to be approximated by generating simpler surrogate models that ease the explanation task by means of a technique known as \textit{post-hoc explainability}~\cite{vale2022explainable}.
    The model complexity is a widely considered aspect in the literature related to \gls{xai} and is generally adopted to classify \gls{xai} approaches~\cite{surv1}.
    \item \textit{Model Agnosticity:} This criterion targets complex \gls{ml}/\gls{dl} models, where \gls{xai} models can be categorized based on the nature of their target explanations~\cite{surv1}\cite{surv2}. In the paradigm of model-agnostic interpretability, the model is regarded as an opaque entity. This conceptualization dissociates interpretability from the specific characteristics and inner workings of the model, thereby liberating the model to exhibit maximum flexibility tailored to the requirements of the task at hand. This approach facilitates the utilization of diverse machine learning methodologies, encompassing even intricate deep neural networks. Furthermore, it affords the opportunity to manage the delicate balance between model complexity and interpretability, a crucial consideration delineated in the subsequent section. Importantly, this methodology allows for graceful handling of situations where achieving an interpretable explanation proves unattainable. Techniques such as SHAP and feature importance scores derived from permutation importance fall into this category. They work by analyzing the input-output relationship of the model without relying on its internal structure.
    
    \item \textit{Explainability Methods:} When \gls{ml}/\gls{dl} models are considered complex models, some techniques should be devised and used to interpret such models. Thus, \gls{xai} models rely on several explanation types, to describe how these \gls{ml}/\gls{dl} models output their predictions for any input data.
    \begin{itemize}
        \item Explanations by simplification refer to the techniques that simplify a complex model and approximate it to an interpretable model, which is easier to explain~\cite{lime}.
        \item Feature relevance explanations study and quantify the impact of each input data, to explain a given \gls{ml} model's prediction~\cite{features}.
        \item Local explanations focus on a single or particular prediction (output) of \gls{ml} models to generate explanations~\cite{lime}.
        \item Visual explanations aim to generate explanations in a visual way, describing the inner functioning of \gls{ml}/\gls{dl} models~\cite{visual}.
        For instance, they could reveal which set of pixels is the most relevant to recognize content in image classification tasks. Visual explanations rely on several tools, e.g., graphs, heatmaps, scatter plots, etc.
        \item Text explanations generate symbol interpretations of learning models using, for example, natural language text to explain their results~\cite{text}. For instance, they could be used to highlight which words (or forms) are leveraged in automatic email spam filtering.
    \end{itemize}
\end{itemize}

{\setlength\arrayrulewidth{0.1pt}
\begin{table*}[]
\caption{\gls{xai} Users}
\label{tab:xai_users}
\centering
\begin{tabular}{|p{.12\textwidth}|p{.35\textwidth}|p{.35\textwidth}|p{.09\textwidth}|}
\hline 
\cellcolor[HTML]{EFEFEF}\textbf{XAI Users} & \cellcolor[HTML]{EFEFEF}\textbf{Needs} & \cellcolor[HTML]{EFEFEF}\textbf{Key Application Areas} & \cellcolor[HTML]{EFEFEF}\textbf{Reference}\tabularnewline
\hline 
\hline 
Data Scientists and Machine Learning Researchers & They require XAI techniques to understand and debug complex models, identify biases, and improve model performance & Model development, debugging, and optimization across various domains such as telecommunications, healthcare, finance, natural language processing, and computer
vision & \centering\cite{lime, intml}\tabularnewline
\hline 
End Users and Consumers & They need explanations to trust and understand AI systems in applications
like recommender systems, personalized marketing, and decision support tools & E-commerce and personalized recommendation systems, healthcare decision support tools, financial advice platforms, and autonomous vehicles & \centering\cite{euca, VC_} \tabularnewline
\hline 
Managers and Decision Makers & They require transparent and interpretable AI models to make informed
decisions, assess risks, and gain insights into the AI system's behavior & Business intelligence and analytics, risk assessment and management, fraud detection, and regulatory compliance across industries such as finance, healthcare, and manufacturing & \centering\cite{xai_decision}\tabularnewline
\hline 
Developers and Engineers & They need tools and methods to troubleshoot networks faults/SLA violations, build explainable AI systems, ensure reliability, and meet regulatory requirements & Building interpretable machine learning models, developing explainable AI frameworks and libraries, ensuring model reliability and security in domains like cybersecurity, telecommunications, and autonomous systems & \centering\cite{eng, sla1} \tabularnewline
\hline 
Auditors and Compliance Officers & They require XAI to assess the fairness, accountability, and compliance
of AI systems and to identify potential biases or risks & Assessing the fairness and legality of AI systems in finance, hiring practices, loan approvals, credit scoring, and regulatory compliance
in sectors such as finance and human resources & \centering\cite{audit} \tabularnewline
\hline 
Legal Professionals and Judges & They need explanations to understand AI decisions, assess legal implications,
and ensure transparency and fairness in legal proceedings & Interpreting AI-driven legal decisions, evaluating algorithmic fairness, ensuring transparency and accountability in legal proceedings, and addressing ethical concerns in areas like criminal justice and civil
rights & \centering\cite{legal} \tabularnewline
\hline 
Regulators and Policy Makers & They require XAI to establish guidelines, standards, and regulations
around AI ethics, transparency, and accountability & Establishing guidelines, standards, and regulations for trustworthy AI in sectors including healthcare, finance, autonomous systems, and data privacy to protect public interests and ensure ethical AI deployment & \centering\cite{xai_regulation} \tabularnewline
\hline 
\end{tabular}
\end{table*}
}

Based on the above taxonomy criteria, several \gls{xai} approaches have been proposed in the literature. In what follows, we present the most popular ones, highlighting their main features:
\begin{itemize}
    \item \textit{\acrfull{shap}:} This approach relies on feature relevance explanation to interpret a particular prediction of supervised \gls{ml}/\gls{dl} models~\cite{DeepShape}. It computes an additive feature importance score with respect to a set of required properties (e.g., accuracy, consistency, and missingness). Hence, \gls{shap} determines feature influence by applying the Shapley values method, which enables estimating the marginal contribution of one feature over the final reward function. In addition, combining several predictions can also be considered to build a global explanation. Several variants of \gls{shap} have been proposed in the literature in order to optimize its computational complexity, such as Deep\gls{shap}\cite{DeepShape} and Tree\gls{shap}~\cite{treeshape}.
    \item \textit{\acrfull{dlift}~\cite{deeplift}:} The purpose of \gls{dlift} is to clarify the output of a neural network by calculating the significance of each input feature to the output. This is accomplished by comparing the activation of each neuron in the network for a particular input to the activation that would have been obtained if a reference input had been used. The difference in the activations between the input and the reference is measured by \gls{dlift} to compute the contribution of each input feature to the output. The contribution score obtained can be utilized to comprehend how the network reached its conclusion and to identify the most relevant input features. \gls{dlift} has been effective in explaining the behavior of different neural network models, such as convolutional neural networks and recurrent neural networks, and has been applied to various fields, including drug discovery, image classification, and speech recognition.
    \item \textit{\acrfull{lime}:} It is one of the most known solutions, that relies on local and simplification explanations, to explain supervised \gls{ml}/\gls{dl} models~\cite{lime}. LIME is a model-agnostic approach targeting different types of data, e.g., tabular, text, graphs, and images. \gls{lime} aims to approximate the learning models by developing locally linear models, which replace the black-box models to explain their individual predictions. 
    \item \textit{Rulefit:} It integrates the benefits of decision trees and linear models. It first consists on the creation of a wide array of rules from an ensemble of decision trees, which capture intricate, non-linear patterns in the data. These rules are then utilized as features in a sparse linear model, combining high predictive performance with clear interpretability \cite{rulefit}.
    \item \textit{\acrfull{ig}:} also known as Path-\acrlong{ig} or Axiomatic Attribution for Deep Networks. \gls{ig} is an \gls{xai} technique that gives an importance value to each feature of the input using the gradients of the model output~\cite{ig}. Specifically, it is a local method that consists of accumulating the gradients by sampling points at a uniform spacing along a straight line between the input and the baseline. This procedure avoids getting null gradients when, e.g., the deep learning model is flat in the proximity of the input feature. This method yields the specific positive or negative attributions of the input features.
    \item \textit{\acrfull{gnn} Explainer:} It is a technique that explains the predictions of \glspl{gnn} for graph-structured data. It identifies the most important nodes and edges contributing to the output by generating explanation vectors using an additional neural network. This generates an attention map that shows the relative importance of each node and edge. \gls{gnn} Explainer can be applied to various \gls{gnn} architectures and input graphs, without requiring changes to the model or training data. It is useful for understanding how \glspl{gnn} perform predictions and identifying potential issues \cite{gnnexp}.
    \item \textit{Reward Shaping:} It entails altering the reward function of the agent to offer supplementary feedback or incentives. This adjustment assists in steering the agent's learning process by molding the reward signal \cite{reward_shaping}.
    \item \textit{Attention Mechanism:} It enhances interpretability by identifying and highlighting the crucial elements in the input that significantly impact the decision-making process of the agent. They shed light on the specific features that capture the agent's attention and influence its decision \cite{attention}.
    \item \textit{\gls{mr}:} It utilizes logical reasoning and inference techniques to offer insights into the decision-making process of AI models, thereby improving transparency and trust. It generates explanations that are easily comprehensible to humans, fostering a deeper understanding and acceptance of AI systems. Nevertheless, applying machine reasoning in XAI necessitates expertise in logic and reasoning, and it may encounter difficulties when dealing with uncertain or probabilistic information. Nonetheless, the incorporation of machine reasoning in \gls{xai} contributes to the advancement of interpretable and accountable AI systems \cite{MR}.
    \item \textit{Attention Flow Analysis:} It assesses the individual contribution of attention heads in the encoder to the overall performance of the transformer's model. Specifically, it examines the roles played by these attention heads, with a particular focus on the most important and confident ones. These heads often exhibit consistent and linguistically interpretable roles, providing valuable insights into the model's decision-making process \cite{attention_analysis}.
    \item \textit{\acrfull{scm}:} It is another method that targets reinforcement learning models, aiming to show the causal link between the data variables. In~\cite{scm}, the authors leverage \gls{scm} method to explain the behavior of the reinforcement learning model. They are based on visual explanations through a \gls{dag}, where the nodes and edges reflect the model states and actions, respectively. By exploring the DAG, it can be extracted which actions take to move from one state to another. Once DAG is created, regression models are built to approximate the relationships using the minimum number of variables. Then, analyzing the \gls{dag}'s variables will help in generating the explanations, in order to answer the question: "Why action X and not Y ?".
    \item \textit{Caption generation:} It is a class of methods that aims to generate text interpretations to explain the outputs of \gls{dl} models. In~\cite{Caption}, the authors combined a \gls{cnn} model and a bidirectional \gls{lstm} encoder/decoder model. The \gls{lstm} encoder helps to extract video features, which are then used by the \gls{lstm} decoder to generate textual video captions.
    \item \textit{Knowledge Graphs:} To produce human-understandable explanations, it is necessary to represent ideas in terms of concepts rather than numeric values. Concepts and the connection between them make what is called \emph{knowledge graph}. It is a powerful way of representing data because Knowledge Graphs can be built automatically and can then be explored to reveal new insights about the domain, especially to find inferred concepts that were not asserted, along with being able to trace back all the steps, making it fully explainable \cite{kg}.
\end{itemize}

\begin{figure}[h]
    \centering
    \scalebox{0.7}{
    \begin{tikzpicture}[mindmap, 
                        grow cyclic,
                        every node/.style={concept, minimum size=0.6cm, text width=1.1cm, align=center, font=\scriptsize},
                        concept color=orange!30, 
                        level 1/.append style={level distance=3.5cm, sibling angle=51},
                        level 2/.append style={level distance=2.0cm, sibling angle=37}]

        \node [concept color=gray!30, minimum size=1.2cm, text width=1.8cm, align=center] {Black-Box Models}
            child [concept color=purple!30, level 1/.append style={distance=6cm, sibling angle=50}] { node {Attributions}
                child { node {Saliency Maps \\ \cite{saliency, saliency2}} }
                child { node {Gradient x Input \\ \cite{inputgrad, inputgrad2}} }
                child { node {Integrated Gradients \\ \cite{intgrad, intgrad2}} }
                child { node {Smooth Gradient \\ \cite{smoothgrad, smoothgrad2}} }
                child { node {Epsilon-LRP \\ \cite{lrp, lrp2}} }
            }
            child [concept color=magenta!20] { node {Perturbation}
                child { node {SHAP \\ \cite{shap, shap2}} }
                child { node {DeepLIFT \\ \cite{deeplift}} }
                child { node {Occlusion \\ \cite{occlusion, occlusion2}} }
            }
            child [concept color=teal!50] { node {Surrogates}
                child { node {LIME \\ \cite{lime, lime2}} }
                child { node {TREPAN \\ \cite{trepan}} }
                child { node {RuleFit \\ \cite{rule_fit, rule_fit1}} }
            }
            child [concept color=blue!30] { node {RL State}
                child { node {Attention Mechanisms \\ \cite{attention, attention2}} }
            }
            child [concept color=blue!50] { node {RL Reward}
                child { node {Reward Shaping \\ \cite{reward_shaping, reward_shaping2}} }
            }
            child [concept color=yellow!50] { node {Text, Visual and Graph}
                child { node {SCM \\ \cite{visual,scm}} }
                child { node {Caption Generation \\ \cite{text, Caption}} }
                child { node {Knowledge Graphs \\ \cite{kg_book}} }
            }
            child [concept color=orange!50] { node {Symbolic}
                child { node {Machine Reasoning \\ \cite{MR, mr2}} }
            };

    \end{tikzpicture}
    }
    \caption{Black-Box Models Mindmap.}
    \label{fig:black_box_models_mindmap}
\end{figure}

{\setlength\arrayrulewidth{0.1pt}
\begin{table*}[h]
\scriptsize
\caption{Taxonomy of \gls{xai} metrics.}
    \begin{center}
        \begin{tabular}{|c|c|c|c|c|}
\hline 
\cellcolor[HTML]{EFEFEF}\textbf{\gls{xai} Method}  & \cellcolor[HTML]{EFEFEF}\textbf{Basis} & \cellcolor[HTML]{EFEFEF}\textbf{Metric}  & \cellcolor[HTML]{EFEFEF}\textbf{Type of Problem} & \cellcolor[HTML]{EFEFEF}\textbf{Reference}\tabularnewline
\hline 
\hline 
\multirow{5}{*}{Attributions-based } & \multirow{4}{*}{Features Mutation/Masking} & Confidence/Faithfulness & \multirow{3}{*}{Classification} & [115] \tabularnewline
\cline{3-3} \cline{5-5} 
 &  & Log-odds &  & [116] \tabularnewline
\cline{3-3} \cline{5-5} 
 &  & Comprehensiveness &  & [117]\tabularnewline
\cline{3-3} \cline{5-5} 
 &  & Sufficiency &  & [117]\tabularnewline
\cline{2-5} 
 & Raw features & Interpretability & Regression & [118]
 \tabularnewline
 \cline{3-5} 
 &  & Ambiguity & Prediction/Decision & [119], [120] \tabularnewline
\hline 
\multirow{5}{*}{Surrogates-based} & \multirow{3}{*}{Perturbation} & Robustness/Sensitivity & \multirow{2}{*}{Regression/Decision} & [100] \tabularnewline
\cline{3-3} \cline{5-5} 
 &  & (in)fidelity &  & [121], [122] \tabularnewline
\cline{2-5}  
 & \multicolumn{2}{c|}{LIME Explainer R2 Score} & \multirow{2}{*}{Classification and Regression} & [75] \tabularnewline
\cline{2-3} \cline{5-5}  
 & \multicolumn{2}{c|}{Relative Consistency} &  & [123]\tabularnewline
\hline 
\multirow{2}{*}{White-box baseline} & \multicolumn{2}{c|}{Explainer Recall } & \multirow{2}{*}{Classification} & [124]\tabularnewline
\cline{2-3} \cline{5-5}
 & \multicolumn{2}{c|}{Explainer Precision } &  & [125] \tabularnewline
\hline 
\end{tabular}
    \end{center}%
    \label{tab:taxonomy_of_metrics}
\end{table*}
}

As anticipated before, the selection of suitable explainability methods depends both on the complexity of the targeted model to be explained and on the target audience. Indeed, the type of explanation exposed and their level of detail depend mainly on the people who are getting such information. In this context, different user profiles may be targeted by \gls{xai} models, and \gls{xai} models' explanations should differ from one user to another~\cite{surv1}. Table.~\ref{tab:xai_users} illustrates the different objectives of \gls{xai} explainability, expected by different user profiles. For instance, users of the models look at trusting as well as understanding how the model works, while users affected by models' decisions aim to understand their decisions and the main reasons for conducting such decisions. Besides, developers and data scientists expect explanations related to the \gls{ai} models' performance, in order to optimize them over time. However, both regulatory and manager users aim to get more details related to the compliance of \gls{ai} models with the legislation in force to check and assess them.

\subsection{\gls{xai} Metrics}
\label{subsec: metrics}

While human-in-the-loop (HITL) approaches can only yield subjective assessment of the trustworthiness of AI, the existence of objective metrics to characterize the transparency of AI models is a requirement to develop explanation-aware AI systems that exploit such \gls{xai} metrics through a feedback loop to assess the confidence of the models in run-time.
We summarize relevant \gls{xai} metrics in Table \ref{tab:taxonomy_of_metrics}, and compile a list of them in the following.
\begin{itemize}
    \item \textit{Confidence/Faithfulness:} A common approach to measuring the confidence of the explanation relies on the notion of feature relevance. Specifically, observing the effect of muting, i.e., replacing a feature with a baseline value---generally zero---helps to measure the effect on the prediction in both classification and regression tasks \cite{abc}. For instance, for a probabilistic classification model, we can obscure or remove features according to a policy defined as follows
    \begin{equation}
\hat{x}_{i,k}= x_{i,k} \times (1 - p),
\end{equation}
where $p$ is a Bernoulli random variable, $p \sim \mathcal{B}(1, \pi_{i,k})$ and $\pi_{i,k}$ is a probability distribution of the features that can be computed as
\begin{equation}
    \label{eq:softatt}
         \pi_{i,k}= \frac{\exp \Bigl\{\left| a_{i,k}/x_{i,k}\right|\Bigr\}}{\sum _{l=1}^{N}{\exp \Bigl\{\left|a_{l,k}/x_{l,k}\right|\Bigr\}}},\,i=1,\ldots,N, 
\end{equation}
where $N$ is the number of features and $a_{i,k}$ is the attribution of feature $i$ in a sample of class $k$. It is obtained using any attribution-based \gls{xai} method, such as \gls{ig} or \gls{shap}. The confidence score in this case is
\begin{equation}
    c_k = \frac{\Delta_k^{(c)}}{\Delta_k},
    \label{eq:confidence}
\end{equation}
where $\Delta_k^{(c)}$ is the number of samples that conserve their class $k$ after the mutation of the dataset and $\Delta_k$ stands for the original count of samples with class label $k$. For regression tasks, however, the classes are replaced with the notion of groups, which are defined by comparing the continuous prediction output with one or several thresholds.

\item \textit{\acrfull{lo}:} Similarly to the confidence, this score is defined as the average difference of the negative logarithmic probabilities on the predicted class before and after masking the top $p\%$ features with zero padding~\cite{logodds}. Given the attribution scores generated by an explanation algorithm, we select the top $p\%$ features based on their attributions and replace them with zero padding. More concretely, for a dataset with $L$ samples, it is defined as:
\begin{equation}
    \mathrm{log}\text{-}\mathrm{odds}(p)=-\frac{1}{L}\sum_{i=1}^{L}\log \frac{\Pr \left(\hat{y}|\mathbf{x}_i^{(p)}\right)}{\Pr\left(\hat{y}|\mathbf{x}_i\right)}
\end{equation}
where $\hat{y}$ is the predicted class, $\mathbf{x}_i$ is the $i$th sample, and $\mathbf{x}_i^{(p)}$ is the modified samples with top $p\%$ features replaced with zero padding. Lower scores are better.

\item \textit{Comprehensiveness:} is the average difference of the change in predicted class probability before and after removing the top $p\%$ features. Similar to Log-odds, this measures the influence of the top-attributed words on the model's prediction. It is defined as~\cite{suff_comp}:
\begin{equation}
    \mathrm{Comp}(p)=\frac{1}{L}\sum_{i=1}^{L}\left[\Pr\left(\hat{y}|\mathbf{x}_i^{(p)}\right) - \Pr\left(\hat{y}|\mathbf{x}_i\right)\right]
\end{equation}
Here $\mathbf{x}_i^{(p)}$ denotes the modified dataset with top $p\%$ samples deleted. Higher scores are better.

\item \textit{Sufficiency:} is defined as the average difference of the change in predicted class probability before and after keeping only the top $p\%$ features. This measures the adequacy of the top $p\%$ attributions for the model's prediction. Its definition follows the one of comprehensiveness, except for the fact that the $x_i^{(p)}$ is defined as the samples containing only the top $p\%$ features. Lower scores are better~\cite{suff_comp}.

\item \textit{Robustness/Sensitivity:} A crucial property that interpretability methods should satisfy to generate meaningful explanations is that of robustness with respect to local perturbations of the input. This is not the case for popular interpretability methods; even adding minimal white noise to the input introduces visible changes in the explanations~\cite{senn}. To formally quantify the stability of an explanation generation model, one can estimate the Lipschitz constant $\lambda$ for a given input $x_i$ and a neighborhood $B_{\epsilon}$ of size $\epsilon$ as,
\begin{equation}
\label{eq:robustness}
     \lambda(x_i) = \argmax_{{x_j\in B_{\epsilon}(x_i)}} \frac{\lVert \Phi(x_i)-\Phi(x_j)\rVert_2}{\lVert x_i - x_j \rVert_2},
\end{equation}
where the evaluation of the explaining function $\Phi$ for methods like \gls{lime} and \gls{shap} is expensive as it involves model estimation for each query. In contrast, gradient-based attribution methods present a lower complexity. On the other hand, computing (\ref{eq:robustness}) for post-hoc explanation frameworks is much more challenging, since they are not end-to-end differentiable. Thus, one needs to rely on black-box optimization instead of gradient ascent.
This continuous notion of local stability in (\ref{eq:robustness}) might be inadequate for discrete inputs or settings where adversarial perturbations are overly restrictive. In such cases, one can instead define a (weaker) sample-based notion of stability. For any x in a finite set $X = \{x_i\}_{i=1}^{n}$ one replace $B_{\epsilon}(x_i)$ with an $\epsilon$-neighborhood within $X$, i.e., 
\begin{equation}
    \mathcal{N}_{\epsilon}(x)=\{x' \in X | \,\,\lVert x -x'\rVert \leq \epsilon\}.
\end{equation}

\item \textit{Ambiguity:} It indicates how concise is the explanation, i.e., characterized by few prominent features, facilitating interpretation and potentially including higher informational value with reduced noise \cite{complexity}, compared to an ambiguous uniform importance distribution. Indeed, let $N$ denote the number of features. If we map the attributions to a probability space (using e.g., Eq. (\ref{eq:softatt})), the resulting entropy,
\begin{equation}
    \mathcal{H}_k = - \sum_{i=1}^{N} \pi_{i,k} \log(\pi_{i,k}), \label{eq:entropy}
\end{equation}
measures the uncertainty of the output (prediction or decision) with respect to the input (features or states) \cite{sliceops}. On the other hand, when the number of features is very high, one can characterize the uncertainty by comparing the distributions of both the attributions and a reference uniform probability density function. This can be done by invoking the discrete \gls{kl} divergence. The larger the \gls{kl} divergence, the higher the certainty yield by the \gls{xai} method.

\item \textit{Infidelity:} In \gls{xai} surrogate methods, i.e., the schemes that consist of approximating the original model with a low-complexity more interpretable surrogate such as \gls{lime}, the fidelity of the surrogate to the original model can be quantified. Indeed, given a black-box function $f$, explanation functional $\Phi$, a random variable $\mathbf{I}\in \mathbb{R}^n$ with probability measure $\mu_{\mathbf{I}}$, which represents meaningful perturbations of interest, the explanation infidelity can be defined as \cite{fidelity}
\begin{equation}
    \mathcal{I}(\Phi, f, \mathbf{x})=\mathbf{E}_{\mathbf{I} \sim \mu_{\mathbf{I}}}\left[\mathbf{I}^{T}\Phi(f,\mathbf{x}) -(f(\mathbf{x}) - f(\mathbf{x}-\mathbf{I}))^2 \right]
\end{equation}
where $\mathbf{I}$ represents significant perturbations around $\mathbf{x}$ and can be specified in various ways, such as the difference to a baseline $\mathbf{I}= \mathbf{x} - \mathbf{x}_0$.

\item \textit{Fidelity and Soundness:} Two metrics can be applied to evaluate fidelity. Firstly, \cite{lime} used recall ($\mathcal{R}$) as a measure of fidelity for this method, which is defined as
follows,
\begin{equation}
    \mathcal{R} = \frac{|\mathcal{T} \cap \mathcal{E}|}{|\mathcal{T}|}
\end{equation}
where the term True Features $\mathcal{T}$ represents the relevant features as extracted directly from the white box model and Explanation Features $\mathcal{E}$ represents the features characterized as most relevant by the explanation \cite{fidelity2}. This measure indicates how well the explanation captures the most relevant features from the predictive model, i.e., as a measure of the completeness of the explanation. Additionally, to understand how well the explanation excludes irrelevant features (soundness of the explanation), precision ($\mathcal{P}$) can be measured,
\begin{equation}
    \mathcal{P} = \frac{|\mathcal{T} \cap \mathcal{E}|}{|\mathcal{E}|}
\end{equation}

\item \textit{\acrfull{r2} Score:} Behind the workings of \gls{lime} lies the assumption that every complex model is linear on a local scale. \gls{lime} tries to fit a simple model around a single observation that will mimic how the global model behaves at that locality. The simple model can then be used to explain the predictions of the more complex model locally. In this respect, \gls{r2} score is used to measure the performance of the surrogate local model.

\item \textit{Relative Consistency:} 
Let $f_i$ denote a predictor trained over dataset $\mathcal{D}_i$. Explanations arising from different predictors are said to be consistent if they are close when the predictions agree with one another, i.e., given the sets
\begin{equation}
\begin{split}
     \mathcal{S}'=\{\delta_{i,j}(x)| f_i(x) = y \cup f_j(x) = y\} \\
     \mathcal{S}''=\{\delta_{i,j}(x)| f_i(x) = y \xor f_j (x)= y\},
\end{split}
\end{equation}
where $\delta_{i,j}(x)$ is a similarity measure of the explanations $f_i(x)$ and $f_j(x)$, and $\gamma$ is a fixed threshold. We aim at making the gap between the set of consistent explanations $\mathcal{S}'$ and inconsistent ones $\mathcal{S}''$ visible. In this respect, we invoke the true positive rate,
\begin{equation}
    \mathrm{TPR}(\gamma) = \frac{|\{\delta \in \mathcal{S}': \delta \leq \gamma\}|}{|\{\delta \in \mathcal{S}: \delta \leq \gamma\}|},
\end{equation}
where $\mathcal{S}= \mathcal{S}' \cup \mathcal{S}''$. In addition, we also consider the true negative rate,
\begin{equation}
    \mathrm{TNR}(\gamma) = \frac{|\{\delta \in \mathcal{S}'': \delta > \gamma\}|}{|\{\delta \in \mathcal{S}: \delta > \gamma\}|}.
\end{equation}
The quality of these explanations can be assessed independently of the accuracy of the predictor via the \emph{\acrfull{reco}} metric \cite{consistency}:
\begin{equation}
\mathrm{ReCo} = \max_\gamma \mathrm{TPR}(\gamma) + \mathrm{TNR}(\gamma) - 1,
\end{equation}
with a score of $1$ indicating perfect consistency of the predictors' explanations, and a score of $0$ indicating complete inconsistency.

\item \emph{BLEU Score:} Bilingual Evaluation Understudy Score \cite{bleu} is used to evaluate the quality of text generated by a language model by comparing it to one or more reference texts. It measures the overlap of $n$-grams between the generated text and the reference texts.

\item \emph{ROUGE Score:} Recall-Oriented Understudy for Gisting Evaluation \cite{rouge} is a set of metrics for evaluating automatic summarization and machine translation by comparing the overlap of $n$-grams, word sequences, and word pairs between the generated summary/translation and reference texts.

\item \emph{Perplexity:} It is a measurement of how well a language model predicts a sample \cite{perp}. It is defined as the exponentiated average negative log-likelihood of a sequence. Lower perplexity indicates better predictive performance.

\end{itemize}

\begin{table*}[t]
\centering
\caption{Benchmarking of XAI methods for a NeSy-based O-RAN CPU usage prediction task}
\label{tab:benchmark}
\begin{tabular}{lcccccc}
\toprule
\textbf{Epoch} & \textbf{Metric} & \textbf{SHAP} & \textbf{Saliency} & \textbf{Grad $\times$ Input} & \textbf{Int. Grad} & \textbf{LRP} \\
\midrule
100 & Confidence & 0.8981 & 0.8681 & 0.8912 & 0.8588 & 0.8681 \\
 & Ambiguity & 1.6092 & 1.4833 & 1.6086 & 1.6076 & 1.6076 \\
 & Time complexity (s) & 2.1999 & 0.0768 & 0.0815 & 0.1207 & 0.1067 \\
\midrule
500 & Confidence & 0.8900 & 0.8923 & 0.8684 & 0.8828 & 0.8732 \\
 & Ambiguity & 1.6092 & 1.4166 & 1.6085 & 1.6084 & 1.6084 \\
 & Time complexity (s) & 1.8113 & 0.0540 & 0.0530 & 0.0935 & 0.0780 \\
\bottomrule
\end{tabular}
\end{table*}

\subsection{Ranking of XAI Methods in O-RAN Prediction Tasks}\label{sec:ranking}
In this subsection, we present a comparative study of the common classes of XAI methods in the specific task of resource prediction in O-RAN, where Neuro-Symbolic (NeSy) models are an efficient solution~\cite{FMR}. Specifically, we make use of \emph{Logic Tensor Network} to predict CPU usage in a virtual BS (vBS) by leveraging well-established O-RAN experimental datasets~\cite{dataset}. We then assess the explanation \emph{ambiguity} (i.e., lack of evidence) and \emph{confidence} metrics; already described in the previous subsection, as well as the processing time for 1 epoch. Table~\ref{tab:benchmark} summarizes the benchmarking results, which reveal that SHAP presents the higher processing time due to its foundation in game theory, which involves calculating the contribution of each feature to the prediction by considering all possible combinations of features. Gradient methods, including saliency maps, Gradient $\times$ Input, and Integrated Gradients, are less time-consuming compared to SHAP due to their computational efficiency. These methods compute feature attributions using straightforward gradient calculations, involving only a single backward pass or a few integration steps, without the need to evaluate the model on all possible feature subsets. This avoids the combinatorial explosion and significantly reduces computation time. As a result, gradient methods leverage efficient backpropagation algorithms, making them much faster while still providing valuable insights into model predictions.

\subsection{\gls{o-ran} Alliance Specifications}\label{sec:oran}
The \gls{o-ran} Alliance aims to lead the telecom Industry toward designing an intelligent and open \gls{ran}~\cite{oran1}\cite{oran2}
leveraging and extending \gls{3gpp} reference \gls{ran} architecture towards greater flexibility in network deployment and scalability for new services. The O-RAN Alliance aims to foster a more modular and flexible RAN ecosystem by disaggregating software from hardware and establishing open and interoperable interfaces. This approach allows for greater compatibility and interchangeability among different vendors' equipment, enabling network operators to avoid vendor lock-in and embrace a wider range of technology solutions.

The new \gls{o-ran} architecture leverages \gls{nfv} and \gls{sdn} technologies to define new open interfaces and disaggregate the \gls{ran} functional blocks, to allow the deployment of new services and applications. \gls{o-ran} divides the \gls{bbu} of \gls{ran} into three functional blocks, \gls{cu}, \gls{du}, and \gls{ru}. To support control user plane separation, the \gls{cu} block is also divided into control plane \gls{cu}-\gls{cp} and user plane \gls{cu}-\gls{up} sub-blocks. The radio frequency signals are received, transmitted, amplified, and digitized at \gls{ru}, which is located near the antenna, while \gls{cu} and \gls{du} represent the base station's computation parts and are in charge of transmitting the digitalized radio signal to the network.\\ 
We note that in Release 15 \cite{3gpp_release15}, 3GPP introduced a flexible architecture for the 5G RAN. This architecture splits the base station (gNodeB or gNB) into three logical nodes: CU, Responsible for higher-layer functions, coordination, and management. DU, Handles mid-layer functions and connects to the Radio Unit (RU). RU, Deals with lower-layer RF functions. The functional split allows network engineers to optimize performance based on factors like latency, cost, and specific use cases. On the other hand, O-RAN defines the Open RAN concept, which aims for horizontal openness through open interfaces connecting various RAN functions (from RU to DU-CU, controller, and orchestrator). Specifically, O-RAN has standardized the Lower Layer Split (LLS) by defining split option 7-2x \cite{o_ran_lls}. This split results in the Open RU (O-RU) and Open DU (O-DU). Additionally, O-RAN integrates 3GPP-defined interfaces (such as F1, W1, E1, and Xn) within its architecture.

The \gls{du} block may be deployed near or at the RU block, while the \gls{cu} block may be deployed near the core network part. It is also worth noting that \gls{3gpp} has defined different \gls{ran} deployment scenarios and functional split options, which are described in~\cite{Our_surv}~\cite{boutiba}. The two main components introduced by the \gls{o-ran} architecture are summarized below:
\begin{itemize}
    \item \textit{\acrfull{non rt ric}:} it supports non-RT functions (i.e., with a time granularity greater than 1s) such as policy-based guidance. The \gls{non rt ric} is located at the \gls{smo} and comprises two sub-functions: \glspl{rapp} and \gls{non rt ric} framework. The latter is in charge of providing all required services to \glspl{rapp} via the R1 interface, whether from \gls{non rt ric} framework or \gls{smo}, while \glspl{rapp} leverage the functionality provided by the \gls{non rt ric} framework, such as data monitoring via O1 interface (stored in a database), to perform intelligent \gls{ran} optimization functions at non-RT scale. Such functionality enables \glspl{rapp} to get information and trigger actions, e.g., re-configuration and policies. Hence, \gls{non rt ric} enables exposing an intelligent \gls{ran} policy to \gls{near rt ric}, through A1 interface, based mainly on data analytics and \gls{ml}/\gls{dl} inference.\\
    We note that SMO plays a crucial role as an intelligent automation platform that simplifies network complexity, enhances performance, and minimizes operational costs for the RAN domain. Specifically, SMO manages RAN as a service by applying automation at scale, SMO abstracts RAN functions and applications, making them easier to handle. Additionally, SMO interfaces with O1, A1, and O2, overseeing orchestration, management, and automation of RAN elements.
    \item \textit{\acrfull{near rt ric}:} it is in charge of controlling and optimizing the \gls{o-ran} nodes (\gls{cu} and \gls{du}) and their resources through fine-grained data monitoring and actions over E2 interface, at a near RT scale (i.e., from $10$ms to $100$ms). It hosts several \glspl{xapp}, which may collect near RT information (e.g., at a \gls{ue} or Cell basis) through E2 interface, and provide value-added services, with respect to the \gls{non rt ric}'s policies received via the A1 interface. \glspl{xapp} include \gls{mm}, \gls{rm}, \gls{sm}, etc.
\end{itemize}

\section{Projects/Standards on \gls{xai} for \gls{o-ran}}
\label{sec:projects}
\gls{xai} is increasingly becoming critical for the adoption of \gls{ml}/\gls{dl} in \gls{o-ran}. To achieve trustworthiness and transparency in \gls{ml}/\gls{dl} models in \gls{o-ran}, there are some ongoing standardization activities and research projects targeting \gls{xai} and \gls{o-ran} aspects. Some of them include:

\begin{itemize}
    \item \textit{\gls{o-ran} Alliance:} As we describe in Subsection.~\ref{sec:oran}, the \gls{o-ran} Alliance is a global organization that is working to promote an intelligent and open \gls{ran} for mobile cellular networks.
    The \gls{o-ran} Alliance comprises $11$ \glspl{wg} and three focus groups dedicated to \gls{ran} cloudification, automation, and disaggregation. In particular, \gls{wg}2 in~\cite{ai_ml} describes lifecycle management of \gls{ai}/\gls{ml} models on \gls{o-ran} including learning model design, composition, training, runtime, and deployment solutions. It also highlights the main criteria for determining multiple \gls{ml} training and inference host deployment options. In this context, the focus of \gls{wg}2 can be extended to implementing \gls{xai} in \gls{o-ran}. To promote \gls{xai} adoption in \gls{o-ran}, the \gls{wg}2 can work on various initiatives, ranging from XAI models' specifications and  requirements to implementation and deployment. This may also include the creation of \gls{xai} platforms and tools, the development of interfaces and standards for \gls{xai}, and the promotion of \gls{xai} best practices.
    
    \item \textit{IEEE P2894 and P2976:} these standards aim to deliver specifications on \gls{xai} in order to facilitate its adoption in real-world scenarios. The IEEE P2894 standard aims to design an architectural framework and define application guidelines for \gls{xai}, including the definition and description of \gls{xai}, the main classes of \gls{xai} techniques, the main application scenarios of \gls{xai} techniques, and performance evaluations of \gls{xai} in real systems such as telecommunication networks~\cite{ieee-p2894}. Besides, the IEEE P2976 standard is working to achieve interoperability and clarity of \gls{ai} systems design through leveraging \gls{xai} techniques~\cite{ieee-p2976}. Specifically, IEEE P2976 defines optional and mandatory constraints and requirements that should be satisfied for an \gls{ai} algorithm, method, or system to be considered explainable. In this context, these specifications can be leveraged by \gls{o-ran} standards such as \gls{o-ran} Alliance in order to develop and advance the adoption of \gls{xai} in the \gls{o-ran} ecosystem.

    \item \textit{ETSI \acrfull{eni}:} The ETSI \acrfull{isg} is working on defining a cognitive network management architecture based on context-aware policies and leveraging \gls{ai} techniques. This effort aspires to adjust provided services in \gls{5g} networks and beyond based on changes in business goals, environmental conditions, and user requirements. Thus, it aims to provide automated service operation, provisioning, and assurance, along with efficient resource management and orchestration. Besides, recently, ETSI has released its first specifications on \gls{o-ran} called "\gls{o-ran} Fronthaul Control, User and Synchronization Plane Specification v7.02"~\cite{etsi-oran}. This specification focuses on Open Fronthaul as one of the interfaces in the \gls{o-ran} Architecture. It specifies the synchronization plane protocols, user plane, and control plane used over the fronthaul interface to link the O-RU and O-RU components. This specification has been submitted to ETSI as a publicly available specification (PAS) produced by the \gls{o-ran} \gls{wg}4 and approved by the ETSI Technical Committee. Therefore, considering this first specification of ETSI about \gls{o-ran}, the ETSI \gls{eni} \gls{isg} can also focus on adopting \gls{xai} on top of the designed cognitive network architecture in order to create an \gls{ai} framework that is explainable, transparent, and thus can be used to ensure the accountability of \gls{ai}-enabled systems in \gls{o-ran}. 

    \item \textit{\gls{6g}-Bricks:} is a Horizon Europe project that explores novel unified control paradigms based on Explainable \gls{ai} and Machine Reasoning, which will be delivered in the form of a reusable component with open \glspl{api}, termed "bricks"~\cite{6g_bricks}. Initial integration with \gls{o-ran} will be performed, aiming for the future-proofing and interoperability of \gls{6g}-BRICKS outcomes.

    \item \textit{NANCY:} it is the acronym of \emph{An Artificial Intelligent Aided Unified Network for Secure Beyond \gls{5g} Long Term Evolution}; a Horizon Europe project which partly investigates the design of an \gls{xai} engine, to provide transparency and trustworthiness~\cite{nancy}. It also aims to identify the key factors that affect the system's local and overall performance.

    \item \textit{Hexa-X:} developed a user-friendly support to Federated Learning (FL) of explainable-by-design models termed OpenFL-XAI which extends the open-source framework OpenFL \cite{OpenFL-XAI}. Specifically, Hexa-X showed the benefits of building XAI models in a federated manner, with a specific focus on an automotive use case, namely Tele-operated Driving (ToD), which is one of the innovative services envisioned in 6G.
\end{itemize}

Overall, these standards and projects are working to promote the adoption of \gls{ai} techniques, particularly machine learning and deep learning in \gls{o-ran}, while ensuring that these technologies are interpretable, accountable, and transparent. By doing so, they can help build trust in \gls{ai} systems deployed in \gls{o-ran}. Thus, they encourage competition and innovation in the telecommunication industry.

\section{\gls{xai} Deployment on \gls{o-ran}}
\label{sec:xai}
In this section, we describe how \gls{xai} methods can be deployed in the \gls{o-ran} framework and architecture by means of three realistic reference scenarios that are derived from XAI literature. 

\subsection{Introduction and Motivation}
\label{subsec:intro_motivation}
As described in Section.~\ref{sec:oran}, the basic idea of \gls{o-ran} is not only to disaggregate \gls{ran} functions exploiting the flexibility brought by virtualization techniques, but also to design \glspl{ric} that locally host specific \gls{ran} applications (e.g., \glspl{rapp} and \glspl{xapp}), addressing several and heterogeneous control tasks such as handover management, energy management, fault detection, and radio resource allocation. The \gls{o-ran} framework has been devised to natively support a heavy usage of machine/deep learning (\gls{ml}/\gls{dl}) techniques to enhance the development and operations of intelligent \gls{ran} applications to pave the road for future \gls{b5g} network services. For instance, as shown in~\cite{tl}, enabling cooperation among several \glspl{xapp} can help to optimize network performance, both in terms of data throughput and packet delivery ratio.
However, one of the main challenges of \gls{ai}-based \gls{o-ran} management is the lack of transparency on the decision-making processes that govern \gls{ai} algorithms, which makes it difficult for network operators and engineers to diagnose problems, and further optimize the network behavior. 

Therefore, there is a pressing need to integrate \gls{xai} into the \gls{o-ran} management operations, as to gain more detailed information about the decision-making processes of \gls{ml} and \gls{dl} algorithms. Specifically, \gls{xai} techniques should be incorporated into the running \gls{ai}-based \glspl{rapp}/\glspl{xapp} to provide transparent explanations of their outputs. This would not only improve the accuracy and transparency of the decisions made by these systems but also increase the trust of network operators and engineers in the performance of the network.

\subsection{Local Interpretable \gls{ai} Deployment}
\label{subsec:local_interpretable}

The availability of open interfaces and the distributed nature of \gls{ran} deployments allows for the design and implementation of advanced federated and distributed schemes that aim to can overcome traditional \gls{ran} management scalability issues.

\begin{figure}[!t]
	\begin{center}
		\includegraphics[scale=0.50]{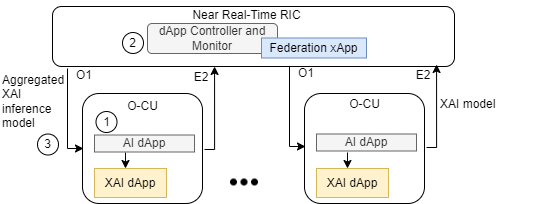}
	\end{center}%
	\caption{Deployment of Federated \gls{ai} and \gls{xai} in \gls{o-ran}. Adapted from \cite{dapps, ai_ml, fuzzyfl, rw9}.}
	\label{fig:fed_models_oran}
\end{figure}
Indeed, to reduce monitoring overhead, reaction time and single point of failure risk, it is always beneficial to process data locally, where they are made available from dedicated monitoring functions. Therefore, raw control plane information generated by end-users at a given cell (or multiple cells) can be processed locally, in the \gls{o-cu}, and used to train \gls{ai}/\gls{ml}-based \emph{dApps} \cite{dapps} and their corresponding local \gls{xai} dApps (\textbf{Step 1}).
As depicted in Fig. \ref{fig:fed_models_oran}, to leverage the distributed nature of \gls{ran} deployments, such local information can be transferred to the \gls{near rt ric}, exploiting the E2 interface, (\textbf{Step 2}).
By combining multiple local models trained over a particular portion of the input space, the \gls{near rt ric} aims to derive more generalized and advanced models to the \gls{o-cu} and the corresponding \gls{dapp}. This information can be provided as feedback (\textbf{Step 3}) via the O1 interface. Hence, leveraging collected data from distributed nodes via the O1 interfaces, predictions along with their corresponding explanations can be performed in real-time. 
Favoured by a continuous learning process, both \gls{ai} and \gls{xai}'s outputs should be considered to perform management decisions and improve network performance. For instance, such outputs can help to update users' scheduling policies or radio resource assignments.
In this context, different \gls{xai} techniques can be leveraged. For instance, RuleFit is one of the most used \gls{xai} techniques~\cite{rule_fit}\cite{rule_fit1}. Its fundamental idea is to capture interactions between the original dataset features in order to create new features in the form of decision rules. Then, RuleFit learns a new transparent learning model using the original features, and also a number of new features that are decision rules.\\
\begin{figure}[!t]
	\begin{center}
		\includegraphics[scale=0.27]{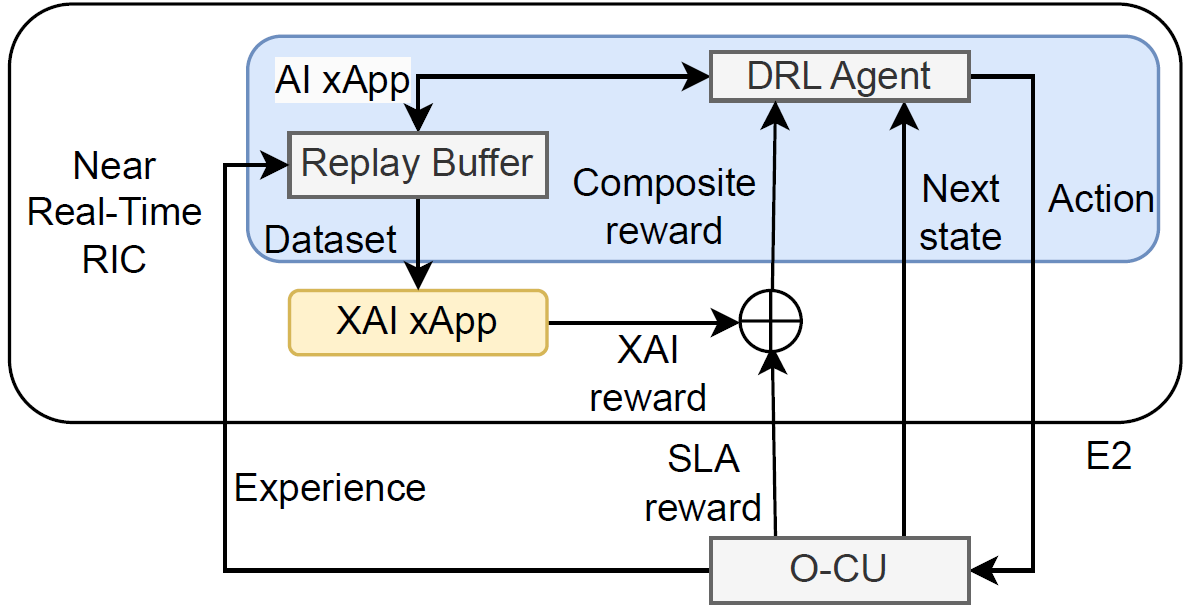}
	\end{center}%
	\caption{Deployment of explanation-guided deep reinforcement learning in \gls{o-ran}. Adapted from \cite{sliceops}.}
	\label{fig:egl}
\end{figure}
Furthermore, the \gls{xai} explainability (outputs) may target different user profiles (\textbf{Step 5}). For instance, users of the models may want to trust and understand how the model works, while explanations related to the \gls{ai} models' performance are sent to developers and data scientists to optimize their behavior over time. In addition, more details about \gls{ai} models' compliance with the legislation in force should be communicated to both regulatory and manager users to check and assess them.

\subsection{Explanation-Guided Deep Reinforcement Learning Deployment}
Undoubtedly, \gls{rl} will play a significant role in enabling smart and automated \gls{o-ran} operations\cite{Sec_ORAN}. In the context of explanation-guided learning \cite{egl, egl2, egfl}, explanation-guided deep reinforcement learning is a branch of artificial intelligence that combines deep learning and reinforcement learning with human-interpretable explanations. The goal of this approach is to enable humans to better understand the decision-making process of \gls{rl} agents. In this method, the \gls{rl} agent learns from its environment while considering human knowledgeable inputs providing explanations for the agent's behavior. The explanations can be in the form of natural language, visualizations, or other means. By providing contextual and external information, a human expert in the field of application can guide the agent toward better decision-making and improve its overall performance.
\begin{figure}[t]
\centering
\includegraphics[scale=0.55]{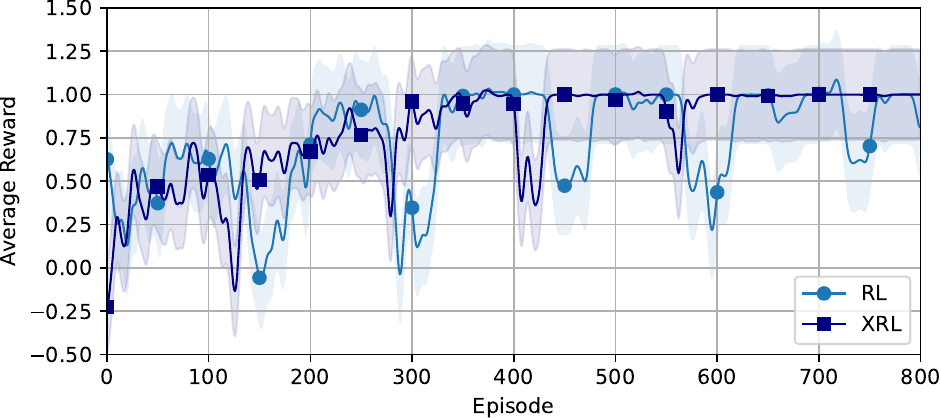}
\caption{Explanation guided DRL maximize the decision confidence compared to DRL. Adapted from \cite{sliceops}.}
\label{fig:rl-xrl-reward}
\end{figure}
As shown in Fig.~\ref{fig:egl}, a \gls{drl} agent at the \gls{near rt ric} performs resource allocation under latency constraints and interacts with the \gls{o-cu} environment through the E2 interface~\cite{oranus}. The agent temporarily stores its experiences and observations in a replay buffer, which is continually updated. Then, the \gls{xai} \gls{xapp} co-located with the \gls{near rt ric} derives the \gls{shap} importance values from a batch state-action dataset extracted from the buffer. To quantify the uncertainty of an action given a specific input state, the obtained \gls{shap} values are afterwards converted to a probability distribution via \emph{softmax} and used to calculate the entropy that measures the uncertainty, as formulated in Eq. (\ref{eq:entropy}). The multiplicative inverse of the maximum entropy value is used as an \gls{xai} reward (to minimize uncertainty and therefore maximize confidence). Combining the \gls{sla} reward (e.g., the multiplicative inverse of the latency) with the \gls{xai} reward results in a composite reward that reduces the uncertainty of state-action pairs and guides the agent to select the best and most explainable actions for specific network state values as illustrated in Fig. \ref{fig:rl-xrl-reward}.

\begin{figure}[!t]
	\begin{center}
		\includegraphics[scale=0.47]{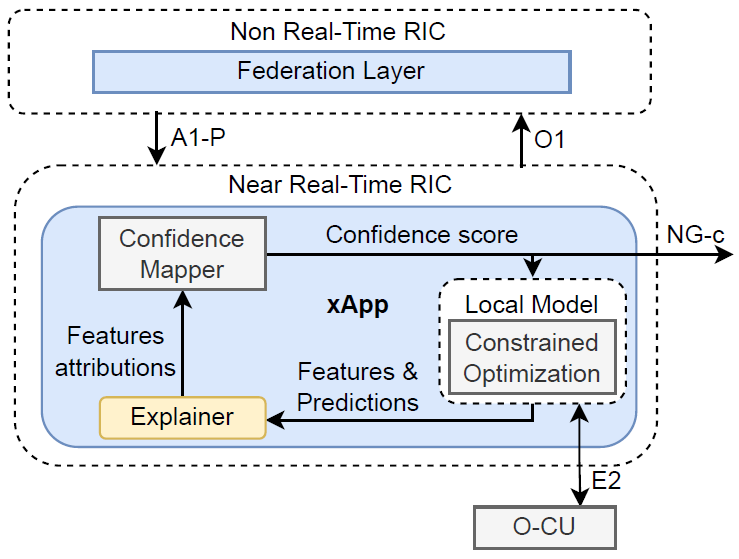}
	\end{center}%
	\caption{Deployment of explanation-aided confident federated learning in \gls{o-ran}. Adapted from \cite{bridging}.}
	\label{fig:robust}
\end{figure}

\subsection{Explanation-Aided Confident Federated Learning Deployment} \label{sec:xai_FL}
Explanation-aided confident \gls{fl} is a type of machine learning that combines federated learning with human-interpretable explanations. In \gls{fl}, data is collected and processed locally on individual devices, and only the necessary information is shared with a central server for model training~\cite{zaka2}\cite{fl_acm}. The goal of explanation-aided confident \gls{fl} is to enable individuals and organizations to collaborate on training models while maintaining privacy and security.

\begin{figure}[t]
    \centering
    \includegraphics[scale=0.65]{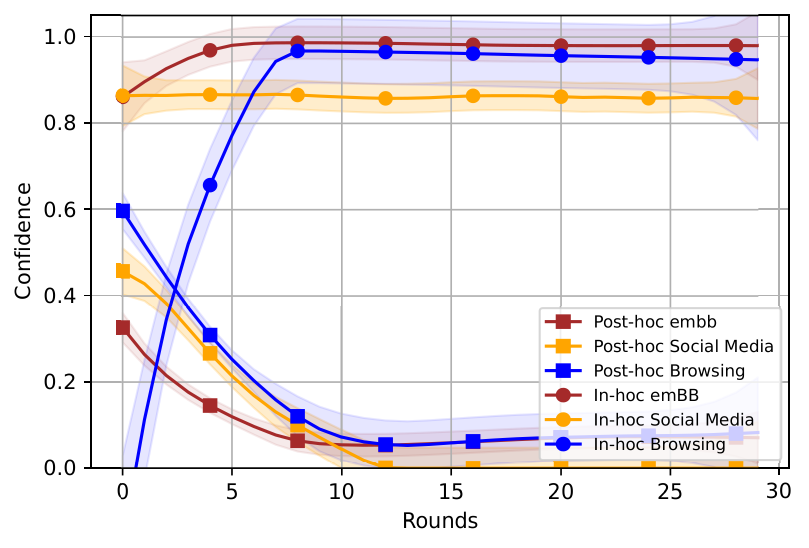}
    \caption{FL confidence vs. rounds. Adapted from \cite{bridging}.}
    \label{fig:confidence}
\end{figure}

To achieve a confident \gls{fl}-based resource allocation/prediction, the local learning is performed iteratively with a run-time explanation as detailed in~\cite{bridging}. The overall working principle of the scheme is manifested in Fig.~\ref{fig:robust}. For each local epoch, the dataset collected through the E2 interface is used to train a local resource allocation model via constrained optimization, which yields the features and the corresponding predictions to the \gls{xai} \gls{xapp} where an \emph{explainer} generates the features attributions using one of the feature attribution \gls{xai} methods (e.g., \gls{shap}, Integrated Gradient, etc.). The \emph{confidence mapper} then converts these attributions to a soft probability distribution and translates it afterwards into a confidence metric according to Eq.~(\ref{eq:confidence}), and feeds it back to the optimizer to include it as an additional constraint in the local optimization. Moreover, the confidence metric is sent via the NG-c interface to the peer \glspl{o-cu}. In this respect, each \gls{o-cu} uses the gathered set of confidence scores to assess its priority, where only the $K$ \glspl{o-cu} with the largest confidence scores out of the available $N$ \glspl{o-cu} take part in the \gls{fl} training to guarantee better confidence. Upon the termination of the local optimization, the model weights are reported to the federation layer---located at the \gls{non rt ric}---to perform model aggregation and broadcast it via the A1-P interface. This iterative procedure results in highly confident FL in O-RAN compared to vanilla post-hoc FL as depicted in Fig. \ref{fig:confidence}.
\begin{figure*}[!t]
	\begin{center}
		\includegraphics[scale=0.65]{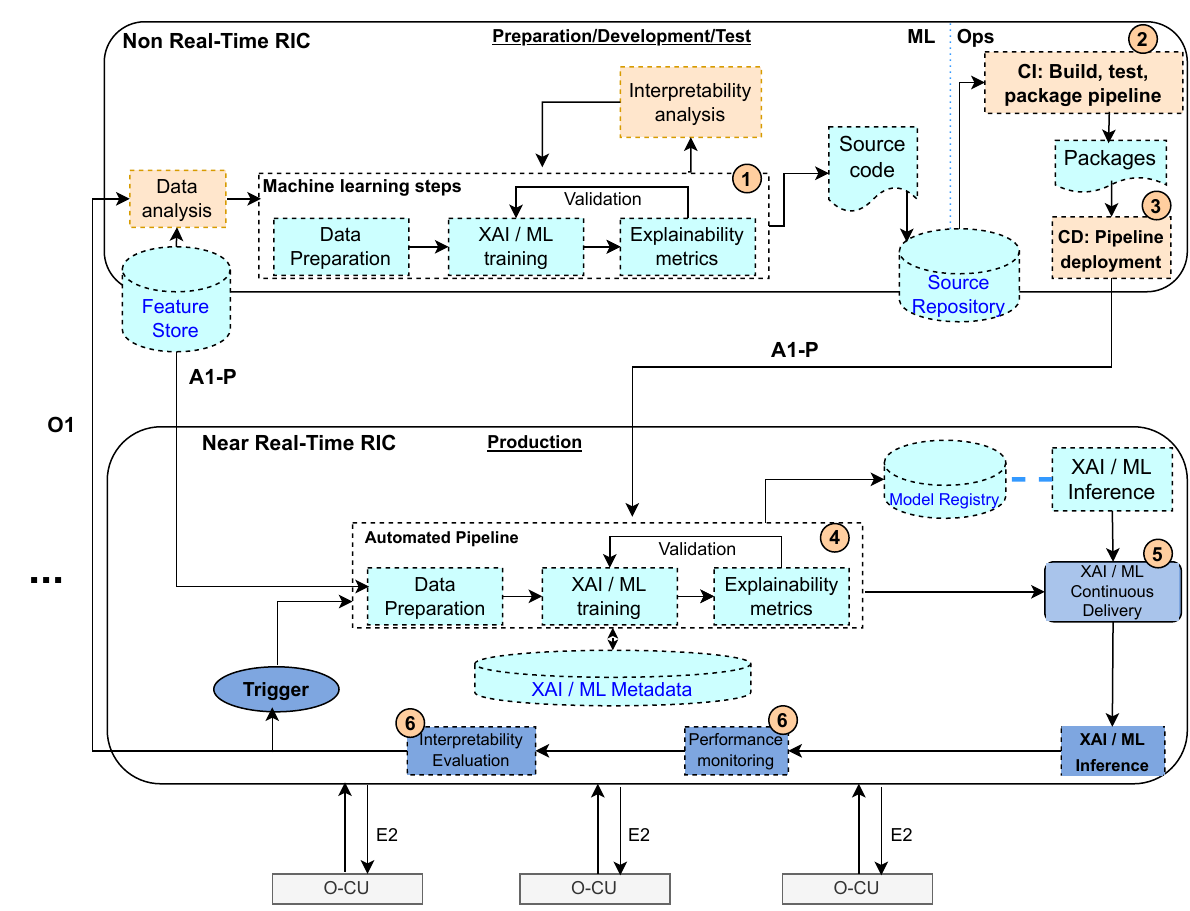}
	\end{center}%
	\caption{XAI-driven Automated Continuous Integration and Delivery Pipeline.}
	\label{fig:lev3}
\end{figure*}

\section{Automation of \gls{ai}/\gls{xai} Pipeline for \gls{o-ran}}
\label{sec:automation}

As discussed in Subsection.~\ref{sec:xai}, \gls{xai} tools can be leveraged to assess the trustworthiness of the \gls{ml}/\gls{dl} models on top of the \gls{o-ran} architecture. In such context, the MLOps pipeline ~\cite{mlop} will be augmented by a model transparency check block in the form of a closed loop that leverages the \gls{xai} objective metrics to evaluate the confidence of O-RAN AI xApps on the fly, as shown in Fig.~\ref{fig:lev3}.

\glsunset{devops}\gls{devops} paradigm includes a set of practices that combines software development (Dev) and IT operations (Ops). \gls{devops} aims not only to reduce the systems' development life cycle but also to provide continuous software delivery with high quality, by leveraging paradigms and concepts like \gls{cicd}. When dealing with machine learning operations, and automation of the learning process, the paradigm can also be called \gls{mlops}~\cite{google1}.

It is worth noting that \gls{o-ran} specification~\cite{ai_ml} introduces three control loops that facilitate the deployment of AI/ML (Artificial Intelligence/Machine Learning) functionalities within the O-RAN framework. 
These control loops are designed to operate at different time scales, enabling efficient integration and utilization of AI/ML capabilities in the network.
Loop $1$ is deployed at \gls{o-du} level to deal with per \gls{tti} scheduling and operates at a timescale of the \gls{tti} or above. Loop $2$ deployed at the \gls{near rt ric} to operate within the range of $10ms-1s$ and above. Loop $3$ at the \gls{non rt ric} at greater than $1$ sec (\gls{ml}/\gls{dl} training, orchestration, etc.). In what follows, we focus more on both loops $3$ and $2$ for \gls{xai} models training, inference, and performance monitoring. Indeed, three main levels of automation have been categorized~\cite{google1}: Manual (no \gls{mlops}), training pipeline automation, and \gls{cicd} pipeline automation. A typical architecture integrating XAI with the MLOps pipeline is introduced in \cite{sliceops}.    

\subsection{Manual Pipeline}
It corresponds to the basic level of maturity, where all the \gls{ml} steps, including data collection and preparation, model training, and validation, are performed manually (cf. Fig.~\ref{fig:lev3}). Hence, it is called no \gls{mlops}. At this level, data scientists usually use a rapid application development tool to build learning models, such as Jupyter Notebooks. In this case, the different steps of \gls{ml} are released at the \gls{non rt ric} module (\gls{ml}), while the trained models are deployed at the \gls{near rt ric} through the A1 interface, to provide prediction services (Ops). Note that the transitions from one step to another are also performed manually, and driven by a source code, developed interactively, till an executable model is created.

In practice, this pipeline corresponds to the learning models, which are rarely updated and often break when they are deployed (models) in the real world. In addition, the performance of learning models at the \gls{ran} environment may degrade, due mainly either to the dynamic evolving of data profiles describing the environment or to the very dynamic changes that may occur in the radio access environment. Hence, automating the whole learning process becomes primordial.

\subsection{Training Pipeline Automation}
This level introduces a continuous training of the models and thus consists of performing the model training steps automatically. In particular, when new data profiles are monitored, the model retraining process is triggered. This process also includes data and model validation phases to achieve continuous delivery of the learning models. This level introduces two new components, named \textit{feature store} as a centralized repository to store features and enable access to new features for training serving, and \textit{machine learning metadata} to store information about the execution of \gls{ml} pipeline (cf. Fig.~\ref{fig:lev3}).\\
We note that the interface A1-P is used to deploy the trained learning model at the near Real-Time RIC. In addition, when new data profiles appeared and thus the learning model should be updated through the automated pipeline, the A1-P is also used to transfer the new data, stored in the Feature Store database, from the Non Real-Time RIC to the Near Real-Time RIC to updated the learning model with its corresponding XAI model in the Near Real-Time RIC entity.

\subsection{Continuous Integration and Delivery Pipeline Automation}
\label{sec:automation_continuous_pipeline}
At this level, a complete \gls{cicd} system is introduced to enable reliable and fast learning model deployments in production. Thus, this level achieves the highest degree of automation in \gls{ml} Ops, by enabling data scientists and developers to efficiently explore new ideas about feature engineering, model hyperparameters, and architecture. The main difference with the previous level is that \gls{cicd} enables building, validating, and deploying the data, learning models, and model training pipeline components automatically. \\
Fig.~\ref{fig:lev3} shows the automation of the \gls{ml} pipeline using \gls{cicd} in \gls{o-ran} context, which mainly features automated both \gls{ml} pipelines and \gls{cicd} routines.\\
In this context, in \cite{RLOps}, the authors introduce principles for applying reinforcement learning (RL) in the O-RAN stack, emphasizing its integration into wireless network research. It reviews current research in this area and applies it to the RAN framework within the O-RAN architecture. The paper proposes a taxonomy to address challenges faced by ML/RL models across their lifecycle--from system specification to production deployment--including data acquisition, model design, testing, and management. To tackle these challenges, the paper integrates existing MLOps principles tailored for RL agents, introducing a systematic model development and validation lifecycle termed RLOps. Key components of RLOps discussed include model specification, development, deployment in production environments, operational monitoring, and ensuring safety/security. The paper concludes by proposing best practices for RLOps to achieve automated and reproducible model development, all within a holistic data analytics platform embedded in O-RAN deployments.



\section{Taxonomy of XAI for 6G O-RAN}
\label{sec:existingXAI-works}

In this section, we give a literature review of existing recent works, which leverage XAI techniques for the 6G O-RAN architecture.\\

\begin{table*}[h]
\centering
\caption{Taxonomy of XAI works for the 6G O-RAN Architecture.}
\label{tab:ExistingXAI-works}

\begin{tabular}{|l|l|l|l|l|}
\hline
\rowcolor[HTML]{EFEFEF} \textbf{\begin{tabular}[c]{@{}l@{}}Ref.\end{tabular}} 
                                             & \textbf{\begin{tabular}[c]{@{}l@{}}XAI Class\end{tabular}}      & \textbf{Target AI Techniques}                                                            & \textbf{O-RAN's challenges}                                                    & \textbf{\begin{tabular}[c]{@{}l@{}}Impacted O-RAN\\  Component\end{tabular}}         \\ \hline \hline 

\cite{T1}                & \multirow{7}{*}{Post-Hoc Explanation} & \multirow{3}{*}{Deep Neural Networks}                                                                          & Security                                                                       & \begin{tabular}[c]{@{}l@{}} rApp, Non Real-Time RIC \end{tabular} \\ \cline{1-1} \cline{4-5}

\cite{T2}                   &       &                                                                           & \begin{tabular}[c]{@{}l@{}}Energy-efficiency and security \end{tabular}  & Near Real-Time RIC                                                            \\ \cline{1-1} \cline{4-5}

\cite{T3}                   &       &                                                                           & \begin{tabular}[c]{@{}l@{}}Misconfiguration and\\ xApps conflict\end{tabular}  & Near Real-Time RIC                                                                   \\ \cline{1-1} \cline{3-5}

\cite{fiandrino2023explora} &       & \begin{tabular}[c]{@{}l@{}}Deep Reinforcement\\ Learning\end{tabular}             & \begin{tabular}[c]{@{}l@{}}Resource \\ Allocation\end{tabular}                 & \begin{tabular}[c]{@{}l@{}}Near Real-Time\\ RIC\end{tabular}                         \\ \cline{1-1} \cline{3-5}

\cite{T4}                   &       & \begin{tabular}[c]{@{}l@{}}Multi-Agent Deep\\ Reinforcement Learning\end{tabular} & \begin{tabular}[c]{@{}l@{}}Resource allocation in\\ O-RAN Slicing\end{tabular} & Non Real-Time RIC                                                                    \\ \cline{1-1} \cline{3-5}

\cite{queries}            &       & \begin{tabular}[c]{@{}l@{}}Convolutional Neural Networks \end{tabular}                & \begin{tabular}[c]{@{}l@{}} O-RAN monitoring overhead \\ reduction \end{tabular}        & \begin{tabular}[c]{@{}l@{}} E2, vBS, Non Real-Time and \\ Near Real-Time RICs \end{tabular}     \\ \cline{1-1} \cline{3-5}

\cite{depth}            &       & \begin{tabular}[c]{@{}l@{}}DLinear, PatchTST and LSTM \end{tabular}                & \begin{tabular}[c]{@{}l@{}} O-RAN traffic forecasting \end{tabular}        & \begin{tabular}[c]{@{}l@{}} Non Real-Time RIC \end{tabular}     \\ \hline

\cite{T5, egfl}             & \begin{tabular}[c]{@{}l@{}}Explanation-Guided\\ Learning\end{tabular} & \multirow{2}{*}{Federated Deep Learning}                & \begin{tabular}[c]{@{}l@{}}Per-slice RAN dropped traffic\\ detection \end{tabular}        & \begin{tabular}[c]{@{}l@{}}Near Real-Time and \\ Non Real-Time RICs\end{tabular}     \\ \cline{1-2} \cline{4-5}

\cite{FMR}            & \begin{tabular}[c]{@{}l@{}}Neuro-Symbolic \\ Reasoning \end{tabular} &                 & \begin{tabular}[c]{@{}l@{}} O-RAN CPU resource \\ provisioning \end{tabular}        & \begin{tabular}[c]{@{}l@{}} vBS, O-CU, O-DU \end{tabular}     \\ \hline

\end{tabular}
\end{table*}

A recent work is leveraging \gls{xai} for \gls{drl} on top of the \gls{o-ran} architecture\cite{fiandrino2023explora}. It addresses resource allocation problems at the O-RAN level and leverages \gls{xai} to provide network-oriented explanations based on an attributed graph, which forms a link between different \gls{drl} agents (graph nodes) and the state space input (the attributes of each graph node). This new scheme, termed EXPLORA, explains the wireless context in which the reinforcement learning agents operate. It shows \gls{xai} can be leveraged in \gls{drl} to perform optimal actions leading to median transmission bitrate and tail improvements of 4\% and 10\%, respectively.\\
In \cite{T1}, the authors discuss XAI-based security architecture for the Open RAN in 6G, named XcARet. This architecture aims to provide transparent and cognitive security solutions for O-RAN while ensuring energy efficiency. They first describe the new security issues of O-RAN due mainly to its open interface and data flow features. Then, they provide recommendations for a dynamic policy of security adjustments, while considering the energy efficiency of the O-RAN architecture. Additionally, they also discussed about how to ensure the transparency of their dynamic security policy by explaining the adjustment decisions. In this context, another work discussed about the security challenges of the O-RAN architecture was proposed in \cite{T2}. The authors discussed about reliable AI and how to design and train rApps and xApps which are robust and secure against attacks. They also discussed on how to prevent, detect, and react to attacks that may target different components of O-RAN. Once an attack is performed and detected, the authors recommend to leverage XAI in order to understand what caused the attack, how it was performed, and eventually learn to recover from it. In this context, the XAI techniques can be applied to provide the non-RT RIC with information on which rApps and xApps were impacted, what type of input caused the attack, and why some applications gave unexpected outputs. This information can then be exploited by the non-RT RIC to re-train the AI models and thus deal with the observed vulnerabilities.\\
In \cite{T3}, the authors address the misconfiguration issues in O-RAN. They present an depth analysis of the potential misconfiguration issues in O-RAN with respect to the use of NFV and SDN, specifically, the use of AI/ML. They investigated how AI/ML can be used to identify the O-RAN misconfigurations. A case study is proposed to show the impact on the UEs of conflicting policies amongst xApps, along with a potential AI-based solution. As AI finds use at different levels of O-RAN, the authors stress the need for XAI for O-RAN, especially for safety-critical use cases such as transportation automation, vital infrastructure operation (e.g., nuclear energy and water), human-machine brain interfaces, and healthcare. \\
In \cite{T5, egfl}, the authors got inspired by XAI and closed-loop automation to design an Explainable Federated deep learning (FDL) model to predict per-slice RAN dropped traffic probability in a non-IID setup, while jointly considering the explainability and sensitivity-aware metrics as constraints. Specifically, they quantitatively validate the explanations of faithfulness through the so-called attribution-based log-odds metric that is included as a constraint in the run-time FL optimization problem.\\
A novel multi-agent deep reinforcement learning (MADRL) framework, named standalone explainable protocol (STEP), for 6G O-RAN slicing was proposed in \cite{T4}. STEP enables Slices' orchestration agents to learn and adapt resource allocation while ensuring their post-hoc explainability, thanks to XAI. 
It is based on an information bottleneck framework to extract the most relevant information from running network slices at the O-RAN level, thus ensuring efficient decision-making and communication.

Moreover, in \cite{queries}, the paper demonstrates how XAI can enhance the design of xApps by presenting a case study on an ML model trained for traffic classification using O-RAN KPIs. Utilizing SHAP, the study identifies the most influential KPIs for the model's predictions. By training the model with these selected KPIs, the research aims to reduce the overhead of transmitting all KPIs while observing the impact on model accuracy. Unlike existing works focusing solely on feature contribution, this paper uniquely leverages XAI to refine the training features based on their contribution. The contributions include: a SHAP-based XAI framework to identify key KPIs for traffic classification; two methods to reduce the number of KPIs-top $K$ overall and top $K$ per class-resulting in only a $7\%$ accuracy drop with fewer KPIs; and an analysis showing a $33\%$ reduction in control traffic data rate while maintaining high accuracy.

Additionally, \cite{depth} advances mobile traffic forecasting by introducing AIChronoLens, which links XAI explanations with temporal input properties. This approach addresses shortcomings in legacy XAI techniques and enables a direct comparison of different AI models on the same dataset, enhancing their integration into the O-RAN architecture. Specifically, the AIChronoLens is used to explain DLinear, PatchTST, and LSTM in a prediction task of vBS mobile traffic and RRC connected users.

Finally, \cite{FMR} introduces the Federated Machine Reasoning (FLMR) framework, a neuro-symbolic approach tailored for federated reasoning. FLMR enhances CPU demand prediction by leveraging contextual data and vBS configuration specifics from local monitoring within a shared O-Cloud platform of O-RAN. The framework ensures transparency in AI/ML decisions, addressing while comparative analysis against the DeepCog baseline demonstrates superior performance, achieving a six-fold reduction in resource under- and over-provisioning.

As we can observe from Table \ref{tab:ExistingXAI-works}, there are multiple works leveraging XAI for the O-RAN architecture to provide more trust, transparency, and robustness to the AI-empowered solutions. The proposed works addressed the resource allocation, security, and misconfiguration of O-RAN xApps at the non Real-Time and Near Real-Time RICs. They leverage mainly post-hoc explanation and explanation-guided learning to explain different AI algorithms such as Deep Reinforcement Learning and supervised Deep Federated Learning. \\
The above works motivate our study and show the need to provide a comprehensive survey of \gls{xai} and its potential in designing the future \gls{o-ran} to guide the practitioners as well as researchers.

\section{Mapping of Existing \gls{ai}-based \gls{o-ran} works to \gls{xai}-enabled solution}
\label{sec:xai_oran}

In this section, we first give a literature review of existing works, which leverage \gls{ai} (\gls{ml}/\gls{dl}) techniques on top of the \gls{o-ran} architecture, in order to optimize \gls{ran} functions. We then discuss how these works can be mapped to \gls{xai} methods.

\subsection{Existing \gls{ai}-driven \gls{o-ran} Works}
\label{sec:xai_oran_existing}
\begin{itemize}
\item \textit{User Access Control:} The user access control or user association challenge is addressed in~\cite{rw1}\cite{rw2}, in order to ensure load balancing among \glspl{bs} and avoid frequent handovers. The authors designed a federated deep reinforcement learning. The \glspl{ue} collaboratively trained their local models and then aggregated them at the \gls{ric} level. The designed model succeeded in maximizing the overall \glspl{ue}' throughput and reducing frequent handovers.
\item \textit{Attack Detection:} In \cite{scalingi2024infocom}, the authors tackle security vulnerabilities in RAN cellular networks, focusing on the lack of integrity protection in the Radio Resource Control (RRC) layer. They propose a real-time anomaly detection framework using distributed applications in 5G Open RAN networks. By leveraging AI, they identify legitimate message sources and detect suspicious activities through Physical Layer features, which generate reliable fingerprints, infer the time of arrival of unprotected uplink packets, and handle cross-layer features. Their approach, validated in emulation environments with over 85\% accuracy in attack prediction, is integrated into a real-world prototype with a large channel emulator. It meets the 2 ms low-latency real-time constraint, making it suitable for real-world deployments.
\item \textit{Energy-Aware RAN scalability:} In \cite{maxenti2024}, the authors introduce ScalO-RAN, an optimization-based control framework designed as an O-RAN rApp to allocate and scale AI-based O-RAN applications (xApps, rApps, dApps). This framework ensures application-specific latency requirements are met, monetizes shared infrastructure, and reduces energy consumption. ScalO-RAN is prototyped on an OpenShift cluster with base stations, RIC, and AI-based xApps deployed as micro-services. Numerical and experimental evaluations show that ScalO-RAN optimally allocates and distributes O-RAN applications within computing nodes to meet stringent latency requirements. The study highlights that scaling O-RAN applications is primarily a time-constrained issue, necessitating policies that prioritize AI applications' inference time over resource consumption.
\item \textit{Channel State Information (CSI):} a novel research platform for real-time inference using AI-enabled CSI feedback, closely simulating real-world scenarios, is designed in \cite{CSI-ORAN}. The framework is validated by integrating a CSI auto-encoder into the OpenAir-Interface (OAI) SG protocol stack. The authors demonstrate real-time functionality with the encoder at the User Equipment (UE) and the decoder at the Next Generation Node Base (gNB). The experiments are conducted on both an Over-the-Air (OTA) indoor testbed platform, ARENA, and on the Colosseum wireless network emulator. 
\item \textit{Total Cell Throughput:} An online training environment of a reinforcement learning model is deployed at the \gls{ric} level in~\cite{rw3}. The developed model controlled function parameters in DU, to maximize total cell throughput. Thanks to the deployed learning model, the total cell throughput increased by $19.4\%$.
\item \textit{SLA-Aware Network Slicing:} The authors in \cite{raftopoulos2024} propose a Deep Reinforcement Learning (DRL) agent for O-RAN applications, specifically for RAN slicing with Service Level Agreements (SLAs) focused on end-to-end latency. Using the OpenRAN Gym environment, the DRL agent adapts to varying SLAs and outperforms state-of-the-art methods, achieving significantly lower SLA violation rates and resource consumption without the need for re-training.
\item \textit{Function Placement:} The \gls{o-ran} architecture leverages virtualization and disaggregation of \gls{ran} functionalities among three key units (RU, DU, and CU). The authors of~\cite{rw4} studied the placement of resource allocation function based on service requirements, by dynamically selecting CU-DU units. Thus, they generated two reinforcement learning models based on Actor-Critic. The first one is used to assign resource blocks to \glspl{ue} according to traffic types, delay budget, and \glspl{ue} priorities, while the second one is leveraged to optimize function placement and hence the decisions of resource allocation. The authors showed that through this dynamic placement, both latency and throughput are highly improved.
\item \textit{RAN Orchestration:} In \cite{OrchestRAN}, the authors present OrchestRAN, a network intelligence orchestration framework for next-generation systems based on the Open Radio Access Network (RAN) paradigm. Designed to function in the non-Real-time (RT) RAN Intelligent Controller (RIC) as an rApp, OrchestRAN allows Network Operators (NOs) to specify high-level control and inference objectives, such as scheduling adjustments and near-RT capacity forecasting for specific base stations. OrchestRAN automatically selects the optimal set of data-driven algorithms and their execution locations (cloud or edge) to fulfill the NOs' objectives, ensuring timing requirements are met and preventing conflicts between algorithms managing the same parameters.
\item \textit{Resource allocation:} In~\cite{rw5}~\cite{rw6}~\cite{rw7}, the authors studied the multi-agent team learning deployment on top of the \gls{o-ran} architecture by deciding about each agent placement and the required \gls{ai} feedback. As a case study, the authors addressed the challenge of how to coordinate several running and independent \glspl{xapp} in \gls{o-ran}. They designed two \glspl{xapp}, called resource allocation \gls{xapp} and power control \gls{xapp}, and then used federated deep reinforcement learning to enhance learning efficiency as well as network performance in terms of throughput and latency. Similarly, in~\cite{rw8}, the authors aimed to deal with the conflicts that may occur among running \glspl{xapp} when deployed by different vendors. Leveraging Q-learning, they proposed a team learning algorithm for resource allocation, to increase cooperation between \glspl{xapp} and hence optimize the performance of the network. \\
Another distributed \gls{rl} model was generated in~\cite{rw9}, to manage \gls{ran} slice resource orchestration on top of the \gls{o-ran} architecture. The distributed \gls{rl} architecture is composed of multiple intelligent agents, one for each network slice, performing local radio allocation decisions. Similarly, in~\cite{rw11}, the authors leveraged federated distributed \gls{rl} to manage the radio resource allocation among multiple \glspl{mvno} for two different network slices (\gls{urllc} and \gls{embb}). In~\cite{rw10}, the challenge of how to optimally assign DU resources for various \glspl{ru} is studied. A deep reinforcement learning model is built to achieve efficient management of \glspl{ru}-\gls{du} resources. Experimental results showed that the proposed scheme improves highly resource usage efficiency. In the same context of resource allocation, in \cite{tsampazi2024pandora}, the authors present PandORA, a framework for automatically designing and training DRL agents for Open RAN applications, packaging them as xApps, and evaluating them in the Colosseum wireless network emulator. They benchmark 23 xApps embedding DRL agents trained with various architectures, reward designs, action spaces, and decision-making timescales, enabling hierarchical control of different network parameters. These agents are tested on the Colosseum testbed under diverse traffic and channel conditions, both static and mobile. The experimental results show that fine-tuning RAN control timers and selecting appropriate reward designs and DRL architectures can significantly enhance network performance based on conditions and demand.

\end{itemize}

\begin{table*}[]
\caption{ Mapping of Existing \gls{ai}-based \gls{o-ran} works to \gls{xai}-enabled Solutions.}
\label{tab:mapping_to_xai}
\centering
\begin{tabular}{|l|l|l|cllll|}
\hline
\rowcolor[HTML]{EFEFEF} 
\multicolumn{1}{|c|}{\cellcolor[HTML]{EFEFEF}} &
  \multicolumn{1}{c|}{\cellcolor[HTML]{EFEFEF}} &
  \multicolumn{1}{c|}{\cellcolor[HTML]{EFEFEF}} &
  \multicolumn{5}{c|}{\cellcolor[HTML]{EFEFEF}\textbf{\gls{xai} Deployment at \gls{ric} as \glspl{xapp}}} \\ \cline{4-8} 
\rowcolor[HTML]{EFEFEF} 
\multicolumn{1}{|c|}{\multirow{-2}{*}{\cellcolor[HTML]{EFEFEF}\textbf{Works}}} &
  \multicolumn{1}{c|}{\multirow{-2}{*}{\cellcolor[HTML]{EFEFEF}\textbf{\begin{tabular}[c]{@{}c@{}}Addressed \gls{ran} \\ Function\end{tabular}}}} &
  \multicolumn{1}{c|}{\multirow{-2}{*}{\cellcolor[HTML]{EFEFEF}\textbf{\gls{ai} Technique}}} &
  \multicolumn{1}{c|}{\cellcolor[HTML]{EFEFEF}\textbf{\begin{tabular}[c]{@{}c@{}}\gls{xai} \\ Technique\end{tabular}}} &
  \multicolumn{1}{c|}{\cellcolor[HTML]{EFEFEF}\textbf{Metrics}} &
  \multicolumn{1}{c|}{\cellcolor[HTML]{EFEFEF}\textbf{\begin{tabular}[c]{@{}c@{}}\gls{o-ran}\\ Module\end{tabular}}} &
  \multicolumn{1}{c|}{\cellcolor[HTML]{EFEFEF}\textbf{\begin{tabular}[c]{@{}c@{}}Functional \\ Blocks\end{tabular}}} &
  \multicolumn{1}{c|}{\cellcolor[HTML]{EFEFEF}\textbf{Interfaces}} \\ \hline\hline
  
\begin{tabular}[c]{@{}l@{}}Yang et al. \\ \cite{rw1}\\ \cite{rw2}\end{tabular} &
  User access control &
  \begin{tabular}[c]{@{}l@{}}Federated deep\\ reinforcement \\ learning\end{tabular} &
  \multicolumn{1}{c|}{} &
  \multicolumn{1}{l|}{Confidence} &
  \multicolumn{1}{l|}{\glsunset{o-cu-cp}\gls{o-cu-cp}} &
  \multicolumn{1}{l|}{\begin{tabular}[c]{@{}l@{}}\gls{ue} and \glsunset{gnb}\gls{gnb} \\ procedure\\ management\end{tabular}} &
   \\ \cline{1-3} \cline{5-7}
\begin{tabular}[c]{@{}l@{}}Hoejoo et al. \\ \cite{rw3}\end{tabular} &
  Total cell throughput &
  \begin{tabular}[c]{@{}l@{}}Deep \\ reinforcement \\ learning\end{tabular} &
  \multicolumn{1}{c|}{} &
  \multicolumn{1}{l|}{State-action certainty} &
  \multicolumn{1}{l|}{\gls{o-du}} &
  \multicolumn{1}{l|}{\begin{tabular}[c]{@{}l@{}}Resource \\ assignment \\ (\glsunset{nr-mac}\gls{nr-mac})\end{tabular}} &
   \\ \cline{1-3} \cline{5-7}
\begin{tabular}[c]{@{}l@{}}Shahram et al. \\ \cite{rw4}\end{tabular} &
  \begin{tabular}[c]{@{}l@{}}Resource allocation \\ and function placement\end{tabular} &
  \begin{tabular}[c]{@{}l@{}}Actor-Critic \\ learning\end{tabular} &
  \multicolumn{1}{c|}{} &
  \multicolumn{1}{l|}{State-action certainty} &
  \multicolumn{1}{l|}{\gls{o-du}} &
  \multicolumn{1}{l|}{\begin{tabular}[c]{@{}l@{}}Resource \\ assignment \\ (\gls{nr-mac})\end{tabular}} & \multirow{-7}{*}{\begin{tabular}[c]{@{}l@{}}\rotatebox{90}{E2 (Realization).}\\ \rotatebox{90}{A1 (Analytics and Policies),}\\ \rotatebox{90}{O1 (Monitoring),}\end{tabular}}
   \\ \cline{1-3} \cline{5-7}
\begin{tabular}[c]{@{}l@{}}Rivera et al.\\ \cite{rw5}\\ \cite{rw6}\\ Han et al. \\ \cite{rw7}\end{tabular} &
  \begin{tabular}[c]{@{}l@{}}Resource allocation\\ and power control\end{tabular} &
  \begin{tabular}[c]{@{}l@{}}Federated deep \\ reinforcement \\ learning\end{tabular} &
  \multicolumn{1}{c|}{\multirow{-7}{*}{\rotatebox{90}{Reactive/Proactive Explanations}}} &
  \multicolumn{1}{l|}{Confidence, Log-odds} &
  \multicolumn{1}{l|}{\gls{o-du}} &
  \multicolumn{1}{l|}{\begin{tabular}[c]{@{}l@{}}Resource \\ assignment \\ (\gls{nr-mac}) and\\ PDSCH \\ (High-PHY)\end{tabular}} & 
   \\ \cline{1-3} \cline{5-7}
\begin{tabular}[c]{@{}l@{}}Han et al.\\ \cite{rw8}\end{tabular} &
  Resource allocation &
  Q-learning &
  \multicolumn{1}{c|}{} &
  \multicolumn{1}{l|}{State-action certainty} &
  \multicolumn{1}{l|}{\gls{o-du}} &
  \multicolumn{1}{l|}{\begin{tabular}[c]{@{}l@{}}Resource \\ assignment \\ (\gls{nr-mac})\end{tabular}} &
   \\ \cline{1-3} \cline{5-7}
\begin{tabular}[c]{@{}l@{}}Farhad et al.\\ \cite{rw9}\end{tabular} &
  Resource allocation &
  \begin{tabular}[c]{@{}l@{}}Distributed deep\\ reinforcement \\ learning\end{tabular} &
  \multicolumn{1}{c|}{} &
  \multicolumn{1}{l|}{Robustness} &
  \multicolumn{1}{l|}{\gls{o-du}} &
  \multicolumn{1}{l|}{\begin{tabular}[c]{@{}l@{}}Resource \\ assignment \\ (\gls{nr-mac})\end{tabular}} &
   \\ \cline{1-3} \cline{5-7}
\begin{tabular}[c]{@{}l@{}}Wang et al.\\ \cite{rw10}\end{tabular} &
  Resource allocation &
  \begin{tabular}[c]{@{}l@{}}Deep \\ reinforcement\\ learning\end{tabular} &
  \multicolumn{1}{c|}{} &
  \multicolumn{1}{l|}{State-action certainty} &
  \multicolumn{1}{l|}{\begin{tabular}[c]{@{}l@{}}\gls{o-du} \\ O-RU\end{tabular}} &
  \multicolumn{1}{l|}{\begin{tabular}[c]{@{}l@{}}Resource \\ assignment \\ (\gls{nr-mac})\end{tabular}} & \\ \cline{1-3} \cline{5-7}
\begin{tabular}[c]{@{}l@{}}Abouaomar et al.\\ \cite{rw11}\end{tabular} &
  Resource allocation &
  \begin{tabular}[c]{@{}l@{}}Federated Deep \\ reinforcement\\ learning\end{tabular} &
  \multicolumn{1}{c|}{} &
  \multicolumn{1}{l|}{Confidence, Log-odds} &
  \multicolumn{1}{l|}{\begin{tabular}[c]{@{}l@{}}\gls{o-du} \\ \end{tabular}} &
  \multicolumn{1}{l|}{\begin{tabular}[c]{@{}l@{}}Resource \\ assignment \\ (\gls{nr-mac})\end{tabular}} & \\  \cline{1-3} \cline{5-7}
 
\begin{tabular}[c]{@{}l@{}}Scalingi et al.\\ \cite{scalingi2024infocom}\end{tabular} &
  Attack Detection &
  \begin{tabular}[c]{@{}l@{}}CNN, LSTM \\ DRL\end{tabular} &
  \multicolumn{1}{c|}{} &
  \multicolumn{1}{l|}{Confidence, Log-odds} &
  \multicolumn{1}{l|}{\begin{tabular}[c]{@{}l@{}}O-RU \\ \end{tabular}} &
  \multicolumn{1}{l|}{\begin{tabular}[c]{@{}l@{}}Low \\ Physical \\ \end{tabular}} & \\    
   \cline{1-3} \cline{5-7}
  
\begin{tabular}[c]{@{}l@{}}Maxenti et al.\\ \cite{maxenti2024} \end{tabular} &
  \begin{tabular}[c]{@{}l@{}}Energy-Aware \\ scalability\end{tabular} & 
  \begin{tabular}[c]{@{}l@{}}CNN, LSTM \\ DRL\end{tabular} &
  \multicolumn{1}{c|}{} &
  \multicolumn{1}{l|}{Confidence, Log-odds} &
  \multicolumn{1}{l|}{\begin{tabular}[c]{@{}l@{}}O-CU\\ and O-DU\\ \end{tabular}} &
  \multicolumn{1}{l|}{\begin{tabular}[c]{@{}l@{}}UE and gNB \\ procedure \\ and NR-MAC \\ \end{tabular}} & \\     \cline{1-3} \cline{5-7}

\begin{tabular}[c]{@{}l@{}}Cheng et al.\\ \cite{CSI-ORAN} \end{tabular} &
  \begin{tabular}[c]{@{}l@{}}Channel State \\ Information\end{tabular} &
  \begin{tabular}[c]{@{}l@{}}Encoder \\ Decoder\end{tabular} &
  \multicolumn{1}{c|}{} &
  \multicolumn{1}{l|}{Confidence, Log-odds} &
  \multicolumn{1}{l|}{\begin{tabular}[c]{@{}l@{}}O-DU\\ \end{tabular}} &
  \multicolumn{1}{l|}{\begin{tabular}[c]{@{}l@{}}PUSCH \\ and PDSCH \\ \end{tabular}} & \\     \cline{1-3} \cline{5-7}

\begin{tabular}[c]{@{}l@{}}Raftopoulos et al.\\ \cite{raftopoulos2024} \end{tabular} &
  \begin{tabular}[c]{@{}l@{}}SLA-Aware \\ Network Slicing\end{tabular} &
  \begin{tabular}[c]{@{}l@{}}DRL\end{tabular} &
  \multicolumn{1}{c|}{} &
  \multicolumn{1}{l|}{Confidence, Log-odds} &
  \multicolumn{1}{l|}{\begin{tabular}[c]{@{}l@{}}O-CU,\\ O-DU\\ \end{tabular}} &
  \multicolumn{1}{l|}{\begin{tabular}[c]{@{}l@{}}Resource \\ assignment \\(NR-MAC) \end{tabular}} & \\     \cline{1-3} \cline{5-7}

\begin{tabular}[c]{@{}l@{}} D'Oro et al.\\ \cite{OrchestRAN} \end{tabular} &
  \begin{tabular}[c]{@{}l@{}}RAN \\ Orchestration\end{tabular} &
  \begin{tabular}[c]{@{}l@{}}General\end{tabular} &
  \multicolumn{1}{c|}{} &
  \multicolumn{1}{l|}{Confidence, Log-odds} &
  \multicolumn{1}{l|}{\begin{tabular}[c]{@{}l@{}}O-CU, \\O-DU,\\ O-RU\\ \end{tabular}} &
  \multicolumn{1}{l|}{\begin{tabular}[c]{@{}l@{}}All \\Blocks \end{tabular}} & \\     \cline{1-3} \cline{5-7}

\begin{tabular}[c]{@{}l@{}} Tsampazi et al.\\ \cite{tsampazi2024pandora} \end{tabular} &
  \begin{tabular}[c]{@{}l@{}} Resource \\ Allocation\end{tabular} &
  \begin{tabular}[c]{@{}l@{}}DRL\end{tabular} &
  \multicolumn{1}{c|}{} &
  \multicolumn{1}{l|}{Confidence, Log-odds} &
  \multicolumn{1}{l|}{\begin{tabular}[c]{@{}l@{}}O-DU\\ \end{tabular}} &
  \multicolumn{1}{l|}{\begin{tabular}[c]{@{}l@{}}Resource \\assignment\\ (NR-MAC) \end{tabular}} & \\     \cline{1-3} \cline{5-7}

\hline

\end{tabular}
\end{table*}
\subsection{How \gls{xai} can Help}
\label{sec:xai_oran_how_can_help}
Integrating \gls{ml}/\gls{dl}-based algorithms with \gls{ran} functionalities has been found to address many challenging tasks: power control, user access control, handover, resources management, etc., which accordingly helps to optimize the performance of the \gls{ran} part. This was highly motivated in \gls{o-ran}, especially with the introduction of \gls{ric} modules. Indeed, the \gls{ran} functions are usually formulated as \gls{mdp}~\cite{mdp}, which explains the wide application of reinforcement learning, e.g., Q-learning, deep \gls{dqn}, and Actor-Critic, either in a centralized or a federated way, in order to derive the optimal policy about the corresponding \gls{ran} function. In addition, team learning is also an emerging paradigm to optimize the coordination and control of the running \glspl{xapp} at the \gls{o-ran}'s \glspl{ric}.\\  
It is worth noting that resource management is the most studied \gls{ran} function using feature engineering approaches, such as feature extraction and feature selection, in addition to reinforcement learning algorithms (\gls{dqn} and Q-learning). With this strategy of \gls{ran} functions analysis, it is possible to determine the contribution of every feature, e.g., higher-order cumulants, related to the \gls{ran} performances. This helps to adjust the features' values to optimize the \gls{ml}/\gls{dl} predictions. In fact, humans/users' understanding of Q-learning models is limited to small scenarios involving a few states and actions. However, these models may become complex, especially with a high number of features, states, and actions, making them less interpretable by humans. The challenge here is the accuracy-interpretability trade-off, which means that the greater the accuracy, the less likely the model is interpretable and vice versa. For example, the Q-learning model can improve the overall performance of radio resource management by exploiting more descriptive frequency and time features, but its complexity increases when considering more features including network density and network/user power, service requirements, and the trust and security of the wireless communications, and thus this will introduce more states and actions in the system. Besides, despite being a more advanced algorithm as compared to Q-learning, \gls{dqn} gives black-box models that output a lack of explainability. For instance, radio resource allocation using a \gls{dqn} can introduce many ambiguous points, which should be explained, such as which layers/neurons in the \gls{dqn} architecture can help to improve the accuracy, and why some \glspl{ue} get the same number of radio block than others, even with different service requirements (\gls{urllc}, \gls{embb}, \gls{mmtc}). In this context, \gls{xai} is highly recommended since it provides profound insights into automated decisions and predictions. These details can help different users, as well as the network operators, to deal with unexpected problems, either related to the \gls{ml}/\gls{dl} models or to the corresponding \glspl{xapp} of the \gls{o-ran}'s \glspl{ric}. Therefore, the performance of the different \gls{ran} functions can be highly enhanced.\\
Within this context, a recent work is leveraging \gls{xai} for \gls{drl} on top of the \gls{o-ran} architecture\cite{fiandrino2023explora}. This work addresses resource allocation and control challenges on top of the \gls{o-ran} architecture. It leverages \gls{xai} to provide network-oriented explanations based on an attributed graph which forms a link between different \gls{drl} agents (graph nodes) and the state space input (the attributes of each graph node). This new scheme explains the wireless context in which the reinforcement learning agents operate. It shows \gls{xai} can be leveraged in \gls{drl} to perform optimal actions leading to median transmission bitrate and tail improvements of 4\% and 10\%, respectively.

\subsection{Mapping to \gls{xai}-enabled Works}
\label{sec:xai_oran_mapping}

In Table~\ref{tab:mapping_to_xai}, we compare the existing \gls{ai}-driven \gls{o-ran} works according to several criteria, including the addressed \gls{ran} function and the leveraged \gls{ai} techniques. In addition, we illustrate how \gls{xai} can be deployed, as \glspl{xapp}, on top of these works, to explain their \gls{ai}-based decisions.
We observe that most of the existing works are based on reinforcement learning to manage \gls{ran} functions, especially power and resource allocation, user access control, function placement, and total cell throughput. 
According to~\cite{xdrl}, two main groups of \gls{xai} techniques can be applied to reinforcement learning strategies, to give both local and global explanations.\\
\begin{itemize}
    \item \textit{Reactive \gls{xai} techniques} imply that explanations are given in an immediate time horizon. It includes three main approaches: $i$) \textit{Policy simplification}, which aims to simplify the policy of transition function into a form that is interpretable/understandable by itself, by using for instance decision trees and fuzzy rule-sets~\cite{pol_sim}. $ii$) \textit{Reward decomposition} into understandable components, which aims to better understand the reasons for certain reward values~\cite{rew_deco}. $iii$) \textit{Feature contribution and visual methods}, to determine features' contribution to the decision-making process, and then generate explanations based on that contribution. Examples of such techniques are both \gls{lime}~\cite{lime} and \gls{shap}~\cite{DeepShape}.\\
    \item \textit{Proactive \gls{xai} techniques} focus on a longer time horizon to provide the required explanations. These techniques can be classified into four main classes: $i$) \textit{Structural causal model}, that aims to learn the relationships between variables/features. This technique generates human explanations since it considers the world as a causal lens~\cite{scm}\cite{scm1}. $ii$) \textit{Explanation in terms of consequences}, which enables agents in reinforcement learning to answer a question about their policy in terms of consequences~\cite{econs}. In other words, it enables to determine what each agent can get from a state, and which outputs it expects from a visited state or corresponding action. $iii$) \textit{Hierarchical policy} by decomposing a given task into multiple sub-tasks, located at different abstraction levels~\cite{hie_rl}. Then, a prerequisite for executing a subsequent task is given as an interpretation of a particular action. $iv$) \textit{Relational reinforcement learning}, which is based on a set of rules to provide background knowledge to the decision agent~\cite{rela_drl}. In this way, actions, states, and policies are represented by relation language, which helps to understand the reinforcement learning model's outputs.
Moreover, the deployed \gls{xai} techniques can be evaluated according to several metrics, as shown in Table \ref{tab:mapping_to_xai}. Specifically, in \gls{drl}-based resource management use-cases, the state-action mapping \emph{certainty} can be measured via the entropy, which is computed from the attributions of input state features. Moreover, the \emph{confidence and Log-odds} metrics serve to quantify the trust in \gls{ai} predictions/decisions by using the input \gls{xai} attributions as a basis to mask the impactful features in either offline or \gls{drl}-based on-the-fly datasets, and measure the corresponding deviation of the output, which is fed back to the optimizer/agent for accountability.
\end{itemize}
It is noteworthy that different data types can be monitored at the \gls{non rt ric} from \gls{o-ru} modules, via the \textbf{O1 interfaces}, in order to feed \gls{ai} and \gls{xai} xApps that are deployed at the \gls{near rt ric} and which collaborate by exchanging model and predictions/decisions XAI metrics, respectively as exemplified in \ref{sec:xai} to achieve transparent AI operation. The resulting trustworthy output AI-based action is enforced at different levels (\gls{o-du} and \gls{o-cu}), via the E2 interface.     
 
\begin{figure}[t]
	\begin{center}
		\includegraphics[width=\columnwidth,clip, trim = 9cm 4.5cm 7cm 2cm ]{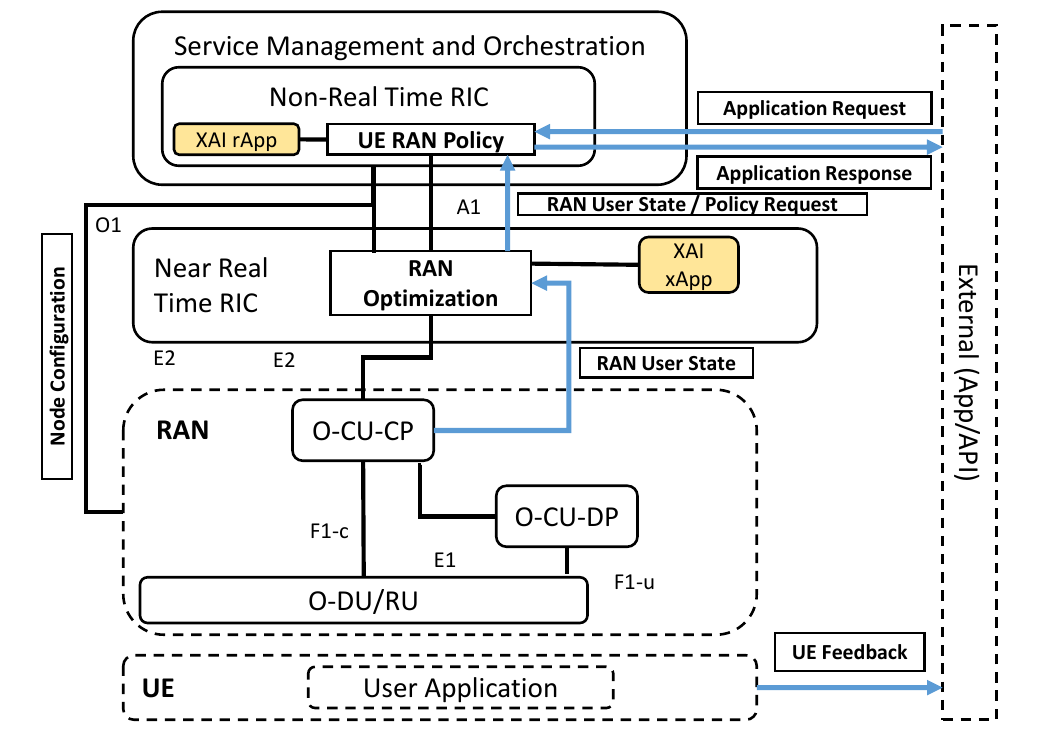}
	\end{center}%
	\caption{Use case: User-Centric \acrshort{qoe} Optimization. Adapted from~\cite{oran_use_cases}}
	\label{fig:usecase_qoe}
\end{figure}

\section{\gls{xai} for \gls{o-ran} Use-cases}
\label{sec:oran_use_cases}
In the following, we collect a list of use-cases in the context of \gls{o-ran} and network slicing, highlighting how they would benefit from the introduction of \gls{xai} methods.

\subsection{\gls{qoe} Optimization}
\label{sec:oran_use_cases_qoe_optim}

Modern applications in the \gls{5g} ecosystem demand large bandwidth and low-latency communication to provide an adequate level of \gls{qoe}, which can hardly be achieved by current semi-static \gls{qos} frameworks devoted to concurrently supporting heterogeneous broadband applications like in the \gls{4g} era.
Radio fluctuations impair radio transmission capabilities, especially when adopting higher carrier frequencies like mm waves, leading to variable application requirements even within the same communication session.
In order to improve \gls{qoe}, estimation and prediction tasks performed at the application level can help in dealing with such a dynamic environment, favouring both user experience and efficient use of \gls{ran} resources~\cite{oran_use_cases}.

Several works have addressed \gls{qoe} modeling with traditional \gls{ml} methods \cite{izima2021survey}. However, the prevalent black-box nature of \gls{ml} models limits insights into \gls{qoe} influence factors. Differently, \gls{xai} tools can provide contextual information for \gls{qoe} assurance xApp \cite{fiandrino2023explora} by e.g., identifying the relevant network environment factors that lead to under-provisioning decisions to underweight them, reducing thereby SLA violation. Moreover, \gls{xai} models such as Fuzzy decision trees have shown their suitability in identifying stall events in data transmissions impairing the resulting \gls{qoe}, while providing interpretability of such events that can be leveraged to identify their cause \cite{renda2021xai}, while \gls{shap} have been used successfully to interpret \gls{dnn} and random forest black-box models specifically trained to the \gls{qoe} modelling task. In this regard, the interpretable output of \gls{xai} approaches becomes a valuable source of information to define strategies to improve \gls{qoe} delivered by the network.

The open interfaces introduced by the \gls{o-ran} architecture significantly ease the per-user flow modification and configuration utilizing proactive closed-loop network optimization. Fig.~\ref{fig:usecase_qoe} depicts a possible deployment addressing this use case. It involves \gls{non rt ric}, \gls{near rt ric}, E2 Nodes, and external applications running on the \gls{ue}. The open interface allows external applications to interface with the \gls{o-ran} domain, which, empowered by ad-hoc optimization logic, would be capable of dynamically re-configuring the networking settings in response to real-time triggering events.
By integrating \gls{xai} tools in the form of a standalone \gls{xapp} into \gls{o-ran}, objective metrics measuring the trustworthiness of \gls{ai} functions, including features attributions, can be computed on the fly  (see Section \ref{subsec: metrics}), providing context to the \gls{o-ran} reconfiguration decisions. This not only enables the transparency of the reconfiguration process but also facilitates the identification of patterns, root causes, and potential improvements in the overall performance of the network.

\subsection{Traffic Steering}
\label{sec:oran_use_cases_traffic_steering}
Imbalances in the traffic load across cells of different access technologies may lead to performance degradation.~\cite{oran_use_cases}.
In this context, \gls{o-ran} A1 interface would allow enforcing desired optimization policies and utilizing the appropriate performance criteria to manage user traffic across different radio access technologies proactively.
The \gls{non rt ric} monitors the user experience by \gls{ue} level performance measurements on a cell level and may decide to relocate one or more users to other cells based on global optimization objectives, e.g., fairness in bandwidth sharing, \gls{qoe} maximization, load-balancing.
In all these scenarios, attribution-based \gls{xai} methods such as SHAP and Integrated-Gradient can provide context to a traffic steering xApp in the form of attributions pointing out the impactful factors to consider in e.g., an offloading decision, which contributes to its transparency \cite{fiandrino2023explora}. Besides, in \cite{surv6}, \gls{lime} is applied to \gls{drl}-driven traffic offloading in wireless networks, aiming at helping radio engineers better understand the consequences of the model choices and better monitor and configure the network. Whereas in \cite{ayoub2022towards}, authors leverage \gls{xai} methods to improve the \gls{qot} estimation process in optical transport networks, which is a key component in driving traffic steering decisions.

\begin{figure*}[t]
	\begin{center}
		\includegraphics[clip, width=1.5\columnwidth, trim = 3cm 7cm 3cm 0cm]{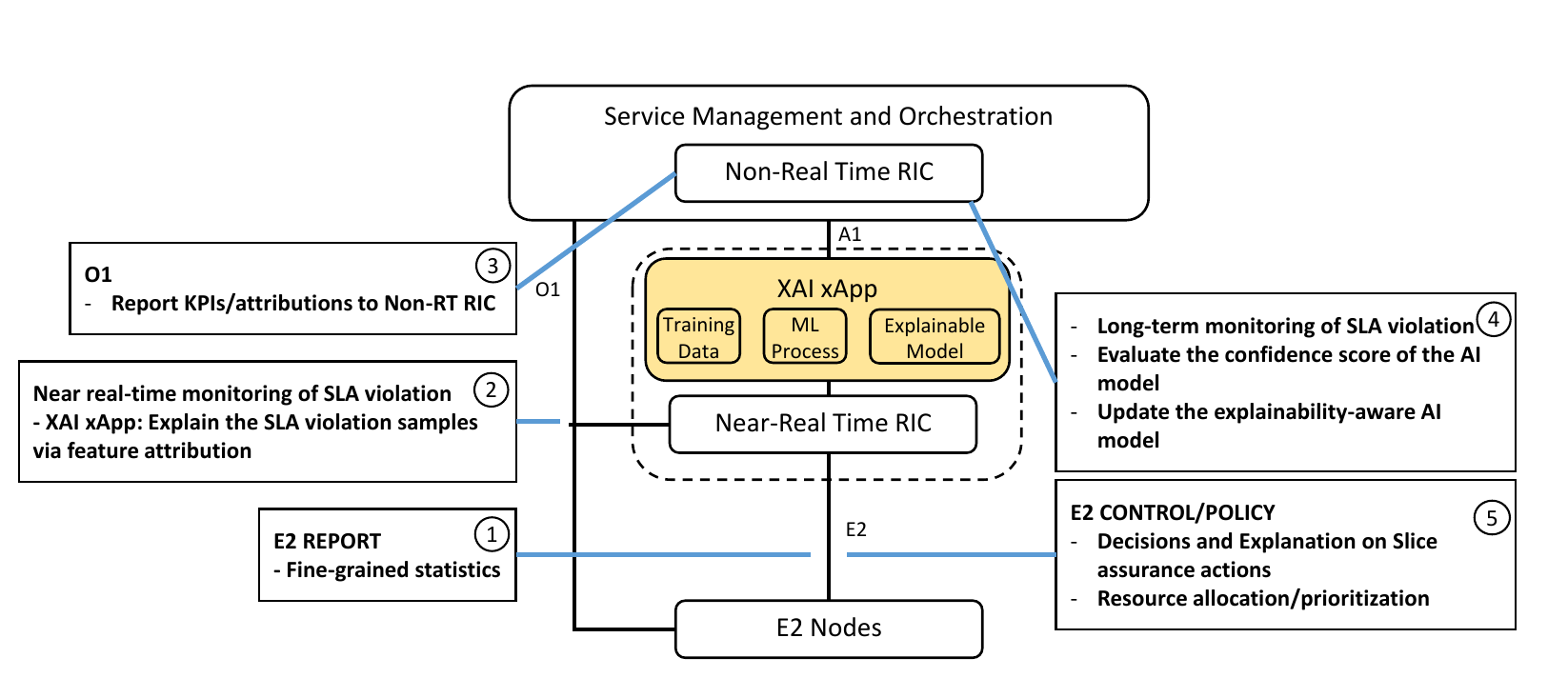}
	\end{center}%
	\caption{Use case: Explainable slice \gls{sla} assurance. Adapted from~\cite{oran_slicing_architecture}}
	\label{fig:usecase_sla}
\end{figure*}

\subsection{\gls{ran} Slice \acrfull{sla} Assurance}
\label{sec:oran_use_cases_sla}

The \gls{5g} infrastructure has been designed to cope with highly diverse performance requirements coming from heterogeneous services and vertical applications. In this context, network slicing arises as a key technology to efficiently support tailored end-to-end connectivity satisfying specific business requirements. In general, the business parties and the infrastructure provider define the set of networking capabilities required to successfully run the service in a \gls{sla}, e.g., in terms of data rate, latency, and resource availability~\cite{sabra1}.
Perhaps not surprisingly, this introduced the need for ad-hoc mechanisms able to efficiently measure and expose such information to $3^{rd}$ party entities traditionally alien to the telecommunication market.
In this context, \gls{o-ran}'s open interfaces and \gls{ai}/\gls{ml}-based architecture will enable such mechanisms, enabling the operators to take full advantage of the business opportunities brought by the network slicing concept. 

More in detail, specific slice configuration settings derived from the \gls{sla} requirements can be easily enforced by initiating the procedure from the \gls{smo} layer, and finely adjusted over time thanks to measurements feedback and zero-touch \gls{xai}-based mechanisms applied at the different layers of the architecture, especially at the \gls{non rt ric} and \gls{near rt ric}. 
\gls{xai} xApp would build context for traditional \gls{sla} assurance xApps \cite{sla,rw9}, by providing feature attributions derived from e.g., SHAP, to point out the network factors that lead to \gls{sla} violation in the event of a resource under-provisioning decision \cite{barnard2022resource}. \gls{xai} would also assess the trustworthiness of AI functions via various metrics (see \ref{subsec: metrics}), which is part of their \gls{sla} \cite{terra2022using, sla1}. Fig.~\ref{fig:usecase_sla} summarizes the main workflow to achieve the solution.

\begin{figure}[t]
	\begin{center}
		\includegraphics[width=\columnwidth]{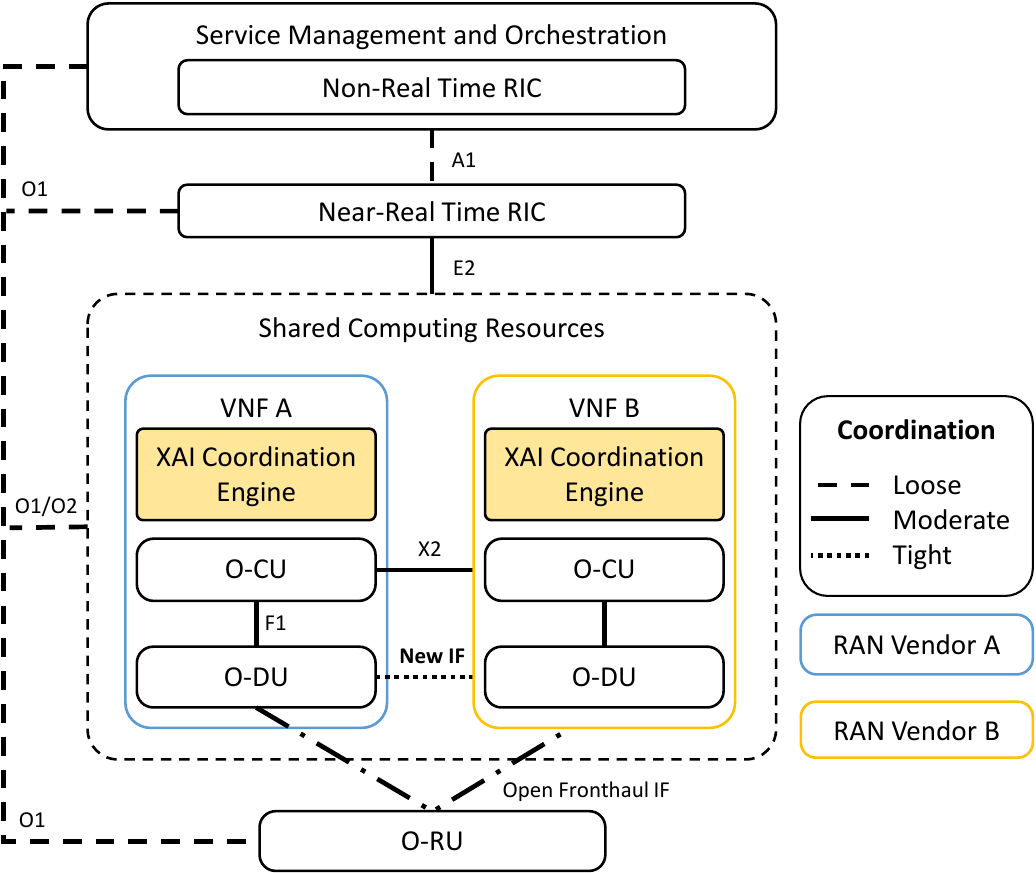}
	\end{center}
    \caption{Use case: Multi-vendor deployment with explainable coordination. Adapted from~\cite{oran_use_cases}.}
	\label{fig:usecase_multi_vendor}
\end{figure}
\subsection{Multi-vendor Slices}
\label{sec:oran_use_multi_vendor}
The coexistence of different network functions provided by different vendors to instantiate operators' services is one of the key enablers for flexible and efficient use of radio resources and \gls{capex}/\gls{opex} optimization. To this extent, the \gls{o-ran} architecture enables the deployment of multiple slices comprising functions provided by different vendors offering a variety of \gls{vo-du} and \gls{vo-cu} options, specifically optimized to meet the requirements of a certain service. This brings several advantages such as a more flexible and time-to-market slice deployment, where operators can select from the available options the most suitable \gls{vo-du} and \gls{vo-cu} to deploy their services and huge business opportunities for the vendors.
\begin{figure*}[!t]
	\centering
	   \begin{subfigure}[t]{.53\textwidth}
		\centering
        \includegraphics[trim = 8cm 2cm 8cm 2cm, clip, width=\columnwidth ]{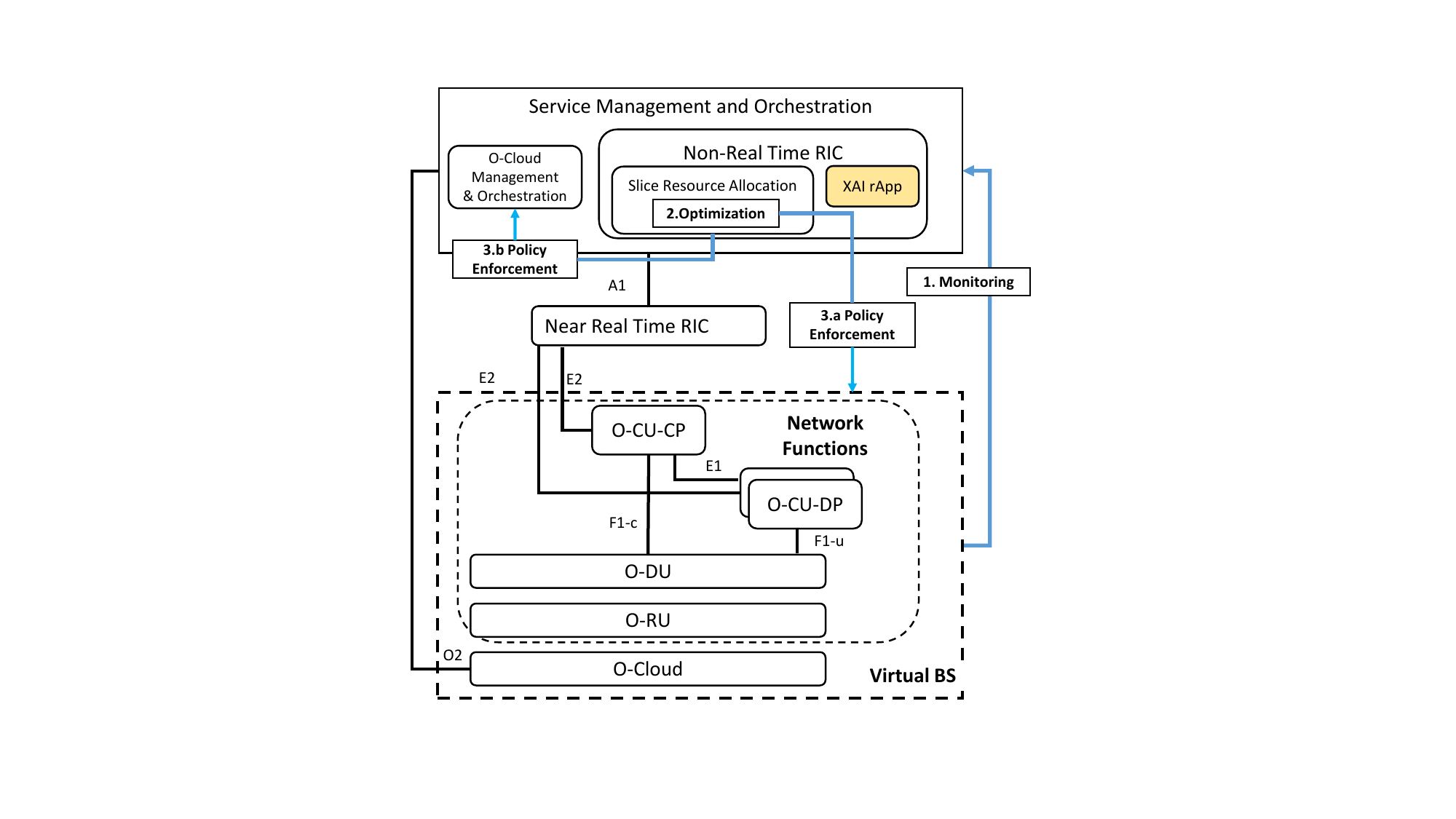} 
        \subcaption{Resource allocation over \gls{non rt ric}.  }
        \label{fig:use_case_rapp}
    \end{subfigure}%
    \begin{subfigure}[t]{.53\textwidth}
    	\centering
        \includegraphics[trim = 8cm 2cm 8cm 2cm, clip, width=\columnwidth]{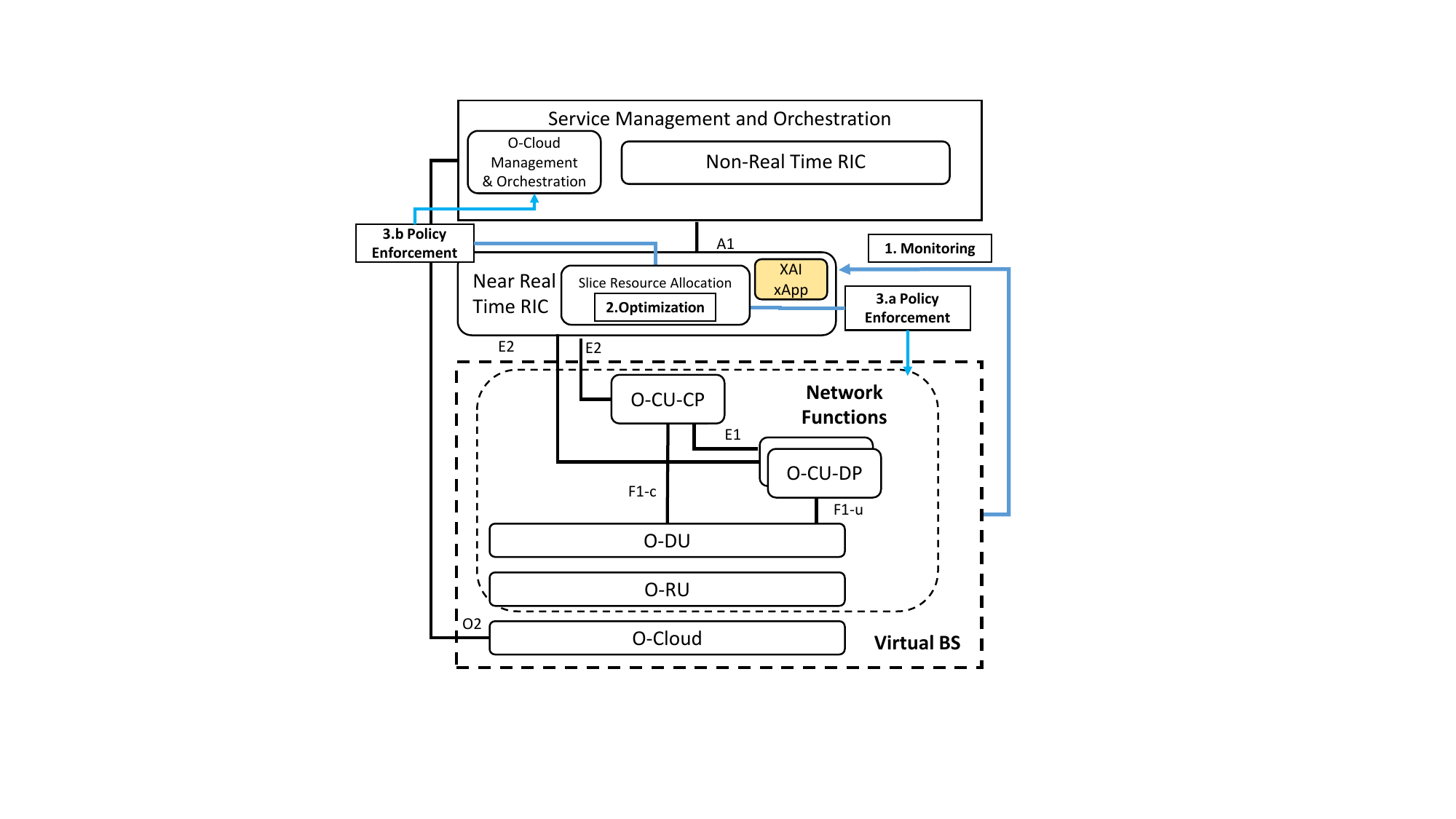}
        \subcaption{Resource allocation over \gls{near rt ric}.}
        \label{fig:use_case_xapp}
    \end{subfigure}%
    \caption{Use case: Explainable \gls{ran} slice resource allocation decisions taken at Non-Real Time and Near-Real Time \glspl{ric}.}
	\label{fig:usecase_resource_allocation}
\end{figure*}%
To deploy multi-vendor slices, \glspl{vo-du} and \glspl{vo-cu} must coordinate to coexist and share the radio environment efficiently and avoid conflicts among the deployed services~\cite{oran_slicing_architecture}.
Fig.~\ref{fig:usecase_multi_vendor} depicts three possible ways of coordination: $i$) \textit{loose coordination} where there is no direct coordination between deployed services, and the radio resource is fully controlled by the \glspl{ric} through the O1, A1, and E1 interfaces. $ii$) \textit{moderate coordination} where different network functions are allowed to communicate with each other through the X2 and the F1 interfaces to negotiate radio resources without directly involving the \glspl{ric}. In this case, the negotiation must cope with the time frame allowed by the X2 interface communication exchange, which is in the order of seconds; $iii$) the \gls{wg}1 and \gls{wg}4 of \gls{o-ran} Alliance envision also a so-called \textit{tight coordination} allowing faster radio resource negotiation among slices, which would require a new interface, dubbed as \textit{New IF} in Fig.~\ref{fig:usecase_multi_vendor}, for direct communication between \glspl{vo-du}.
In this context, distributed \gls{ai}/\gls{ml} models are particularly suitable to smartly perform the negotiation task~\cite{yeh2023deep, hamdan2023recent}. In this regard, an \gls{xai}-enabled component can be deployed to take control of the coordination and negotiation of resources between different vendors, while the heightened transparency and interpretability offered by \gls{xai} enhance the efficacy of resource management and coordination among vendors in this complex scenario. We report in the figure an example of a deployment of this component, suitable for both the moderate and the tight coordination case.

\subsection{Resource Allocation Optimization}
\label{sec:oran_use_res_allocation}
The need to concurrently support multiple heterogeneous slices characterized by service-tailored networking requirements exacerbates the setup of efficient and dynamic resource allocation mechanisms able to cope with highly different spatiotemporal traffic distributions.

For example, end-user mobility towards public events causes spatially localized peaks of traffic in \gls{embb} kind of slices, or \gls{iot} smart sensors sporadically generating data volumes from highly distributed deployments in \gls{mmtc} settings.
Compared to traditional \gls{ran} deployments characterized by monolithic architecture and private management interfaces, the \gls{o-ran}'s paradigm would allow for easier and more flexible control of the radio resources. In addition, the possibility to devise a data-driven ML-based optimization algorithm would help to automatize the process, exploiting the closed-loop management framework and previous historical information to perform the best allocation decisions.
Additionally, \gls{ai}/\gls{ml} model can be used to perform proactive management of the radio resources, predicting recurrent traffic demand patterns of \gls{5g} networks in different time epochs and spatial locations, and for each network slice, therefore anticipating the slice's networking needs favoring a better end-users \gls{qoe}, and limiting the overall energy consumption.
All these methods are traditionally based on RL algorithms and agents interacting with the environment and learning by trial and error. More advanced solutions adopt federated learning techniques to improve the performances of the agents, gaining global knowledge of the system from the collection of multiple locally-trained models~\cite{rw9}. Such enriched information is then sent back to the single agents, improving their training, and speed, and allowing more general management policy definitions.

In both these scenarios, \gls{xai} methods can further extend the potential of the \gls{rl} management solutions. On the one side, they will allow for better control of the learning procedure, and guide the agent towards the definition of a safe decision process by adding confidence and trust in the devised management policies, as demonstrated in~\cite{khan2023explainable}. On the other side, they may help in limiting the information exchange required by the federated learning approach~\cite{zaka3}\cite{fl_lcn}. Only being able to map the context-decision space uniquely, would allow sharing to the federation layer only local models carrying insightful information while filtering out erroneous or redundant items.
Fig.~\ref{fig:usecase_resource_allocation} depicts two possible deployment options, one assuming the main optimization and computing effort running within the \gls{non rt ric} entity, and the other envisioning such task running within the \gls{near rt ric}. The final deployment choice would depend on multiple factors, including the type of use-case and machine-learning model to be run and its timescale and complexity, also considering the different computing capabilities of the \glspl{ric}. An interesting example of integration of \gls{xai} engine running as an \gls{xapp} on an \gls{o-ran} compliant \gls{near rt ric} is provided in~\cite{fiandrino2023explora}.

\subsection{User Access Control}
The \gls{o-ran} vision aims at evolving the current \gls{ran} architecture providing open interfaces and \gls{ml}-based optimization to attract new business entities and ease overall management and reduce operational costs. Current \gls{ran} deployments are composed of thousands of nodes~\cite{pi_road}. In such a complicated deployment, it is expected that the network assigns each \gls{ue} to a serving BS, maximizing the overall throughput and the single end-user \gls{qoe}. This problem is also known as \emph{user access control}. Traditional user access control schemes imply that user associations are based on networking metrics such as the \gls{rss}, which guides the \gls{ue} towards the base station providing the best channel. Handover ping-pong effect and load balancing have been identified as two main issues brought by \gls{rss}-based schemes~\cite{9600616}.

\section{\gls{o-ran} Security Aspects and \gls{xai}} 
\label{sec:security}
Due to the central role of the \gls{5g} network in providing communication to backbone society infrastructures, security, and security risk awareness play a key role in network deployment.
The \gls{o-ran} Alliance has recently created a Security Working Group (\gls{wg}11) which identified a list of stakeholders responsible for ensuring the security of the \gls{ran}. This goes beyond the parties involved in traditional \gls{4g} and \gls{5g} networks, such as vendors, operators, and system integrators. In fact, operators will play a central role in securing the infrastructure, given the platform's openness and the use of multi-vendor components, which allows them to customize and secure the infrastructure. This also enables them to evaluate and verify the security of the open components that are introduced in the network, which may not be possible in fully vendor-driven closed architectures. In addition, according to~\cite{oran_security_req}, network functions and virtualization platform vendors, as well as third-party \gls{xapp} and \gls{rapp} developers, \gls{o-cloud} providers, and administrator profiles, that manage virtualized and disaggregated components, are all new stakeholders. 
However, due to the plethora of heterogeneous components forming the \gls{o-ran} ecosystem, and the high exploitation of \gls{ai}-driven network management and third parties components running services, securing the \gls{o-ran} infrastructure is still a challenge. In this regard, \gls{xai} can strongly enhance security in \gls{o-ran} deployments by providing insights, explanations, and transparency into the decision-making process of \gls{ai} models. Moreover, \gls{xai} helps with threat detection, model transparency, accountability, analysis of training data, and human-in-the-loop security, leading to improved threat detection, increased trust, and compliance with security regulations. However, it could also be the target of cyberattacks that could vanish its benefits.\footnote{Section \ref{sec:security} provides a high-level overview of security in O-RAN, tailored to enhance understanding of related \gls{xai} approaches. For a more detailed survey on generic O-RAN security aspects, we refer readers to \cite{Sab}.}\label{foot:sab}

\subsection{Distributed Architecture}
\label{sec:security_distributed}
The open architecture defined in the \gls{o-ran} specifications has been identified as a possible security issue due to its distributed nature, which expands the attack surface to malicious entities.
The \gls{wg}11 has recently identified the possible vulnerabilities coming from the openness of the platform and classified them into different threat categories~\cite{oran_threat_model}. Such categories include \emph{i)} threats against the \gls{o-ran} system including architectural elements and interfaces that can compromise the availability, data, infrastructure integrity, and data confidentiality of the infrastructure \emph{ii)} Threats against the \gls{o-cloud}, which could compromise virtual network functions, misuse containers or virtual machines, or spoof underlying networking or auxiliary services~\cite{Hexa_X_vision} \emph{iii)} Threats in open-source code, which could potentially contain backdoors~\cite{senevirathna2022survey,Sab}, \emph{iv)}  Physical threats against the hardware infrastructure, \emph{v)} Threats against the protocol stack and threats against \gls{ai}/\gls{ml}, including poisoning attacks that exploit unregulated access to the data stored in the \gls{o-ran} system to inject altered and misleading data. 
To counteract such threats, different security principles have been defined~\cite{oran_threat_model} to provide requirements, recommendations, and potential countermeasures, including mutual authentication (embracing the \emph{zero-trust} paradigm), access control, robust cryptography, trusted communication, secure storage, secure boot and self-configuration, secure update processes, recoverability and backup mechanisms, and effective management of security risks posed by open-source components. As of the latest update, there is no explicit reference to leveraging eXplainable Artificial Intelligence (\gls{xai}) to bolster security within the context of Open Radio Access Network (\gls{o-ran}). However, \gls{xai} holds the potential to offer transparency and insights into the decision-making processes of Artificial Intelligence (\gls{ai}) models, empowering stakeholders with a deeper understanding of how these security principles manifest in practice. This heightened clarity can facilitate enhanced validation, compliance, and trust in the security protocols deployed within \gls{o-ran} setups, thereby bolstering the overall security posture of the network.

\begin{figure}[t]
		\includegraphics[width=\columnwidth]{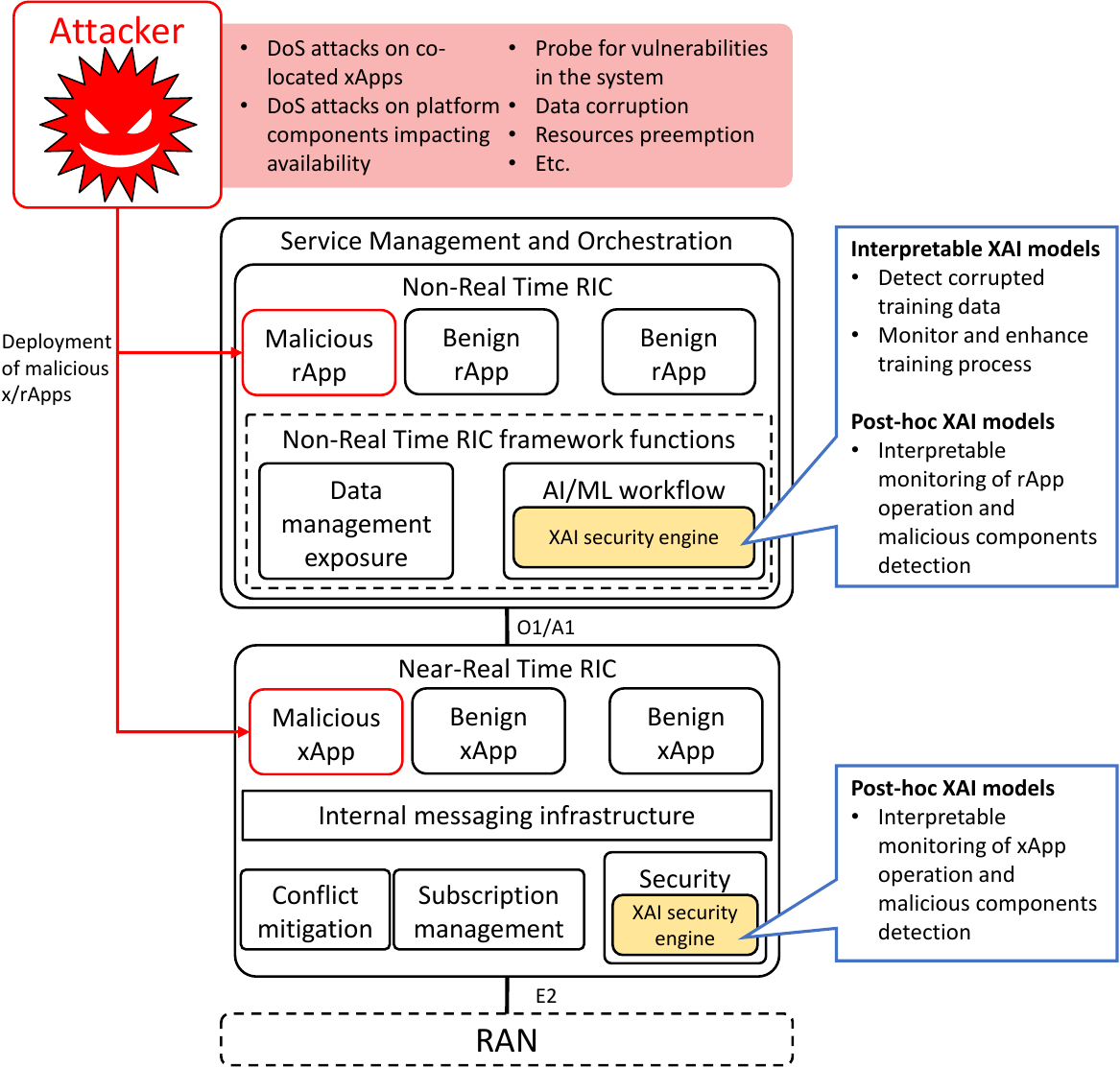}
	\caption{\gls{o-ran} architecture with additional \gls{xai}-based security components. Adapted from~\cite{senevirathna2022survey}.}
	\label{fig:security}
\end{figure}

\subsection{Risk Assessments} 
The Security Working Group (\gls{wg}11) has conducted comprehensive analyses on the \gls{non rt ric} (\gls{non rt ric}), the \gls{o-cloud}, and the Near Real-Time RIC (\gls{near rt ric}) frameworks~\cite{oran_security_risk_Non_RT, oran_security_risk_O_Cloud, oran_security_risk_Near_RT}. These assessments have evaluated the likelihood of attacks and their potential impact on protection goals, which can be categorized as follows:
$i$) \textit{Confidentiality}: Ensuring that sensitive information remains inaccessible to unauthorized entities.
$ii$) \textit{Integrity}: Safeguarding data from unauthorized manipulation, ensuring its integrity and preventing corruption or outdatedness.
$iii$) \textit{Availability}: Guaranteeing the availability of data, information, and services to authorized entities.
Specifically, the Security Technical Report identifies $26$ distinct threats to the \gls{non rt ric} Framework, \glspl{rapp}, R1 interface, and A1 interface, along with corresponding recommended security controls~\cite{oran_security_risk_Non_RT}.

The security analysis of the \gls{o-cloud} encompasses critical services, cloud service and deployment models, stakeholder roles and responsibilities, threat models, and best practices for mitigating threats~\cite{oran_security_risk_O_Cloud}.
Similarly, the Security Technical Report for the \gls{near rt ric} and \glspl{xapp} addresses $11$ key security issues and proposes $13$ solutions, including modifications to existing documents and specifications maintained by \gls{wg}3, with a mapping table illustrating how each solution corresponds to the identified issues~\cite{oran_security_risk_Near_RT}.
With the increasing interest in deploying \gls{o-ran}, third-party entities and government bodies have conducted parallel and independent security risk assessments. ~\cite{oran_risk_DE} evaluates the integrity, availability, and confidentiality aspects alongside two additional protection goals: Accountability, concerning the traceability of actions to specific entities, and privacy, safeguarding sensitive data through anonymity, unlinkability, and unobservability. The analysis reveals a deficiency in adopting a "security/privacy by design/default" approach within \gls{o-ran} specifications, leading to multiple security vulnerabilities. This underscores the urgent need for a thorough revision of \gls{o-ran} specifications with a stronger security emphasis before productive applications are deployed. Another study~\cite{oran_risk_UE} highlights risks associated with multiple suppliers, new network functions, and expanded interfaces, thereby increasing the attack surface. Furthermore, it identifies potential risks arising from the integration of \gls{ai} and \gls{ml}) in network functions, which could compromise network integrity. Additionally, reliance on cloud platforms for hosting base station software in \gls{o-ran} deployments could heighten dependency on cloud service providers, potentially exposing vulnerabilities, especially if multiple \glspl{mno} utilize the same cloud provider.

At the time of writing, and to the best of our knowledge, there are no risk assessments specifically targeting security threats introduced by the use of \gls{xai}, nor suggesting \gls{xai}-based solutions/recommendations to enhance security in open networks. Nonetheless, we will discuss them shortly in the upcoming subsections.

\subsection{\gls{xai} to Improve \gls{o-ran} Security}
\label{sec:security_improvements}
The utilization of \gls{xai} in the security domain of \gls{o-ran} could help enhance the transparency and comprehensibility of the operations and decision-making processes of third-party deployed components as to enable stakeholders to fully understand the decision process of such elements, finally helping to catch malicious behaviors and ensuring accountability, therefore reducing the risk of errors or malicious actions.
This is particularly relevant when considering the high number of \gls{ai}- and \gls{ml}-driven components that will be deployed in the network, which due to their black-box nature pose a significant challenge to reveal malicious behavior and security threats~\cite{senevirathna2022survey}. 
By implementing \gls{xai} techniques, the complex algorithms used in \gls{ml}-/\gls{ai}-based systems can be made more interpretable. This enhances transparency and enables stakeholders to identify potential biases or shortcomings in the system, allowing for continuous improvement and optimization.

The \gls{o-ran} architecture foresees embedding the so-called \textit{security subsystem} in the \gls{near rt ric}. Such a component is in charge of detecting malicious \glspl{xapp} \cite{near-rt-ric}, e.g., preventing them from data leakage or overall network performance reduction. 
Differently, at the time of writing, the \gls{non rt ric} architecture
does not include a functional block solely dedicated to monitoring and detecting malicious rApps for security purposes.
However, although not yet detailed, security is mentioned among the non-functional requirements of the \gls{ai}/\gls{ml} workflow module, which supports the development and implementation of \gls{ai}/\gls{ml} models for tasks such as self-optimization, automation, and data-driven decision-making within the \gls{non rt ric}~\cite{ai_ml}.
In this context, the surface of attacks extends to the \gls{ai}/\gls{ml} models running in \glspl{ric}, i.e., the models that are used for intelligent operations for inference and control in the deployed \glspl{rapp} and \gls{xapp}~\cite{polese}. 

In the \gls{o-ran} architecture, \gls{ai}/\gls{ml} models are mainly deployed in the \gls{ric} as \glspl{xapp}/\glspl{rapp}~\cite{ai_ml}, as depicted in Fig.~\ref{fig:security}. Such elements bring in the autonomous operation of several vital network functions including mobility management, resource allocation, etc. Hence, the deployment of malicious \gls{ai}/\gls{ml} models or the manipulation of benign ones by attackers could disrupt \gls{ran} node functionalities, resulting in severe network failures~\cite{polese}.

To overcome this issue, the use of \gls{xai}-enabled security engines has been recently proposed in~\cite{senevirathna2022survey}. Regarding the \gls{ai}/\gls{ml} workflow module, the work suggests embedding transparent \gls{xai} models such as \gls{pca} and clustering to characterize input data and filter out corrupted or misleading samples. Other self-explanatory transparent models such as decision trees and random forests, are proposed to monitor and enhance the training process. Finally, post-hoc models are employed to further refine threat detection during validation. Similarly, the use of post-hoc \gls{xai} models is proposed to facilitate interpretable monitoring of  \gls{xapp}/\gls{rapp} operation and detection of malicious components deployed in the \glspl{ric}.

Nonetheless, employing \gls{xai} techniques \gls{o-ran} will require additional effort to build and define operational pipelines to generate feedback resulting in explanations. 
The integration of both \gls{ml} and \gls{xai} models will most likely impact computational power requirements. Nevertheless, these additional expenses are warranted by the bolstering of \gls{o-ran} security and management functionalities~\cite{wang2021applications}.

\subsection{Security threats related to \gls{xai}}
\label{sec:security_threats_xai}

As O-RAN continues to gain traction in the telecommunications industry, it becomes imperative to address the various security challenges inherent in its open and disaggregated framework. From concerns surrounding data confidentiality and integrity, to the need for robust authentication mechanisms, the security landscape of O-RAN is both complex and dynamic.
Several attacks targeting \gls{ml}/\gls{ai}-enabled functions can be found in the literature. For example, in \cite{habler2022adversarial} authors realize and demonstrate an \gls{aml} attack on the traffic steering function, exploiting the query-based evasion attack technique proposed in \cite{chen2020hopskipjumpattack}. In particular, the \gls{aml} provides corrupted received signal strength samples to hinder the \gls{qoe} classification and in turn, perform wrong traffic steering decisions. Whereas in \cite{chiejina2024system}, authors develop a malicious xApp designed to execute \gls{aml} attacks on the data stored in the \gls{o-ran} \gls{ric} database, threatening the operation of \gls{ml}-based interference classification \gls{xapp}. In this regard, the authors design also a data distillation technique to mitigate cyber threats effectively.

In this context, \gls{xai} methods can be employed to improve security in \gls{o-ran} deployments. 
However, as the adoption of \gls{xai} gains momentum, so does the prevalence of cyberattacks targeting these models, as noted in recent research~\cite{mirsky2022threat}.
\cite{kuppa2020black} evaluates two different attacks to the \gls{xai} layer: the first wherein the underlying \gls{ml} model and \gls{xai} interpreted are simultaneously corrupted, while the second aims to construct adversarial samples that are sparse, interpretable, and close to the model's training data distribution. This is achieved by using a manifold approximation algorithm on a small set of test data to find data and explanation distributions and inducing minimal distortions on the input to steer the output toward the target (distorted) explanation. 

Similar attacks can vanish the security advantage of an overlaying \gls{xai} layer~\cite{zhao2021exploiting}. For example, Adversarial and Data poisoning attacks involve intentional tampering of input data to mislead an \gls{ai} system's predictions and explanation, aiming to bias the outcome of the model and leading to misinterpretation of the model's behavior~\cite{kuppa2020black}. Evasion attacks, involve revealing sensitive information through the explanations or interpretations provided by an \gls{ai} system. This can be achieved by using the explanations to infer sensitive information about individuals, such as their health status, financial situation, or other private information, even if the \gls{ai} system was not explicitly trained on such data~\cite{zhao2021exploiting}. Finally, social engineering attacks, involve manipulating users or human interpreters of the \gls{ai} system's explanations to make incorrect or biased interpretations. This can be achieved by providing misleading or persuasive explanations that influence the human interpreter's decision-making or perception of the \gls{ai} system's behavior~\cite{giudici2022explainable}. However, the research in the area of security attacks specifically targeting \gls{xai} models is still in its infancy.

\section{Open Challenges and Future Research Directions}
\label{sec:open_challenges}

In this section, we analyze different open questions and research topics for efficient deployment of XAI in future-proof 6G O-RAN architecture, while pointing out various challenges in this area.
\begin{figure}[!t]
	\begin{center}    
	\includegraphics[scale=0.6]{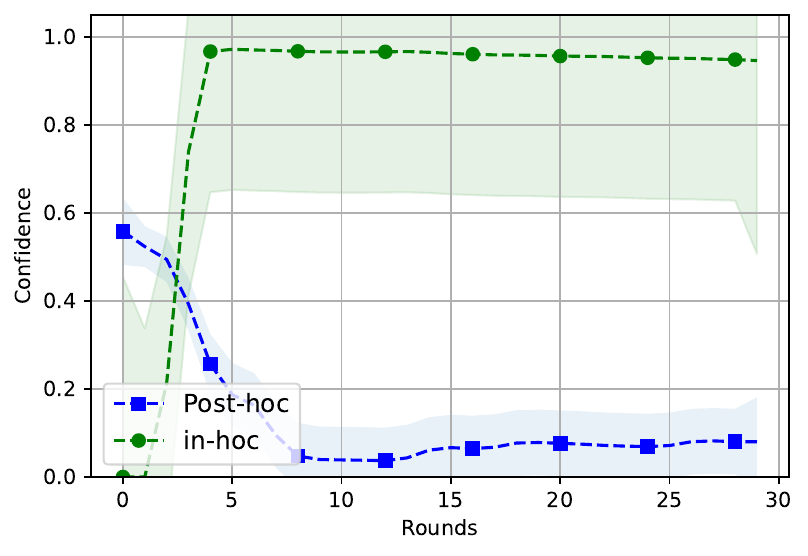}
	\end{center}
	\caption{Confidence vs. Federated Learning rounds \cite{bridging}.}
	\label{fig:tradeoff}
\end{figure}
\subsection{Explainability-Performance Trade-Off}
The flip side of the highly performing yet complex AI models, such as DNNs and Transformers, is being ill-disposed to direct interpretation. They consist of numerous layers and billions of parameters, making it difficult to explain how specific inputs are transformed into outputs. In contrast, simpler models like decision trees or linear regressors are more interpretable, as their decision-making processes are more transparent. Therefore, it turns out that there is a trade-off between performance and explainability, especially when the type of training data justifies the use of complex models \cite{surv1,tradeoff1}. Such observation is also valid when it comes to model optimization. To exemplify this, we plot the confidence metric described in Subsection. \ref{subsec: metrics} in two scenarios, \emph{i}) In each round, an FL  model is optimized via a vanilla loss function such as mean square error, and then a post-hoc explanation in the form of attributions is generated and used to calculate the confidence metric and \emph{ii}) In each round, an FL model is trained through a constrained optimization approach, where the confidence metric is evaluated in run time and jointly enforced as a constraint along with the original loss during training. According to Fig.~\ref{fig:tradeoff}, the model confidence degrades as it gradually converges in the post-hoc scenario, which highlights the abovementioned trade-off. Interestingly, the \emph{in-hoc} strategy can maintain the model confidence across the training rounds, as thoroughly studied in~\cite{bridging}. Striking a balance between performance and interpretability is therefore an open research direction, especially to guarantee a successful deployment of AI in critical 6G use cases under O-RAN architectures.

\subsection{LLMs in O-RAN: An Explainability Perspective}
As anticipated in \cite{llm2}, the potential of \gls{llm}s to transform the Telecom domain lies in their ability to harness generative capabilities and leverage the multimodal nature of wireless network data, thereby enhancing contextual, situational, and temporal awareness. This advancement holds the promise of significantly improving wireless network operations, including localization, beamforming, power allocation, handover, and spectrum management, while also eliminating the requirement for task-specific AI models. Although this is still a far cry from achievement, future-proof 6G O-RAN might incorporate \gls{llm}s into the design of the different radio functions. In such a scenario, the complex architecture of \gls{llm}s that is mainly based on transformers with billions of parameters raises new challenges concerning their explainability and opens a new research line to tackle it.

\subsection{Lack of Standardization}
One of the still open challenges related to the adoption of \gls{xai} to \gls{o-ran} is the lack of standardization efforts across different components and interfaces, except the \gls{o-ran} Alliance and some ongoing projects initiated by network operators~\cite{oran1}\cite{oran2}\cite{nokia-oran} that are more focusing on the main \gls{o-ran}'s components. This makes it difficult to design \gls{xai} tools that can be deployed across different \gls{o-ran} components. Indeed, standardization is critical for enabling interoperability among various components of \gls{o-ran} and facilitating the development and deployment of \gls{xai} tools, on top of \gls{o-ran}. In this context, the research and industry communities should work towards developing common standards for open interfaces, data formats, \glspl{api}, etc.
\subsection{Privacy Concerns of Distributed \gls{xai} Models for the Complex Multi-Vendor \gls{o-ran}}
As mentioned before, one of the main features of \gls{o-ran} is to disaggregate the \gls{ran} functions and manage them through the running \glspl{xapp} at the \gls{near rt ric}, to fulfil the QoS requirements of the envisioned \gls{b5g} network services. In addition, the different \gls{o-ran} components are supplied and supported by various isolated vendors/operators. However, \gls{xai} tools typically need large amounts of data to train and test their models for \gls{o-ran} systems, and the availability of data may be limited or difficult to access, due to security and privacy concerns in a multi-vendor context. Therefore, these vendors/operators should collaborate to ensure stable \gls{ran} performance/cost and deal with the limited available data.\\
In this context, distributed/collaborative deep learning is expected to be widely leveraged. For instance, \gls{fl} is one of the promising collaborative learning techniques that consists of generating learning models collaboratively while preserving the privacy of involved learners such as vendors/operators~\cite{fl}. However, generating learning models (\gls{xai} or \gls{ai}) in a federated way is still challenging since it still presents privacy concerns. Indeed, even \gls{fl} avoids sharing learners' private data, however, it was demonstrated that \gls{fl} is still vulnerable to some attacks, such as poisoning and Byzantine attacks, and sharing model updates during the training process can reveal private information~\cite{challenge_fl}. Privacy-preserving \gls{xai} techniques can be used to provide explanations without revealing sensitive information. These techniques include methods such as differential privacy, blockchain, and homomorphic encryption~\cite{sec_fl}.
\subsection{Interoperable \gls{xai} Models for the Complex Multi-Vendor \gls{o-ran}}
\label{sec:interoperable_xai}

Introducing open interfaces and \gls{ran} functions disaggregation have led to split \gls{gnb} into multiple \glspl{cu} and \glspl{du}, which may belong to different vendors-operators and are connected through the F1 interface. However, designing \gls{xai} models while considering the interoperability among multiple vendors can be extremely challenging as they may have different implementations, capabilities, and requirements.
In this context, a potential solution is leveraging collaborative and distributed learning techniques in building \gls{xai} models, as described in Subsection.~\ref{sec:xai_FL}. This will enable the generation of local \gls{xai} models specific to each vendor, aggregating them at the central \gls{non rt ric} level to compose a global and interoperable \gls{xai} model. This can be achieved by leveraging and extending the \gls{wg}5 activities 
through the definition of open interfaces between base stations (such as E1 and F1), and between \glspl{gnb}s (i.e., Xn interface ), and between \gls{gnb}s and \gls{enb}s (i.e, X2 interface).

\subsection{Complexity of the \gls{o-ran} Systems}
\gls{o-ran} systems can be highly complex, including several layers of software and hardware components. This complexity can complicate the development of efficient \gls{xai} tools, that can provide a concise interpretation and explanation of the decisions made by the \gls{ai}-based \gls{o-ran} systems. One way to deal with the complexity of \gls{ai} models is to simplify them by applying techniques such as rule-based systems, decision trees, or linear models. These models are easier to explain and more transparent than complex deep neural networks. However, this may come at the cost of lower performance and accuracy. Furthermore, model-agnostic \gls{xai} approaches can also be leveraged to provide explanations for any \gls{ai} model, regardless of its structure and complexity. These approaches include local surrogate models, partial dependence plots, and feature importance. The development and deployment of such approaches will enable \gls{xai} for a wider range of \gls{ai}-based models deployed on top of the \gls{o-ran} system.

\subsection{Real-Time Constraints}
\label{sec:real_time_constr}
Some components of the \gls{o-ran} system are designed to operate in real-time, emphasizing the need of tailored \gls{xai} tools for this time-sensitive context. For instance, perturbation methods like SHAP are often computationally intensive, involving iterative evaluations of feature perturbations, which can be impractical for real-time applications with strict latency requirements. In contrast, gradient-based methods offer more efficient computation by leveraging model gradients with respect to input features. This efficiency makes gradient methods better suited for real-time deployment, providing interpretable explanations with reduced computational overhead. While perturbation methods excel in comprehensive feature importance analysis, their practicality in real-time settings is limited compared to gradient-based approaches, which prioritize responsiveness and low latency. These approaches include model compression, feature selection, and online learning. The in-depth development of such approaches will enable the deployment of real-time \gls{xai} models on top of the \gls{o-ran} systems without compromising their performance.

\subsection{Heterogeneity of target audiences in \gls{xai}}
One of the main challenges related to \gls{xai} models is to provide understandable explanations and interpretations to different user profiles (data scientists, developers, managers, etc.). One way to deal with this issue is to design human-centered \gls{xai} models, which are easy to understand and use for different end-users. This can be achieved by developing interactive explanations, visualization tools, and user-friendly interfaces. Moreover, human-centred \gls{xai} can also involve co-design and user studies with end-users to ensure that \gls{xai} models' outputs meet their expectations and needs.

Overall, the future of \gls{xai} for \gls{o-ran} will require interdisciplinary collaboration and research between network engineers, human-computer, \gls{ai} researchers, and data scientists interaction experts. Potential solutions will consist of developing further data formats and standardized interfaces, of the use of simpler and more interpretable/explainable \gls{ai} models, and the development of real-time \gls{xai} models that can operate within the time constraints of \gls{o-ran} systems. By addressing these open challenges of \gls{xai} for \gls{o-ran}, we can ensure that these systems are accountable, trustworthy, and transparent.

\subsection{Security}
\label{subsec:security}
As discussed in Section~\ref{sec:security}, in \gls{o-ran} security enforcement mechanisms are crucial for strengthening the resilience of network deployments to heterogeneous types of security threads. In this regard, the usage of \gls{xai} can be explored to develop and deploy intelligent and interpretable security monitoring systems, which enable more effective detection and mitigation of malicious components. In \gls{o-ran}, \gls{xai}-aided security components can be deployed in different \gls{o-ran} architectural layers. Research efforts in this direction are needed to optimize the design of such components enabling them to provide transparent insights into third-party decision-making, and effectively facilitating the identification of malicious behaviors. At the same time, efforts should be made in the definition of Human-in-the-Loop Security enforcement protocols, that thanks to \gls{xai} will empower human operators to interpret automated decisions, ensuring robust security oversight.

\section{Conclusion}
\label{sec:conclusion}
By providing insights into how these systems work and making their decisions more transparent, \gls{xai} can help to improve the reliability, performance, and security of future \gls{o-ran} networks, playing
a crucial role in their development and helping mobile network operators build and manage more effective and efficient networks.
In this survey paper, we presented a comprehensive overview of \gls{xai} techniques in the context of \gls{o-ran} networks. First, we describe how \gls{xai} methods can be deployed in the \gls{o-ran} framework and architecture by means of three realistic reference scenarios. We then give a literature review of existing works, which leverage \gls{ai} (\gls{ml}/\gls{dl}) techniques on top of the \gls{o-ran} architecture, in order to optimize \gls{ran} functions. We also discuss how these works can be mapped to \gls{xai}-enabled solutions. In addition, we collect a list of use-cases in the context of \gls{o-ran} and network slicing, highlighting how they would benefit from the introduction of \gls{xai} methods.
Besides, to ensure good performance of the intelligent \gls{ran} functions over time, we show how to perform continuous monitoring of both model and data profiles, and how to automate the whole \gls{ai}/\gls{xai} learning models development, including data collection/extraction, model training, validation, and deployment.
Moreover, we explore the potential of \gls{xai} to significantly improve the security layer of \gls{o-ran} and envision how it could be used to build interpretable security threat detection mechanisms. Furthermore, we describe ongoing standardization activities and research projects targeting \gls{xai} and \gls{o-ran} aspects. Finally, we discuss the main open challenges related to \gls{xai} for \gls{o-ran} in addition to suggesting potential solutions to such challenges. With this work, we aim to foster and inspire the research on \gls{xai}, aiming at a new level of interpretable and human-understandable \gls{o-ran} network management operations.

\section*{Acknowledgment}
This work has been partially supported by the Competitive Research Project at Sharjah university in UAE under the DeepORAN project (Grant No. 2402150256), the European Union's Horizon Program under the 6G-Bricks project (Grant No. 101096954), Sunrise-6G project (Grant No. 101139257), NANCY (Grant No. 101096456) and 6G-GOALS (Grant No. 101139232). This paper is also part of the Grant TSI-063000-2021-10 funded by the Ministry for Digital Transformation and of Civil Service and by the ``European Union NextGenerationEU/PRTR''.

\bibliographystyle{IEEEtran}
\bibliography{IEEEabrv,mybibfile.bib}

\begin{thebibliography}{100}
\providecommand{\url}[1]{#1}
\csname url@samestyle\endcsname
\providecommand{\newblock}{\relax}
\providecommand{\bibinfo}[2]{#2}
\providecommand{\BIBentrySTDinterwordspacing}{\spaceskip=0pt\relax}
\providecommand{\BIBentryALTinterwordstretchfactor}{4}
\providecommand{\BIBentryALTinterwordspacing}{\spaceskip=\fontdimen2\font plus
\BIBentryALTinterwordstretchfactor\fontdimen3\font minus \fontdimen4\font\relax}
\providecommand{\BIBforeignlanguage}[2]{{%
\expandafter\ifx\csname l@#1\endcsname\relax
\typeout{** WARNING: IEEEtran.bst: No hyphenation pattern has been}%
\typeout{** loaded for the language `#1'. Using the pattern for}%
\typeout{** the default language instead.}%
\else
\language=\csname l@#1\endcsname
\fi
#2}}
\providecommand{\BIBdecl}{\relax}
\BIBdecl

\bibitem{chowdhury20206g}
M.~Z. Chowdhury, M.~Shahjalal, S.~Ahmed, and Y.~M. Jang, ``{6G wireless communication systems: Applications, requirements, technologies, challenges, and research directions},'' \emph{IEEE Open Journal of the Communications Society}, vol.~1, pp. 957--975, 2020.

\bibitem{xie20216g}
X.~Xie, B.~Rong, and M.~Kadoch, ``{6G Wireless Communications and Mobile Networking},'' \emph{Bentham Books: Sharjah, United Arab Emirates}, 2021.

\bibitem{6g}
W.~Saad, M.~Bennis, and M.~Chen, ``{A vision of 6G wireless systems: Applications, trends, technologies, and open research problems},'' \emph{IEEE network}, vol.~34, no.~3, pp. 134--142, 2019.

\bibitem{6g1}
F.~Fang, Y.~Xu, Q.-V. Pham, and Z.~Ding, ``{Energy-Efficient Design of IRS-NOMA Networks},'' \emph{IEEE Transactions on Vehicular Technology}, vol.~69, no.~11, pp. 14\,088--14\,092, 2020.

\bibitem{6g_tuto_survey1}
C.-X. Wang, X.~You, X.~Gao, X.~Zhu, Z.~Li, C.~Zhang, H.~Wang, Y.~Huang, Y.~Chen, H.~Haas, J.~S. Thompson, E.~G. Larsson, M.~D. Renzo, W.~Tong, P.~Zhu, X.~Shen, H.~V. Poor, and L.~Hanzo, ``{On the Road to 6G: Visions, Requirements, Key Technologies and Testbeds},'' \emph{IEEE Communications Surveys \& Tutorials}, pp. 1--1, 2023.

\bibitem{6g_tuto_survey2}
D.~Zhou, M.~Sheng, J.~Li, and Z.~Han, ``{Aerospace Integrated Networks Innovation for Empowering 6G: A Survey and Future Challenges},'' \emph{IEEE Communications Surveys \& Tutorials}, pp. 1--1, 2023.

\bibitem{6g_tuto_survey3}
B.~Mao, J.~Liu, Y.~Wu, and N.~Kato, ``{Security and Privacy on 6G Network Edge: A Survey},'' \emph{IEEE Communications Surveys \& Tutorials}, pp. 1--1, 2023.

\bibitem{zsm}
E.~Coronado, R.~Behravesh, T.~Subramanya, A.~Fern{\`a}ndez-Fern{\`a}ndez, M.~S. Siddiqui, X.~Costa-P{\'e}rez, and R.~Riggio, ``{Zero touch management: A survey of network automation solutions for 5G and 6G networks},'' \emph{IEEE Communications Surveys \& Tutorials}, vol.~24, no.~4, pp. 2535--2578, 2022.

\bibitem{6g4}
C.~D. Alwis, A.~Kalla, Q.-V. Pham, P.~Kumar, K.~Dev, W.-J. Hwang, and M.~Liyanage, ``{Survey on 6G Frontiers: Trends, Applications, Requirements, Technologies and Future Research},'' \emph{IEEE Open Journal of the Communications Society}, vol.~2, pp. 836--886, 2021.

\bibitem{6g5}
L.~U. Khan, I.~Yaqoob, M.~Imran, Z.~Han, and C.~S. Hong, ``{6G Wireless Systems: A Vision, Architectural Elements, and Future Directions},'' \emph{IEEE Access}, vol.~8, pp. 147\,029--147\,044, 2020.

\bibitem{6g7}
F.~Tariq, M.~R.~A. Khandaker, K.-K. Wong, M.~A. Imran, M.~Bennis, and M.~Debbah, ``{A Speculative Study on 6G},'' \emph{IEEE Wireless Communications}, vol.~27, no.~4, pp. 118--125, 2020.

\bibitem{GE_}
B.~Brik, K.~Dev, Y.~Xiao, G.~Han, and A.~Ksentini, ``{Guest Editorial Introduction to the Special Section on AI-Powered Internet of Everything (IoE) Services in Next-Generation Wireless Networks},'' \emph{IEEE Transactions on Network Science and Engineering}, vol.~9, no.~5, pp. 2952--2954, 2022.

\bibitem{ran_5g}
M.~A. {Habibi}, M.~{Nasimi}, B.~{Han}, and H.~D. {Schotten}, ``{A Comprehensive Survey of RAN Architectures Toward 5G Mobile Communication System},'' \emph{IEEE Access}, vol.~7, pp. 70\,371--70\,421, 2019.

\bibitem{ns_adl}
I.~Afolabi, T.~Taleb, K.~Samdanis, A.~Ksentini, and H.~Flinck, ``{Network Slicing and Softwarization: A Survey on Principles, Enabling Technologies, and Solutions},'' \emph{IEEE Communications Surveys \& Tutorials}, vol.~20, no.~3, pp. 2429--2453, 2018.

\bibitem{ns_wu}
Y.~Wu, H.-N. Dai, H.~Wang, Z.~Xiong, and S.~Guo, ``{A Survey of Intelligent Network Slicing Management for Industrial IoT: Integrated Approaches for Smart Transportation, Smart Energy, and Smart Factory},'' \emph{IEEE Communications Surveys \& Tutorials}, vol.~24, no.~2, pp. 1175--1211, 2022.

\bibitem{zakka}
Z.~A. El~Houda, B.~Brik, and L.~Khoukhi, ``{Ensemble Learning for Intrusion Detection in SDN-Based Zero Touch Smart Grid Systems},'' in \emph{2022 IEEE 47th Conference on Local Computer Networks (LCN)}, 2022, pp. 149--156.

\bibitem{SDN}
W.~Xia, Y.~Wen, C.~H. Foh, D.~Niyato, and H.~Xie, ``{A Survey on Software-Defined Networking},'' \emph{IEEE Communications Surveys \& Tutorials}, vol.~17, no.~1, pp. 27--51, 2015.

\bibitem{net_sli}
S.~Wijethilaka and M.~Liyanage, ``{Survey on Network Slicing for Internet of Things Realization in 5G Networks},'' \emph{IEEE Communications Surveys \& Tutorials}, vol.~23, no.~2, pp. 957--994, 2021.

\bibitem{ns_b5g}
S.~Ben~Saad, A.~Ksentini, and B.~Brik, ``{An end-to-end trusted architecture for network slicing in 5G and beyond networks},'' \emph{SECURITY AND PRIVACY}, vol.~5, no.~1, p. e186, 2022.

\bibitem{oran1}
O.~R. Alliance, ``{O-RAN: Towards an open and smart RAN},'' \emph{White paper}, vol.~19, 2018.

\bibitem{oran2}
S.~Abeta, T.~Kawahara, A.~Umesh, and R.~Matsukawa, ``{O-RAN Alliance Standardization Trends.}'' \emph{NTT DOCOMO Technical Journal}, vol.~21, no.~1, 2019.

\bibitem{oran_architecture}
{O-RAN Alliance}, ``{O-RAN-WG1-O-RAN Architecture Description (ORAN.WG1.O-RAN-Architecture-Description-v04.00)},'' {Technical Specification}, 2020.

\bibitem{non-rt-ric}
O.~Alliance, ``{O-RAN Non-RT RIC: Functional Architecture 1.01-March 2021 (O-RAN. WG2. Non-RT-RIC-ARCH-TR-v01. 01)},'' \emph{Technical Specification}, 2021.

\bibitem{near-rt-ric}
O.-R. W.~G. 3, ``{O-RAN near-RT RAN intelligent controller near-RT RIC architecture 2.00},'' \emph{O-RAN. WG3. RICARCH-v02. 00}, 2021.

\bibitem{ai_networking}
N.~C. Luong, D.~T. Hoang, S.~Gong, D.~Niyato, P.~Wang, Y.-C. Liang, and D.~I. Kim, ``{Applications of Deep Reinforcement Learning in Communications and Networking: A Survey},'' \emph{IEEE Communications Surveys \& Tutorials}, vol.~21, no.~4, pp. 3133--3174, 2019.

\bibitem{fl_iot}
D.~C. Nguyen, M.~Ding, P.~N. Pathirana, A.~Seneviratne, J.~Li, and H.~Vincent~Poor, ``{Federated Learning for Internet of Things: A Comprehensive Survey},'' \emph{IEEE Communications Surveys \& Tutorials}, vol.~23, no.~3, pp. 1622--1658, 2021.

\bibitem{ai_networking1}
M.~Shen, K.~Ye, X.~Liu, L.~Zhu, J.~Kang, S.~Yu, Q.~Li, and K.~Xu, ``{Machine Learning-Powered Encrypted Network Traffic Analysis: A Comprehensive Survey},'' \emph{IEEE Communications Surveys \& Tutorials}, vol.~25, no.~1, pp. 791--824, 2023.

\bibitem{bonati2022intelligent}
L.~Bonati, M.~Polese, S.~D'Oro, S.~Basagni, and T.~Melodia, ``{Intelligent Closed-loop RAN Control with xApps in OpenRAN Gym},'' in \emph{European Wireless 2022; 27th European Wireless Conference}, 2022, pp. 1--6.

\bibitem{tl}
H.~Zhang, H.~Zhou, and M.~Erol-Kantarci, ``{Team Learning-Based Resource Allocation for Open Radio Access Network (O-RAN)},'' in \emph{ICC 2022 - IEEE International Conference on Communications}, 2022, pp. 4938--4943.

\bibitem{sulaiman2022coordinated}
M.~Sulaiman, A.~Moayyedi, M.~Ahmadi, M.~A. Salahuddin, R.~Boutaba, and A.~Saleh, ``{Coordinated slicing and admission control using multi-agent deep reinforcement learning},'' \emph{IEEE Transactions on Network and Service Management}, 2022.

\bibitem{european2021fosterinapproachtoai}
\BIBentryALTinterwordspacing
E.~Commission, ``{Communication on Fostering a European approach to Artificial Intelligence},'' 2021. [Online]. Available: \url{https://digital-strategy.ec.europa.eu/en/library/communication-fostering-european-approach-artificial-intelligence}
\BIBentrySTDinterwordspacing

\bibitem{MaintainingAmericanLeadershipinArtificialIntelligence}
\BIBentryALTinterwordspacing
E.~O. of~the President of United States~of America, ``{Maintaining American Leadership in Artificial Intelligence - Executive Order 13859 of February 11, 2019},'' 2019. [Online]. Available: \url{https://www.federalregister.gov/d/2019-02544}
\BIBentrySTDinterwordspacing

\bibitem{UkNAtionalAIStrategyActionPlan2022}
\BIBentryALTinterwordspacing
D.~for Digital, Culture, M.~. Sport, D.~for Business, E.~.~I. Strategy, and O.~for Artificial Intelligence of~the United~Kingdom, ``{National AI Strategy - AI Action Plan},'' 2022. [Online]. Available: \url{https://www.gov.uk/government/publications/national-ai-strategy-ai-action-plan/national-ai-strategy-ai-action-plan}
\BIBentrySTDinterwordspacing

\bibitem{AlanTuringCommonRegCapAI}
M.~Aitken, D.~Leslie, F.~Ostmann, J.~Pratt, H.~Margetts, and C.~Dorobantu, ``{Common Regulatory Capacity for AI},'' 2022.

\bibitem{xai-iot}
H.~Elayan, M.~Aloqaily, F.~Karray, and M.~Guizani, ``{Internet of Behavior (IoB) and Explainable AI Systems for Influencing IoT Behavior},'' \emph{IEEE Network}, pp. 1--8, 2022.

\bibitem{sla}
M.~A. Habibi, B.~Han, M.~Nasimi, and H.~D. Schotten, ``{The structure of service level agreement of slice-based 5G network},'' \emph{arXiv preprint arXiv:1806.10426}, 2018.

\bibitem{sla1}
A.~Terra, R.~Inam, S.~Baskaran, P.~Batista, I.~Burdick, and E.~Fersman, ``{Explainability Methods for Identifying Root-Cause of SLA Violation Prediction in 5G Network},'' in \emph{IEEE Global Communications Conference (GLOBECOM)}, 2020, pp. 1--7.

\bibitem{surv5}
S.~Wang, M.~A. Qureshi, L.~Miralles-Pechu{\'a}n, T.~Huynh-The, T.~R. Gadekallu, and M.~Liyanage, ``{Explainable AI for B5G/6G: Technical Aspects, Use Cases, and Research Challenges},'' 2021.

\bibitem{sabra}
S.~Ben~Saad, B.~Brik, and A.~Ksentini, ``{A Trust and Explainable Federated Deep Learning Framework in Zero Touch B5G Networks},'' in \emph{GLOBECOM 2022 - 2022 IEEE Global Communications Conference}, 2022, pp. 1037--1042.

\bibitem{surv2}
V.~Belle and I.~Papantonis, ``{Principles and Practice of Explainable Machine Learning},'' 2020.

\bibitem{polese}
M.~Polese, L.~Bonati, S.~D'Oro, S.~Basagni, and T.~Melodia, ``{Understanding O-RAN: Architecture, Interfaces, Algorithms, Security, and Research Challenges},'' \emph{IEEE Communications Surveys \& Tutorials}, pp. 1--1, 2023.

\bibitem{sur-oran}
A.~S. Abdalla, P.~S. Upadhyaya, V.~K. Shah, and V.~Marojevic, ``{Toward Next Generation Open Radio Access Networks--What O-RAN Can and Cannot Do!}'' \emph{IEEE Network}, pp. 1--8, 2022.

\bibitem{niknam2020}
S.~Niknam, A.~Roy, H.~S. Dhillon, S.~Singh, R.~Banerji, J.~H. Reed, N.~Saxena, and S.~Yoon, ``{Intelligent O-RAN for Beyond 5G and 6G Wireless Networks},'' 2020.

\bibitem{dis_oran}
A.~Garcia-Saavedra and X.~Costa-Perez, ``{O-RAN: Disrupting the Virtualized RAN Ecosystem},'' \emph{IEEE Communications Standards Magazine}, pp. 1--8, 2021.

\bibitem{oran_evolu1}
S.~K. {Singh}, R.~{Singh}, and B.~{Kumbhani}, ``{The Evolution of Radio Access Network Towards Open-RAN: Challenges and Opportunities},'' in \emph{IEEE Wireless Communications and Networking Conference Workshops (WCNCW)}, 2020, pp. 1--6.

\bibitem{oran_evolu2}
W.~{Diego}, ``{Evolution Toward the Next Generation Radio Access Network},'' in \emph{IFIP Networking Conference (Networking)}, 2020, pp. 685--685.

\bibitem{oran_mec_son_ns}
C.~L. {I}, S.~{Kuklinski}, T.~{Chen}, and L.~L. {Ladid}, ``{A Perspective of O-RAN Integration with MEC, SON, and Network Slicing in the 5G Era},'' \emph{IEEE Network}, vol.~34, no.~6, pp. 3--5, 2020.

\bibitem{Our_surv}
B.~Brik, K.~Boutiba, and A.~Ksentini, ``{Deep Learning for B5G Open Radio Access Network: Evolution, Survey, Case Studies, and Challenges},'' \emph{IEEE Open Journal of the Communications Society}, pp. 1--1, 2022.

\bibitem{surv1}
A.~B. Arrieta, N.~D{\'\i}az-Rodr{\'\i}guez, J.~Del~Ser, A.~Bennetot, S.~Tabik, A.~Barbado, S.~Garc{\'\i}a, S.~Gil-L{\'o}pez, D.~Molina, R.~Benjamins \emph{et~al.}, ``{Explainable Artificial Intelligence (XAI): Concepts, taxonomies, opportunities and challenges toward responsible AI},'' \emph{Information fusion}, vol.~58, pp. 82--115, 2020.

\bibitem{surv4}
A.~Das and P.~Rad, ``{Opportunities and Challenges in Explainable Artificial Intelligence (XAI): A Survey},'' 2020.

\bibitem{surv3}
R.~R. Hoffman, S.~T. Mueller, G.~Klein, and J.~Litman, ``{Metrics for Explainable AI: Challenges and Prospects},'' 2019.

\bibitem{surv6}
W.~Guo, ``{Explainable Artificial Intelligence for 6G: Improving Trust between Human and Machine},'' \emph{IEEE Communications Magazine}, vol.~58, no.~6, pp. 39--45, 2020.

\bibitem{comcom_sur}
C.~Fiandrino, G.~Attanasio, M.~Fiore, and J.~Widmer, ``{Toward native explainable and robust AI in 6G networks: Current state, challenges and road ahead},'' \emph{Computer Communications}, vol. 193, pp. 47--52, 2022.

\bibitem{thulitha}
T.~Senevirathna, Z.~Salazar, V.~H. La, S.~Marchal, B.~Siniarski, M.~Liyanage, and S.~Wang, ``{A Survey on XAI for Beyond 5G Security: Technical Aspects, Use Cases, Challenges and Research Directions},'' 2022.

\bibitem{mec2}
B.~Brik and A.~Ksentini, ``{Toward Optimal MEC Resource Dimensioning for a Vehicle Collision Avoidance System: A Deep Learning Approach},'' \emph{IEEE Network}, vol.~35, no.~3, pp. 74--80, 2021.

\bibitem{mec1}
B.~Brik, P.~A. Frangoudis, and A.~Ksentini, ``{Service-Oriented MEC Applications Placement in a Federated Edge Cloud Architecture},'' in \emph{ICC 2020 - 2020 IEEE International Conference on Communications (ICC)}, 2020, pp. 1--6.

\bibitem{mec3}
U.~Fattore, M.~Liebsch, B.~Brik, and A.~Ksentini, ``{AutoMEC: LSTM-based user mobility prediction for service management in distributed MEC resources},'' in \emph{Proceedings of the 23rd International ACM Conference on Modeling, Analysis and Simulation of Wireless and Mobile Systems}, 2020, pp. 155--159.

\bibitem{darpa}
D.~Gunning and D.~Aha, ``{DARPA's Explainable Artificial Intelligence (XAI) Program},'' \emph{AI Magazine}, vol.~40, no.~2, pp. 44--58, Jun. 2019.

\bibitem{saliency}
\BIBentryALTinterwordspacing
K.~Simonyan, A.~Vedaldi, and A.~Zisserman, ``{Deep Inside Convolutional Networks: Visualising Image Classification Models and Saliency Maps},'' 2013. [Online]. Available: \url{https://arxiv.org/abs/1312.6034}
\BIBentrySTDinterwordspacing

\bibitem{saliency2}
R.~R. Selvaraju, M.~Cogswell, A.~Das, R.~Vedantam, D.~Parikh, and D.~Batra, ``{Grad-CAM: Visual Explanations from Deep Networks via Gradient-Based Localization},'' in \emph{2017 IEEE International Conference on Computer Vision (ICCV)}, 2017, pp. 618--626.

\bibitem{inputgrad}
A.~Shrikumar, P.~Greenside, and A.~Kundaje, ``{Learning Important Features through Propagating Activation Differences},'' in \emph{Proceedings of the 34th International Conference on Machine Learning - Volume 70}, ser. ICML'17.\hskip 1em plus 0.5em minus 0.4em\relax JMLR.org, 2017, pp. 3145--3153.

\bibitem{inputgrad2}
\BIBentryALTinterwordspacing
{Luisa M Zintgraf and Taco S Cohen and Tameem Adel and Max Welling}, ``{Visualizing Deep Neural Network Decisions: Prediction Difference Analysis},'' in \emph{International Conference on Learning Representations}, 2017. [Online]. Available: \url{https://openreview.net/forum?id=BJ5UeU9xx}
\BIBentrySTDinterwordspacing

\bibitem{intgrad}
M.~Sundararajan, A.~Taly, and Q.~Yan, ``{Axiomatic attribution for deep networks},'' in \emph{International conference on machine learning}.\hskip 1em plus 0.5em minus 0.4em\relax PMLR, 2017\color{black}, pp. 3319--3328.

\bibitem{intgrad2}
\BIBentryALTinterwordspacing
{Yi Huang and Adams Wai-Kin Kong}, ``{Transferable Adversarial Attack based on Integrated Gradients},'' in \emph{International Conference on Learning Representations}, 2022. [Online]. Available: \url{https://openreview.net/forum?id=DesNW4-5ai9}
\BIBentrySTDinterwordspacing

\bibitem{smoothgrad}
D.~Smilkov, N.~Thorat, B.~Kim, F.~Vi{\'e}gas, and M.~Wattenberg, ``{Smoothgrad: removing noise by adding noise},'' \emph{arXiv preprint arXiv:1706.03825}, 2017.

\bibitem{smoothgrad2}
\BIBentryALTinterwordspacing
{Alexander Levine and Sahil Singla and Soheil Feizi}, ``{Certifiably Robust Interpretation in Deep Learning},'' 2020. [Online]. Available: \url{https://openreview.net/forum?id=rkxVz1HKwB}
\BIBentrySTDinterwordspacing

\bibitem{lrp}
S.~Bach, A.~Binder, G.~Montavon, F.~Klauschen, K.-R. M{\"u}ller, and W.~Samek, ``On pixel-wise explanations for non-linear classifier decisions by layer-wise relevance propagation,'' \emph{PloS one}, vol.~10, no.~7, p. e0130140, 2015.

\bibitem{lrp2}
M.~P. Ayyar, J.~Benois-Pineau, and A.~Zemmari, ``{Review of white box methods for explanations of convolutional neural networks in image classification tasks},'' \emph{Journal of Electronic Imaging}, vol.~30, pp. 050\,901 -- 050\,901, 2021.

\bibitem{shap}
J.~Castro, D.~G\'omez, and J.~Tejada, ``{Polynomial Calculation of the Shapley Value Based on Sampling},'' \emph{Computers \& Operations Research}, vol.~36, no.~5, pp. 1726--1730, 2009, selected papers presented at the Tenth International Symposium on Locational Decisions (ISOLDE X).

\bibitem{shap2}
N.~Jethani, M.~Sudarshan, I.~C. Covert, S.-I. Lee, and R.~Ranganath, ``{FastSHAP: Real-Time Shapley Value Estimation},'' in \emph{International Conference on Learning Representations}, 2022.

\bibitem{deeplift}
A.~Shrikumar, P.~Greenside, and A.~Kundaje, ``{Learning important features through propagating activation differences},'' in \emph{International conference on machine learning}.\hskip 1em plus 0.5em minus 0.4em\relax PMLR, 2017\color{black}, pp. 3145--3153.

\bibitem{occlusion}
M.~D. Zeiler and R.~Fergus, ``{Visualizing and understanding convolutional networks},'' in \emph{Computer Vision--ECCV 2014: 13th European Conference, Zurich, Switzerland, September 6-12, 2014, Proceedings, Part I 13}.\hskip 1em plus 0.5em minus 0.4em\relax Springer, 2014\color{black}, pp. 818--833.

\bibitem{occlusion2}
X.~Kong and X.~Zhang, ``{Understanding Masked Image Modeling via Learning Occlusion Invariant Feature},'' 2023.

\bibitem{gnnexp}
R.~Ying, D.~Bourgeois, J.~You, M.~Zitnik, and J.~Leskovec, \emph{{GNNExplainer: Generating Explanations for Graph Neural Networks}}.\hskip 1em plus 0.5em minus 0.4em\relax Red Hook, NY, USA: Curran Associates Inc., 2019.

\bibitem{lime}
M.~T. Ribeiro, S.~Singh, and C.~Guestrin, ``{"Why should I trust you?" Explaining the predictions of any classifier},'' in \emph{Proceedings of the 22nd ACM SIGKDD international conference on knowledge discovery and data mining}, 2016, pp. 1135--1144.

\bibitem{lime2}
H.~Jia, H.~Chen, J.~Guan, A.~S. Shamsabadi, and N.~Papernot, ``{A Zest of LIME: Towards Architecture-Independent Model Distances},'' in \emph{International Conference on Learning Representations}, 2022.

\bibitem{trepan}
M.~W. Craven and J.~W. Shavlik, ``{Extracting Tree-Structured Representations of Trained Networks},'' in \emph{NIPS}, 1995.

\bibitem{rule_fit}
Z.~A. El~Houda, B.~Brik, and S.-M. Senouci, ``{A Novel IoT-Based Explainable Deep Learning Framework for Intrusion Detection Systems},'' \emph{IEEE Internet of Things Magazine}, vol.~5, no.~2, pp. 20--23, 2022.

\bibitem{rule_fit1}
Z.~A.~E. Houda, B.~Brik, and L.~Khoukhi, ``{"Why Should I Trust Your IDS?": An Explainable Deep Learning Framework for Intrusion Detection Systems in Internet of Things Networks},'' \emph{IEEE Open Journal of the Communications Society}, vol.~3, pp. 1164--1176, 2022.

\bibitem{reward_shaping}
A.~Gupta, A.~Pacchiano, Y.~Zhai, S.~Kakade, and S.~Levine, ``{Unpacking Reward Shaping: Understanding the Benefits of Reward Engineering on Sample Complexity},'' in \emph{Advances in Neural Information Processing Systems}, S.~Koyejo, S.~Mohamed, A.~Agarwal, D.~Belgrave, K.~Cho, and A.~Oh, Eds., vol.~35.\hskip 1em plus 0.5em minus 0.4em\relax Curran Associates, Inc., 2022, pp. 15\,281--15\,295.

\bibitem{reward_shaping2}
Y.~Hu, W.~Wang, H.~Jia, Y.~Wang, Y.~Chen, J.~Hao, F.~Wu, and C.~Fan, ``{Learning to Utilize Shaping Rewards: A New Approach of Reward Shaping},'' in \emph{Advances in Neural Information Processing Systems}, H.~Larochelle, M.~Ranzato, R.~Hadsell, M.~Balcan, and H.~Lin, Eds., vol.~33.\hskip 1em plus 0.5em minus 0.4em\relax Curran Associates, Inc., 2020, pp. 15\,931--15\,941.

\bibitem{attention}
A.~Mott, D.~Zoran, M.~Chrzanowski, D.~Wierstra, and D.~Jimenez~Rezende, ``{Towards Interpretable Reinforcement Learning Using Attention Augmented Agents},'' in \emph{Advances in Neural Information Processing Systems}, H.~Wallach, H.~Larochelle, A.~Beygelzimer, F.~d\textquotesingle Alch\'{e}-Buc, E.~Fox, and R.~Garnett, Eds., vol.~32.\hskip 1em plus 0.5em minus 0.4em\relax Curran Associates, Inc., 2019.

\bibitem{attention2}
A.~Manchin, E.~Abbasnejad, and A.~van~den Hengel, ``{Reinforcement Learning with Attention that Works: A Self-Supervised Approach},'' in \emph{Neural Information Processing}, T.~Gedeon, K.~W. Wong, and M.~Lee, Eds.\hskip 1em plus 0.5em minus 0.4em\relax Cham: Springer International Publishing, 2019, pp. 223--230.

\bibitem{MR}
K.~Cyras, R.~Badrinath, S.~K. Mohalik, A.~Mujumdar, A.~Nikou, A.~Previti, V.~Sundararajan, and A.~V. Feljan, ``{Machine Reasoning Explainability},'' in \emph{AAMAS 2021}.\hskip 1em plus 0.5em minus 0.4em\relax IFAAMAS, 2021.

\bibitem{mr2}
\BIBentryALTinterwordspacing
D.~A. Hudson and C.~D. Manning, ``{Compositional Attention Networks for Machine Reasoning},'' in \emph{International Conference on Learning Representations}, 2018. [Online]. Available: \url{https://openreview.net/forum?id=S1Euwz-Rb}
\BIBentrySTDinterwordspacing

\bibitem{attention_analysis}
E.~Voita, D.~Talbot, F.~Moiseev, R.~Sennrich, and I.~Titov, ``{{Analyzing Multi-Head Self-Attention: Specialized Heads Do the Heavy Lifting, the Rest Can Be Pruned}},'' in \emph{Annual Meeting of the Association for Computational Linguistics}, Jul. 2019, pp. 5797--5808.

\bibitem{AttnRLP}
\BIBentryALTinterwordspacing
R.~Achtibat, S.~M.~V. Hatefi, M.~Dreyer, A.~Jain, T.~Wiegand, S.~Lapuschkin, and W.~Samek, ``{AttnLRP: Attention-Aware Layer-wise Relevance Propagation for Transformers},'' \emph{ArXiv}, vol. abs/2402.05602, \color{black} 2024. [Online]. Available: \url{https://api.semanticscholar.org/CorpusID:267547813}
\BIBentrySTDinterwordspacing

\bibitem{visual}
J.~Krause, A.~Perer, and K.~Ng, ``{Interacting with Predictions: Visual Inspection of Black-Box Machine Learning Models},'' in \emph{ACM Conference on Human Factors in Computing Systems}, 2016, pp. 5686--5697.

\bibitem{scm}
P.~Madumal, T.~Miller, L.~Sonenberg, and F.~Vetere, ``{Explainable Reinforcement Learning Through a Causal Lens},'' 2019.

\bibitem{text}
A.~Bennetot, J.-L. Laurent, R.~Chatila, and N.~Diaz-Rodriguez, ``{Towards Explainable Neural-Symbolic Visual Reasoning},'' 2019.

\bibitem{Caption}
Y.~Dong, H.~Su, J.~Zhu, and B.~Zhang, ``{Improving Interpretability of Deep Neural Networks with Semantic Information},'' in \emph{2017 IEEE Conference on Computer Vision and Pattern Recognition (CVPR)}, 2017, pp. 975--983.

\bibitem{kg_book}
I.~Tiddi, F.~L{\'{e}}cu{\'{e}}, and P.~Hitzler, Eds., \emph{{Knowledge Graphs for eXplainable Artificial Intelligence: Foundations, Applications and Challenges}}, ser. Studies on the Semantic Web.\hskip 1em plus 0.5em minus 0.4em\relax {IOS} Press, 2020, vol.~47.

\bibitem{kg}
F.~Bianchi, G.~Rossiello, L.~Costabello, M.~Palmonari, and P.~Minervini, ``Knowledge graph embeddings and explainable {AI},'' \emph{CoRR}, vol. abs/2004.14843, 2020.

\bibitem{ml_book}
\BIBentryALTinterwordspacing
A.~M{\"u}ller and S.~Guido, \emph{{Introduction to Machine Learning with Python: A Guide for Data Scientists}}.\hskip 1em plus 0.5em minus 0.4em\relax O'Reilly Media, Incorporated, 2016. [Online]. Available: \url{https://books.google.es/books?id=q5pnAQAACAAJ}
\BIBentrySTDinterwordspacing

\bibitem{xgboost}
\BIBentryALTinterwordspacing
T.~Chen and C.~Guestrin, ``{XGBoost},'' in \emph{ACM International Conference on Knowledge Discovery and Data Mining (SIGKDD)}.\hskip 1em plus 0.5em minus 0.4em\relax {ACM}, aug 2016. [Online]. Available: \url{https://doi.org/10.11452F2939672.2939785}
\BIBentrySTDinterwordspacing

\bibitem{knear}
\BIBentryALTinterwordspacing
P.~Cunningham and S.~J. Delany, ``{k-Nearest Neighbour Classifiers - A Tutorial},'' \emph{ACM Computing Surveys}, vol.~54, no.~6, pp. 1--25, Jul. 2021. [Online]. Available: \url{https://doi.org/10.11452F3459665}
\BIBentrySTDinterwordspacing

\bibitem{rule}
Z.~Wang, W.~Zhang, N.~Liu, and J.~Wang, ``{Scalable rule-based representation learning for interpretable classification},'' \emph{Advances in Neural Information Processing Systems}, vol.~34, pp. 30\,479--30\,491, 2021\color{black}.

\bibitem{gen}
R.~Agarwal, L.~Melnick, N.~Frosst, X.~Zhang, B.~Lengerich, R.~Caruana, and G.~E. Hinton, ``Neural additive models: Interpretable machine learning with neural nets,'' \emph{Advances in neural information processing systems}, vol.~34, pp. 4699--4711, 2021\color{black}.

\bibitem{bayes}
\BIBentryALTinterwordspacing
M.~C.~A. Clare, M.~Sonnewald, R.~Lguensat, J.~Deshayes, and V.~Balaji, ``{Explainable Artificial Intelligence for Bayesian Neural Networks: Toward Trustworthy Predictions of Ocean Dynamics},'' \emph{Journal of Advances in Modeling Earth Systems}, vol.~14, no.~11, oct 2022. [Online]. Available: \url{https://doi.org/10.10292F2022ms003162}
\BIBentrySTDinterwordspacing

\bibitem{senn}
D.~Alvarez-Melis and T.~Jaakkola, ``{Towards Robust Interpretability with Self-Explaining Neural Networks},'' \emph{ArXiv}, vol. abs/1806.07538, 2018.

\bibitem{vale2022explainable}
D.~Vale, A.~El-Sharif, and M.~Ali, ``{Explainable artificial intelligence (XAI) post-hoc explainability methods: Risks and limitations in non-discrimination law},'' \emph{AI and Ethics}, pp. 1--12, 2022.

\bibitem{features}
B.~Zhou, A.~Khosla, A.~Lapedriza, A.~Oliva, and A.~Torralba, ``{Learning Deep Features for Discriminative Localization},'' 2015.

\bibitem{intml}
W.~J. Murdoch, C.~Singh, K.~Kumbier, R.~Abbasi-Asl, and B.~Yu, ``{Interpretable machine learning: definitions, methods, and applications},'' \emph{ArXiv}, vol. abs/1901.04592, 2019.

\bibitem{euca}
W.~Jin, J.~Fan, D.~Gromala, P.~Pasquier, and G.~Hamarneh, ``{EUCA: A Practical Prototyping Framework towards End-User-Centered Explainable Artificial Intelligence},'' \emph{ArXiv}, vol. abs/2102.02437, 2021.

\bibitem{VC_}
N.~Tamani, B.~Brik, N.~Lagraa, and Y.~Ghamri-Doudane, ``{Vehicular Cloud Service Provider Selection: A Flexible Approach},'' in \emph{GLOBECOM 2017 - 2017 IEEE Global Communications Conference}, 2017, pp. 1--6.

\bibitem{xai_decision}
\BIBentryALTinterwordspacing
J.~Gerlach, P.~Hoppe, S.~Jagels, L.~Licker, and M.~H. Breitner, ``{Decision support for efficient XAI services - A morphological analysis, business model archetypes, and a decision tree},'' \emph{Electron. Mark.}, vol.~32, no.~4, pp. 2139--2158, 2022. [Online]. Available: \url{https://doi.org/10.1007/s12525-022-00603-6}
\BIBentrySTDinterwordspacing

\bibitem{eng}
A.~H. Mohammadkhani, N.~S. Bommi, M.~Daboussi, O.~Sabnis, C.~Tantithamthavorn, and H.~Hemmati, ``{A Systematic Literature Review of Explainable AI for Software Engineering},'' 2023.

\bibitem{audit}
C.~A. Zhang, S.~Cho, and M.~Vasarhelyi, ``{Explainable Artificial Intelligence (XAI) in auditing},'' \emph{International Journal of Accounting Information Systems}, vol.~46, p. 100572, 2022.

\bibitem{legal}
J.~Collenette, K.~Atkinson, and T.~Bench-Capon, ``{Explainable AI tools for legal reasoning about cases: A study on the European Court of Human Rights},'' \emph{Artificial Intelligence}, vol. 317, p. 103861, 2023.

\bibitem{xai_regulation}
\BIBentryALTinterwordspacing
N.~A. Smuha, ``{The EU Approach to Ethics Guidelines for Trustworthy Artificial Intelligence},'' \emph{Computer Law Review International}, vol.~20, no.~4, pp. 97--106, 2019. [Online]. Available: \url{https://ssrn.com/abstract=3443537}
\BIBentrySTDinterwordspacing

\bibitem{DeepShape}
S.~M. Lundberg and S.-I. Lee, ``{A Unified Approach to Interpreting Model Predictions},'' in \emph{International Conference on Neural Information Processing Systems (NIPS)}, 2017, pp. 4768--4777.

\bibitem{treeshape}
S.~M. Lundberg, G.~G. Erion, and S.-I. Lee, ``{Consistent Individualized Feature Attribution for Tree Ensembles},'' 2019.

\bibitem{rulefit}
J.~H. Friedman and B.~E. Popescu, ``{Predictive Learning via Rule Ensembles},'' \emph{The Annals of Applied Statistics}, vol.~2, no.~3, pp. 916--954, 2008.

\bibitem{ig}
\BIBentryALTinterwordspacing
M.~Sundararajan, A.~Taly, and Q.~Yan, ``{Axiomatic Attribution for Deep Networks},'' \emph{CoRR}, vol. abs/1703.01365, 2017. [Online]. Available: \url{http://arxiv.org/abs/1703.01365}
\BIBentrySTDinterwordspacing

\bibitem{abc}
S.~Jha, S.~Raj, S.~L. Fernandes, S.~K. Jha, S.~Jha, B.~Jalaian, G.~Verma, and A.~Swami, \emph{{Attribution-Based Confidence Metric for Deep Neural Networks}}.\hskip 1em plus 0.5em minus 0.4em\relax Red Hook, NY, USA: Curran Associates Inc., 2019.

\bibitem{logodds}
A.~Shrikumar, P.~Greenside, and A.~Kundaje, ``{Learning Important Features Through Propagating Activation Differences},'' \emph{CoRR}, vol. abs/1704.02685, 2017.

\bibitem{suff_comp}
J.~DeYoung, S.~Jain, N.~F. Rajani, E.~Lehman, C.~Xiong, R.~Socher, and B.~C. Wallace, ``{ERASER: A Benchmark to Evaluate Rationalized NLP Models},'' \emph{CoRR}, vol. abs/1911.03429, 2019.

\bibitem{complexity}
\BIBentryALTinterwordspacing
U.~Bhatt, A.~Weller, and J.~M.~F. Moura, ``{Evaluating and Aggregating Feature-based Model Explanations},'' in \emph{International Joint Conference on Artificial Intelligence}, 2020 \color{black}. [Online]. Available: \url{https://api.semanticscholar.org/CorpusID:218486810}
\BIBentrySTDinterwordspacing

\bibitem{sliceops}
\BIBentryALTinterwordspacing
F.~Rezazadeh, H.~Chergui, L.~Alonso, and C.~Verikoukis, ``{SliceOps: Explainable MLOps for Streamlined Automation-Native 6G Networks},'' 2023. [Online]. Available: \url{https://arxiv.org/abs/2307.01658}
\BIBentrySTDinterwordspacing

\bibitem{fidelity}
C.-K. Yeh, C.-Y. Hsieh, A.~S. Suggala, D.~I. Inouye, and P.~Ravikumar, ``{On the (in)Fidelity and Sensitivity of Explanations},'' in \emph{Proceedings of the 33rd International Conference on Neural Information Processing Systems}.\hskip 1em plus 0.5em minus 0.4em\relax Red Hook, NY, USA: Curran Associates Inc., 2019.

\bibitem{fidelity2}
M.~Velmurugan, C.~Ouyang, C.~Moreira, and R.~Sindhgatta, ``{Developing a Fidelity Evaluation Approach for Interpretable Machine Learning},'' \emph{CoRR}, vol. abs/2106.08492, 2021.

\bibitem{consistency}
T.~Fel and D.~Vigouroux, ``{Representativity and Consistency Measures for Deep Neural Network Explanations},'' \emph{CoRR}, vol. abs/2009.04521, 2020.

\bibitem{bleu}
K.~Papineni, S.~Roukos, T.~Ward, and W.-J. Zhu, ``{Bleu: a method for automatic evaluation of machine translation},'' in \emph{Proceedings of the 40th annual meeting of the Association for Computational Linguistics}, 2002 \color{black}, pp. 311--318.

\bibitem{rouge}
\BIBentryALTinterwordspacing
C.-Y. Lin, ``{ROUGE: A Package for Automatic Evaluation of Summaries},'' in \emph{Text Summarization Branches Out}.\hskip 1em plus 0.5em minus 0.4em\relax Barcelona, Spain: Association for Computational Linguistics, Jul 2004\color{black}, pp. 74--81. [Online]. Available: \url{https://aclanthology.org/W04-1013}
\BIBentrySTDinterwordspacing

\bibitem{perp}
P.~F. Brown, J.~Cocke, S.~A. Della~Pietra, V.~J. Della~Pietra, F.~Jelinek, R.~L. Mercer, and P.~Roossin, ``{A statistical approach to language translation},'' in \emph{Coling Budapest 1988 Volume 1: International Conference on Computational Linguistics}, 1988 \color{black}.

\bibitem{FMR}
\BIBentryALTinterwordspacing
S.~Roy, H.~Chergui, A.~Ksentini, and C.~Verikoukis, ``{Federated Machine Reasoning for Resource Provisioning in 6G O-RAN},'' 2024\color{black}. [Online]. Available: \url{https://api.semanticscholar.org/CorpusID:270371415}
\BIBentrySTDinterwordspacing

\bibitem{dataset}
\BIBentryALTinterwordspacing
J.~X. Salvat~Lozano, J.~A. Ayala-Romero, L.~Zanzi, A.~Garcia-Saavedra, and X.~Costa-Perez, ``{O-RAN experimental evaluation datasets},'' 2022. [Online]. Available: \url{https://dx.doi.org/10.21227/64s5-q431}
\BIBentrySTDinterwordspacing

\bibitem{3gpp_release15}
\BIBentryALTinterwordspacing
3rd Generation Partnership Project~(3GPP), ``{Technical Specification Group Radio Access Network; Study on New Radio Access Technology: Radio Access Architecture and Interfaces},'' 3GPP, Technical Specification (TS) 38.300, 2018. [Online]. Available: \url{https://www.3gpp.org/ftp/Specs/html-info/38300.htm}
\BIBentrySTDinterwordspacing

\bibitem{o_ran_lls}
\BIBentryALTinterwordspacing
O.-R. Alliance, ``{O-RAN Alliance Specification: Lower Layer Split (LLS) Option 7-2x},'' Specification, 2023. [Online]. Available: \url{https://www.o-ran.org/specifications}
\BIBentrySTDinterwordspacing

\bibitem{boutiba}
K.~Boutiba, A.~Ksentini, B.~Brik, Y.~Challal, and A.~Balla, ``{NRflex: Enforcing network slicing in 5G New Radio},'' \emph{Computer Communications}, vol. 181, pp. 284--292, 2022.

\bibitem{ai_ml}
O.-R. W.~G. 2, ``{O-RAN AI/ML workflow description and requirements 1.03},'' \emph{O-RAN. WG2. AIML-v01. 03 Technical Specification}, 2021.

\bibitem{ieee-p2894}
{IEEE P2894}, ``{Guide for an Architectural Framework for Explainable Artificial Intelligence},'' \url{https://standards.ieee.org/ieee/2894/10284/}.

\bibitem{ieee-p2976}
{IEEE P2976}, ``{Standard for XAI, eXplainable Artificial Intelligence, for Achieving Clarity and Interoperability of AI Systems Design},'' \url{https://standards.ieee.org/ieee/2976/10522/}.

\bibitem{etsi-oran}
O.~Alliance, ``{O-RAN fronthaul control, user and synchronization plane specification 8.0},'' \emph{Specification WG4: Open Fronthaul Interfaces Workgroup}, 2022\color{black}.

\bibitem{6g_bricks}
\BIBentryALTinterwordspacing
6G-BRICKS, 2023. [Online]. Available: \url{https://6g-bricks.eu/}
\BIBentrySTDinterwordspacing

\bibitem{nancy}
\BIBentryALTinterwordspacing
NANCY, 2023. [Online]. Available: \url{https://cordis.europa.eu/project/id/101096456}
\BIBentrySTDinterwordspacing

\bibitem{OpenFL-XAI}
\BIBentryALTinterwordspacing
H.-X. project, ``{OpenFL-XAI},'' 2023 \color{black}. [Online]. Available: \url{https://github.com/Unipisa/OpenFL-XAI}
\BIBentrySTDinterwordspacing

\bibitem{dapps}
S.~D'Oro, M.~Polese, L.~Bonati, H.~Cheng, and T.~Melodia, ``{dApps: Distributed Applications for Real-Time Inference and Control in O-RAN},'' \emph{IEEE Communications Magazine}, vol.~60, no.~11, pp. 52--58, 2022\color{black}.

\bibitem{fuzzyfl}
J.~L.~C. B{\'a}rcena, P.~Ducange, A.~Ercolani, F.~Marcelloni, and A.~Renda, ``An approach to federated learning of explainable fuzzy regression models,'' in \emph{2022 IEEE International Conference on Fuzzy Systems (FUZZ-IEEE)}.\hskip 1em plus 0.5em minus 0.4em\relax IEEE, 2022, pp. 1--8.

\bibitem{rw9}
F.~Rezazadeh, L.~Zanzi, F.~Devoti, H.~Chergui, X.~Costa-P\'erez, and C.~Verikoukis, ``{On the Specialization of FDRL Agents for Scalable and Distributed 6G RAN Slicing Orchestration},'' \emph{IEEE Transactions on Vehicular Technology}, pp. 1--15, 2022.

\bibitem{Sec_ORAN}
Z.~A.~E. Houda, H.~Moudoud, and B.~Brik, ``Federated deep reinforcement learning for efficient jamming attack mitigation in o-ran,'' \emph{IEEE Transactions on Vehicular Technology}, vol.~73, no.~7, pp. 9334--9343, 2024.

\bibitem{egl}
A.~A. Ismail, H.~Corrada~Bravo, and S.~Feizi, ``{Improving deep learning interpretability by saliency guided training},'' \emph{Advances in Neural Information Processing Systems}, vol.~34, pp. 26\,726--26\,739, 2021.

\bibitem{egl2}
J.~Sun, S.~Lapuschkin, W.~Samek, Y.~Zhao, N.-M. Cheung, and A.~Binder, ``{Explanation-Guided Training for Cross-Domain Few-Shot Classification},'' \emph{2020 25th International Conference on Pattern Recognition (ICPR)}, pp. 7609--7616, 2020.

\bibitem{egfl}
S.~Roy, H.~Chergui, and C.~Verikoukis, ``Explanation-guided fair federated learning for transparent 6g ran slicing,'' \emph{IEEE Transactions on Cognitive Communications and Networking}, pp. 1--1, 2024.

\bibitem{oranus}
O.~Adamuz-Hinojosa, L.~Zanzi, V.~Sciancalepore, A.~Garcia-Saavedra, and X.~Costa-P\'erez, ``{ORANUS: Latency-tailored Orchestration via Stochastic Network Calculus in 6G O-RAN},'' \emph{IEEE International Conference on Computer Communications (INFOCOM)}, 2024\color{black}.

\bibitem{bridging}
S.~Roy, H.~Chergui, and C.~Verikoukis, ``{Towards Bridging the FL Performance-Explainability Trade-Off: A Trustworthy 6G RAN Slicing Use-Case},'' \emph{IEEE Transactions on Vehicular Technology}, pp. 1--11, 2024.

\bibitem{zaka2}
Z.~A.~E. Houda, B.~Brik, A.~Ksentini, L.~Khoukhi, and M.~Guizani, ``{When Federated Learning Meets Game Theory: A Cooperative Framework to Secure IIoT Applications on Edge Computing},'' \emph{IEEE Transactions on Industrial Informatics}, vol.~18, no.~11, pp. 7988--7997, 2022.

\bibitem{fl_acm}
\BIBentryALTinterwordspacing
B.~Brik, M.~Messaadia, M.~Sahnoun, B.~Bettayeb, and M.~A. Benatia, ``{Fog-Supported Low-Latency Monitoring of System Disruptions in Industry 4.0: A Federated Learning Approach},'' \emph{ACM Trans. Cyber-Phys. Syst.}, vol.~6, no.~2, may 2022. [Online]. Available: \url{https://doi.org/10.1145/3477272}
\BIBentrySTDinterwordspacing

\bibitem{mlop}
D.~Sculley, G.~Holt, D.~Golovin, E.~Davydov, T.~Phillips, D.~Ebner, V.~Chaudhary, and M.~Young, ``{Machine Learning: The High Interest Credit Card of Technical Debt},'' in \emph{SE4ML: Software Engineering for Machine Learning (NIPS 2014 Workshop)}, 2014.

\bibitem{google1}
\BIBentryALTinterwordspacing
``{MLOps: Continuous delivery and automation pipelines in machine learning},'' Dec. 2022. [Online]. Available: \url{https://cloud.google.com/architecture/mlops-continuous-delivery-and-automation-pipelines-in-machine-learning}
\BIBentrySTDinterwordspacing

\bibitem{RLOps}
P.~Li, J.~Thomas, X.~Wang, A.~Khalil, A.~Ahmad, R.~Inacio, S.~Kapoor, A.~Parekh, A.~Doufexi, A.~Shojaeifard, and R.~J. Piechocki, ``{RLOps: Development Life-Cycle of Reinforcement Learning Aided Open RAN},'' \emph{IEEE Access}, vol.~10, pp. 113\,808--113\,826, 2022.

\bibitem{T1}
P.~Porambage, J.~Pinola, Y.~Rumesh, C.~Tao, and J.~Huusko, ``{XcARet: XAI based Green Security Architecture for Resilient Open Radio Access Networks in 6G},'' in \emph{2023 Joint European Conference on Networks and Communications \& 6G Summit (EuCNC/6G Summit)}, 2023, pp. 699--704.

\bibitem{T2}
J.~Groen, S.~DOro, U.~Demir, L.~Bonati, M.~Polese, T.~Melodia, and K.~Chowdhury, ``{Implementing and Evaluating Security in O-RAN: Interfaces, Intelligence, and Platforms},'' 2023.

\bibitem{T3}
N.~M. Yungaicela-Naula, V.~Sharma, and S.~Scott-Hayward, ``{Misconfiguration in O-RAN: Analysis of the impact of AI/ML},'' \emph{Computer Networks}, vol. 247, p. 110455, 2024.

\bibitem{fiandrino2023explora}
C.~Fiandrino, L.~Bonati, S.~D'Oro, M.~Polese, T.~Melodia, and J.~Widmer, ``{EXPLORA: AI/ML EXPLainability for the Open RAN},'' \emph{Proceedings of the ACM on Networking}, vol.~1, no. CoNEXT3, pp. 1--26, 2023\color{black}.

\bibitem{T4}
F.~Rezazadeh, H.~Chergui, S.~Siddiqui, J.~Mangues, H.~Song, W.~Saad, and M.~Bennis, ``{Intelligible Protocol Learning for Resource Allocation in 6G O-RAN Slicing},'' 2023.

\bibitem{queries}
C.~Tassie, B.~Kim, J.~Groen, M.~Belgiovine, and K.~R. Chowdhury, ``{Leveraging Explainable AI for Reducing Queries of Performance Indicators in Open RAN},'' in \emph{IEEE International Conference on Communications (ICC)}, June 2024\color{black}.

\bibitem{depth}
P.~F. P\'{e}rez, C.~Fiandrino, M.~Fiore, and J.~Widmer, ``{An In-Depth Analysis of Advanced Time Series Forecasting Models for the Open RAN},'' in \emph{IEEE INFOCOM 2024}, 2024 \color{black}.

\bibitem{T5}
S.~Roy, F.~Rezazadeh, H.~Chergui, and C.~Verikoukis, ``{Joint Explainability and Sensitivity-Aware Federated Deep Learning for Transparent 6G RAN Slicing},'' in \emph{ICC 2023 - IEEE International Conference on Communications}, 2023, pp. 1238--1243.

\bibitem{rw1}
Y.~Cao, S.-Y. Lien, Y.-C. Liang, K.-C. Chen, and X.~Shen, ``{User Access Control in Open Radio Access Networks: A Federated Deep Reinforcement Learning Approach},'' \emph{IEEE Transactions on Wireless Communications}, vol.~21, no.~6, pp. 3721--3736, 2022.

\bibitem{rw2}
Y.~Cao, S.-Y. Lien, Y.-C. Liang, and K.-C. Chen, ``{Federated Deep Reinforcement Learning for User Access Control in Open Radio Access Networks},'' in \emph{ICC 2021 - IEEE International Conference on Communications}, 2021, pp. 1--6.

\bibitem{scalingi2024infocom}
A.~Scalingi, S.~D'Oro, F.~Restuccia, T.~Melodia, and D.~Giustiniano, ``{Det-RAN: Data-Driven Cross-Layer Real-Time Attack Detection in 5G Open RANs},'' in \emph{{IEEE INFOCOM 2024 - IEEE Conference on Computer Communications}}, May 2024.

\bibitem{maxenti2024}
\BIBentryALTinterwordspacing
S.~Maxenti, S.~D'Oro, L.~Bonati, M.~Polese, A.~Capone, and T.~Melodia, ``{ScalO-RAN: Energy-aware Network Intelligence Scaling in Open RAN},'' 2024. [Online]. Available: \url{https://arxiv.org/abs/2312.05096}
\BIBentrySTDinterwordspacing

\bibitem{CSI-ORAN}
H.~Cheng, P.~Johari, M.~A. Arfaoui, F.~Periard, P.~Pietraski, G.~Zhang, and T.~Melodia, ``{Real-Time AI-Enabled CSI Feedback Experimentation with Open RAN},'' in \emph{2024 19th Wireless On-Demand Network Systems and Services Conference (WONS)}, 2024, pp. 121--124.

\bibitem{rw3}
H.~Lee, Y.~Jang, J.~Song, and H.~Yeon, ``{O-RAN AI/ML Workflow Implementation of Personalized Network Optimization via Reinforcement Learning},'' in \emph{2021 IEEE Globecom Workshops (GC Wkshps)}, 2021, pp. 1--6.

\bibitem{raftopoulos2024}
\BIBentryALTinterwordspacing
R.~Raftopoulos, S.~D'Oro, T.~Melodia, and G.~Schembra, ``{DRL-based Latency-Aware Network Slicing in O-RAN with Time-Varying SLAs},'' 2024. [Online]. Available: \url{https://arxiv.org/abs/2401.05042}
\BIBentrySTDinterwordspacing

\bibitem{rw4}
S.~Mollahasani, M.~Erol-Kantarci, and R.~Wilson, ``{Dynamic CU-DU Selection for Resource Allocation in O-RAN Using Actor-Critic Learning},'' in \emph{2021 IEEE Global Communications Conference (GLOBECOM)}, 2021, pp. 1--6.

\bibitem{OrchestRAN}
S.~D'Oro, L.~Bonati, M.~Polese, and T.~Melodia, ``{OrchestRAN: Network automation through orchestrated intelligence in the open RAN},'' in \emph{IEEE INFOCOM 2022-IEEE Conference on Computer Communications}.\hskip 1em plus 0.5em minus 0.4em\relax IEEE, 2022, pp. 270--279.

\bibitem{rw5}
\BIBentryALTinterwordspacing
P.~E. Iturria-Rivera, H.~Zhang, H.~Zhou, S.~Mollahasani, and M.~Erol-Kantarci, ``{Multi-Agent Team Learning in Virtualized Open Radio Access Networks (O-RAN)},'' \emph{Sensors}, vol.~22, no.~14, 2022. [Online]. Available: \url{https://www.mdpi.com/1424-8220/22/14/5375}
\BIBentrySTDinterwordspacing

\bibitem{rw6}
\BIBentryALTinterwordspacing
P.~E.~I. Rivera, S.~Mollahasani, and M.~Erol{-}Kantarci, ``{Multi Agent Team Learning in Disaggregated Virtualized Open Radio Access Networks (O-RAN)},'' \emph{CoRR}, vol. abs/2012.04861, 2020. [Online]. Available: \url{https://arxiv.org/abs/2012.04861}
\BIBentrySTDinterwordspacing

\bibitem{rw7}
\BIBentryALTinterwordspacing
H.~Zhang, H.~Zhou, and M.~Erol-Kantarci, ``{Federated Deep Reinforcement Learning for Resource Allocation in O-RAN Slicing},'' 2022. [Online]. Available: \url{https://arxiv.org/abs/2208.01736}
\BIBentrySTDinterwordspacing

\bibitem{rw8}
------, ``{Team Learning-Based Resource Allocation for Open Radio Access Network (O-RAN)},'' in \emph{ICC 2022 - IEEE International Conference on Communications}, 2022, pp. 4938--4943.

\bibitem{rw11}
\BIBentryALTinterwordspacing
A.~Abouaomar, A.~Taik, A.~Filali, and S.~Cherkaoui, ``{Federated Deep Reinforcement Learning for Open RAN Slicing in 6G Networks},'' 2022. [Online]. Available: \url{https://arxiv.org/abs/2206.11328}
\BIBentrySTDinterwordspacing

\bibitem{rw10}
X.~Wang, J.~D. Thomas, R.~J. Piechocki, S.~Kapoor, R.~Santos-Rodr{\'\i}guez, and A.~Parekh, ``{Self-play learning strategies for resource assignment in Open-RAN networks},'' \emph{Computer Networks}, vol. 206, p. 108682, 2022.

\bibitem{tsampazi2024pandora}
M.~Tsampazi, S.~D'Oro, M.~Polese, L.~Bonati, G.~Poitau, M.~Healy, and T.~Melodia, ``{PandORA: Automated Design and Comprehensive Evaluation of Deep Reinforcement Learning Agents for Open RAN},'' \emph{ResearchGate Preprint}, pp. 1--18, May 2024.

\bibitem{mdp}
Y.~{Chen}, Y.~{Gao}, C.~{Jiang}, and K.~J.~R. {Liu}, ``{Game theoretic Markov decision processes for optimal decision making in social systems},'' in \emph{2014 IEEE Global Conference on Signal and Information Processing (GlobalSIP)}, 2014, pp. 268--272.

\bibitem{xdrl}
A.~Krajna, M.~Brcic, T.~Lipic, and J.~Doncevic, ``{Explainability in reinforcement learning: perspective and position},'' 2022.

\bibitem{pol_sim}
R.~S. Sutton, D.~McAllester, S.~Singh, and Y.~Mansour, ``{Policy Gradient Methods for Reinforcement Learning with Function Approximation},'' in \emph{Advances in Neural Information Processing Systems}, S.~Solla, T.~Leen, and K.~M\"{u}ller, Eds., vol.~12.\hskip 1em plus 0.5em minus 0.4em\relax MIT Press, 1999.

\bibitem{rew_deco}
Z.~Juozapaitis, A.~Koul, A.~Fern, M.~Erwig, and F.~Doshi-Velez, ``{Explainable Reinforcement Learning via Reward Decomposition},'' in \emph{in proceedings at the International Joint Conference on Artificial Intelligence. A Workshop on Explainable Artificial Intelligence.}, 2019.

\bibitem{scm1}
S.~Sloman, \emph{{Causal Models: How People Think about the World and Its Alternatives}}.\hskip 1em plus 0.5em minus 0.4em\relax Oxford University Press, 08 2005.

\bibitem{econs}
J.~van~der Waa, J.~van Diggelen, K.~v.~d. Bosch, and M.~Neerincx, ``{Contrastive explanations for reinforcement learning in terms of expected consequences},'' \emph{arXiv preprint arXiv:1807.08706}, 2018.

\bibitem{hie_rl}
T.~Shu, C.~Xiong, and R.~Socher, ``{Hierarchical and Interpretable Skill Acquisition in Multi-task Reinforcement Learning},'' 2017.

\bibitem{rela_drl}
V.~Zambaldi, D.~Raposo, A.~Santoro, V.~Bapst, Y.~Li, I.~Babuschkin, K.~Tuyls, D.~Reichert, T.~Lillicrap, E.~Lockhart, M.~Shanahan, V.~Langston, R.~Pascanu, M.~Botvinick, O.~Vinyals, and P.~Battaglia, ``{Relational Deep Reinforcement Learning},'' 2018.

\bibitem{oran_use_cases}
{O-RAN Alliance}, ``{O-RAN Working Group (O-RAN.WG1.Use-Cases-Analysis-Report-v09.00)},'' {Technical Specification}, Oct. 2022.

\bibitem{izima2021survey}
O.~Izima, R.~de~Fr{\'e}in, and A.~Malik, ``{A Survey of Machine Learning Techniques for Video Quality Prediction from Quality of Delivery Metrics},'' \emph{Electronics}, vol.~10, no.~22, p. 2851, 2021\color{black}.

\bibitem{renda2021xai}
A.~Renda, P.~Ducange, G.~Gallo, and F.~Marcelloni, ``{XAI models for quality of experience prediction in wireless networks},'' in \emph{IEEE International Conference on Fuzzy Systems (FUZZ-IEEE)}.\hskip 1em plus 0.5em minus 0.4em\relax IEEE, 2021\color{black}, pp. 1--6.

\bibitem{ayoub2022towards}
O.~Ayoub, S.~Troia, D.~Andreoletti, A.~Bianco, M.~Tornatore, S.~Giordano, and C.~Rottondi, ``{Towards explainable artificial intelligence in optical networks: the use case of lightpath QoT estimation},'' \emph{Journal of Optical Communications and Networking}, vol.~15, no.~1, pp. A26--A38, 2022\color{black}.

\bibitem{oran_slicing_architecture}
{O-RAN Alliance}, ``{O-RAN-WG1-O-RAN Slicing Architecture v03.00},'' {Technical Specification}, 2021.

\bibitem{sabra1}
S.~B. Saad, A.~Ksentini, and B.~Brik, ``{A Trust architecture for the SLA management in 5G networks},'' in \emph{ICC 2021 - IEEE International Conference on Communications}, 2021, pp. 1--6.

\bibitem{barnard2022resource}
P.~Barnard, I.~Macaluso, N.~Marchetti, and L.~A. DaSilva, ``{Resource reservation in sliced networks: an explainable artificial intelligence (XAI) approach},'' in \emph{IEEE International Conference on Communications (IEEE ICC)}.\hskip 1em plus 0.5em minus 0.4em\relax IEEE, 2022, pp. 1530--1535.

\bibitem{terra2022using}
A.~Terra, R.~Inam, P.~Batista, and E.~Fersman, ``{Using Counterfactuals to Proactively Solve Service Level Agreement Violations in 5G Networks},'' in \emph{IEEE International Conference on Industrial Informatics (INDIN)}.\hskip 1em plus 0.5em minus 0.4em\relax IEEE, 2022\color{black}, pp. 552--559.

\bibitem{yeh2023deep}
S.-p. Yeh, S.~Bhattacharya, R.~Sharma, and H.~Moustafa, ``{Deep Learning for Intelligent and Automated Network Slicing in 5G Open RAN (ORAN) Deployment},'' \emph{IEEE Open Journal of the Communications Society}, 2023\color{black}.

\bibitem{hamdan2023recent}
M.~Q. Hamdan, H.~Lee, D.~Triantafyllopoulou, R.~Borralho, A.~Kose, E.~Amiri, D.~Mulvey, W.~Yu, R.~Zitouni, R.~Pozza \emph{et~al.}, ``{Recent advances in machine learning for network automation in the o-ran},'' \emph{Sensors}, vol.~23, no.~21, p. 8792, 2023\color{black}.

\bibitem{khan2023explainable}
N.~Khan, S.~Coleri, A.~Abdallah, A.~Celik, and A.~M. Eltawil, ``{Explainable and Robust Artificial Intelligence for Trustworthy Resource Management in 6G Networks},'' \emph{IEEE Communications Magazine}, 2023\color{black}.

\bibitem{zaka3}
Z.~A. El~Houda, L.~Khoukhi, and B.~Brik, ``{A Low-Latency Fog-based Framework to secure IoT Applications using Collaborative Federated Learning},'' in \emph{2022 IEEE 47th Conference on Local Computer Networks (LCN)}, 2022, pp. 343--346.

\bibitem{fl_lcn}
B.~Brik and A.~Ksentini, ``{On Predicting Service-oriented Network Slices Performances in 5G: A Federated Learning Approach},'' in \emph{2020 IEEE 45th Conference on Local Computer Networks (LCN)}, 2020, pp. 164--171.

\bibitem{pi_road}
A.~{Okic}, L.~{Zanzi}, V.~{Sciancalepore}, A.~{Redondi}, and X.~{Costa-P\'erez}, ``{$\pi$-ROAD: a Learn-as-You-Go Framework for On-Demand Emergency Slices in V2X Scenarios},'' in \emph{{IEEE International Conference on Computer Communications}}, Jul. 2021.

\bibitem{9600616}
Y.~Cao, S.-Y. Lien, Y.-C. Liang, K.-C. Chen, and X.~Shen, ``{User Access Control in Open Radio Access Networks: A Federated Deep Reinforcement Learning Approach},'' \emph{IEEE Transactions on Wireless Communications}, vol.~21, no.~6, pp. 3721--3736, 2022.

\bibitem{oran_security_req}
{O-RAN Alliance}, ``{O-RAN-WG11-O-RAN Security Requirements Specifications v04.00)},'' {Technical Specification}, 2022.

\bibitem{Sab}
S.~B. Saad, B.~Brik, and A.~Ksentini, ``{Toward Securing Federated Learning against Poisoning Attacks in Zero Touch B5G networks},'' \emph{IEEE Transactions on Network and Service Management}, pp. 1--1, 2023.

\bibitem{oran_threat_model}
{O-RAN Alliance}, ``{O-RAN-WG11-O-RAN Security Threat Modeling and Remediation Analysis (O-RAN.WG11.Threat-Model-v04.00)},'' {Technical Specification}, 2022.

\bibitem{Hexa_X_vision}
J.~P\'erez-Valero, A.~Virdis, A.~G. Sanchez, C.~Ntogkas, P.~Serrano, G.~Landi, S.~Kuklinski, C.~Morin, I.~L. Pavon, and B.~Sayadi, ``{AI-driven Orchestration for 6G Networking: the Hexa-X vision},'' in \emph{IEEE Globecom Workshops (GC Wkshps)}, 2022\color{black}, pp. 1335--1340.

\bibitem{senevirathna2022survey}
T.~Senevirathna, Z.~Salazar, V.~H. La, S.~Marchal, B.~Siniarski, M.~Liyanage, and S.~Wang, ``{A survey on XAI for beyond 5G security: technical aspects, use cases, challenges and research directions},'' \emph{arXiv preprint arXiv:2204.12822}, 2022.

\bibitem{oran_security_risk_Non_RT}
{O-RAN Security Focus Group}, ``{Study on Security for Non-RT-RIC (O-RAN.WG11.Security-Requirements-Specifications-v04.00)},'' {Technical Specification}, 2022.

\bibitem{oran_security_risk_O_Cloud}
------, ``{Study on Security for O-Cloud (O-RAN.WG11.Security-Requirements-Specifications-v04.00)},'' {Technical Specification}, 2022.

\bibitem{oran_security_risk_Near_RT}
------, ``{Study on Security for Near Real-Time RIC and xApps (O-RAN.SFG.Security-for-Near-RT-RIC-xApps-v01.00)},'' {Technical Specification}, 2022.

\bibitem{oran_risk_DE}
{K\"opsell, Stefan and Ruzhanskiy, Andrey and Hecker, Andreas and Stachorra, Dirk and Franchi, Norman}, ``{Open RAN Risk Analysis},'' {Report}, Feb. 2022.

\bibitem{oran_risk_UE}
{NIS Cooperation Group}, ``{Report on the cybersecurity of Open RAN},'' {Report}, May 2022.

\bibitem{wang2021applications}
S.~Wang, M.~A. Qureshi, L.~Miralles-Pechu{\'a}n, T.~Huynh-The, T.~R. Gadekallu, and M.~Liyanage, ``{Applications of Explainable AI for 6G: Technical Aspects, Use Cases, and Research Challenges},'' \emph{arXiv preprint arXiv:2112.04698}, 2021.

\bibitem{habler2022adversarial}
E.~Habler, R.~Bitton, D.~Avraham, D.~Mimran, E.~Klevansky, O.~Brodt, H.~Lehmann, Y.~Elovici, and A.~Shabtai, ``{Adversarial Machine Learning Threat Analysis and Remediation in Open Radio Access Network (O-RAN)},'' \emph{arXiv preprint arXiv:2201.06093}, 2022\color{black}.

\bibitem{chen2020hopskipjumpattack}
J.~Chen, M.~I. Jordan, and M.~J. Wainwright, ``{Hopskipjumpattack: A query-efficient decision-based attack},'' in \emph{IEEE Symposium on Security and Privacy (SP)}.\hskip 1em plus 0.5em minus 0.4em\relax IEEE, 2020\color{black}, pp. 1277--1294.

\bibitem{chiejina2024system}
A.~Chiejina, B.~Kim, K.~Chowhdury, and V.~K. Shah, ``{System-level Analysis of Adversarial Attacks and Defenses on Intelligence in O-RAN based Cellular Networks},'' \emph{arXiv preprint arXiv:2402.06846}, 2024\color{black}.

\bibitem{mirsky2022threat}
Y.~Mirsky, A.~Demontis, J.~Kotak, R.~Shankar, D.~Gelei, L.~Yang, X.~Zhang, M.~Pintor, W.~Lee, Y.~Elovici \emph{et~al.}, ``{The threat of offensive AI to organizations},'' \emph{Computers \& Security}, p. 103006, 2022.

\bibitem{kuppa2020black}
A.~Kuppa and N.-A. Le-Khac, ``{Black box attacks on explainable artificial intelligence (XAI) methods in cyber security},'' in \emph{2020 International Joint Conference on Neural Networks (IJCNN)}.\hskip 1em plus 0.5em minus 0.4em\relax IEEE, 2020, pp. 1--8.

\bibitem{zhao2021exploiting}
X.~Zhao, W.~Zhang, X.~Xiao, and B.~Lim, ``{Exploiting explanations for model inversion attacks},'' in \emph{Proceedings of the IEEE/CVF international conference on computer vision}, 2021, pp. 682--692.

\bibitem{giudici2022explainable}
P.~Giudici and E.~Raffinetti, ``{Explainable AI methods in cyber risk management},'' \emph{Quality and reliability engineering international}, vol.~38, no.~3, pp. 1318--1326, 2022.

\bibitem{tradeoff1}
M.~Naylor, C.~French, S.~Terker, and U.~Kamath, ``{Quantifying Explainability in {NLP} and Analyzing Algorithms for Performance-Explainability Tradeoff},'' in \emph{International Conference on Machine Learning}, 2021.

\bibitem{llm2}
\BIBentryALTinterwordspacing
L.~Bariah, Q.~Zhao, H.~Zou, Y.~Tian, F.~Bader, and M.~Debbah, ``{Large Language Models for Telecom: The Next Big Thing?}'' 2023. [Online]. Available: \url{https://arxiv.org/abs/2306.10249}
\BIBentrySTDinterwordspacing

\bibitem{nokia-oran}
{Nokia}, ``{Nokia opens new O-RAN Collaboration and Testing Center in the U.S.}'' https://www.nokia.com/about-us/news/releases/2021\\ /06/16/nokia-opens-new-o-ran-collaboration-and-testing-center-in-the-us/.

\bibitem{fl}
W.~Hammedi, B.~Brik, and S.~M. Senouci, ``{Federated Deep Learning-Based Framework to Avoid Collisions Between Inland Ships},'' in \emph{2022 International Wireless Communications and Mobile Computing (IWCMC)}, 2022, pp. 967--972.

\bibitem{challenge_fl}
H.~B. McMahan, D.~Ramage, K.~Talwar, and L.~Zhang, ``{Learning Differentially Private Recurrent Language Models},'' in \emph{International Conference on Learning Representations}, 2017.

\bibitem{sec_fl}
K.~Zhang, X.~Song, and C.~e.~a. Zhang, ``{Challenges and future directions of secure federated learning: a survey.}'' \emph{Front. Comput. Sci.}, vol.~16, no.~5, Dec 2022.

\end{thebibliography}

\end{document}